\documentclass{JHEP}
\usepackage{times}
\usepackage{epsfig}
\usepackage{amssymb}

\title{Bits and Pieces in\\ Logarithmic Conformal Field Theory}
\author{Michael Flohr\thanks{Work supported by 
  the DFG String network (SPP no.\ 1096), Fl 259/2-1.} \\
  Institute for Theoretical Physics, University of Hannover \\
  Appelstra\ss e 2, D-30167 Hannover, Germany \\
  E-mail: \email{flohr@itp.uni-hannover.de}}
\abstract{These are notes of my lectures held at the first
  {\it School \& Workshop on Logarithmic Conformal Field Theory and
  its Applications}, September 2001 in Tehran, Iran.\\
  These notes cover only selected parts of the by now quite extensive
  knowledge on logarithmic conformal field theories. In particular,
  I discuss the proper generalization of null vectors towards the
  logarithmic case, and how these can be used to compute correlation 
  functions. My other main topic is modular invariance,
  where I discuss the problem of the generalization of characters in
  the case of indecomposable representations, a proposal for a
  Verlinde formula for fusion rules and identities relating the partition
  functions of logarithmic conformal field theories to such of well known
  ordinary conformal field theories.\\
  These two main topics are complemented by some remarks on ghost systems,
  the Haldane-Rezayi fractional quantum Hall state, and the relation of
  these two to the logarithmic $c=-2$ theory.}
\preprint{hep-th/0111228}

\newcommand{\be}{\begin{equation}}
\newcommand{\ee}{\end{equation}}
\newcommand{\bea}{\begin{eqnarray}}
\newcommand{\eea}{\end{eqnarray}}
\newenvironment{qq}{\footnotesize%
  \list{}{%
    \listparindent 1cm%
    \leftmargin   -1cm%
    \rightmargin\leftmargin%
    \baselineskip=10pt}%
  \item\relax}
{\endlist}
\newcommand{\bq}{\begin{qq}}
\newcommand{\eq}{\end{qq}}
\newcommand{\WEPSFIGURE}[4][v]{\begin{floatingfigure}[#1]\centerline{%
       \noindent\parbox{.55\textwidth}{\centerline{\epsfig{file=#2}}}~~~
                \parbox{.55\textwidth}{\centerline{\epsfig{file=#3}}}}
                \caption{#4}\end{floatingfigure}}

\def\vec#1{{\rm\bf#1}}
\def\bra{\langle}
\def\ket{\rangle}
\def\vak#1{|{\bf#1}\rangle}

\def\vac#1{|{#1}\rangle}
\def\Vac#1{\left|{#1}\right\rangle}

\def\avac#1{\langle{#1}|}
\def\Avac#1{\left\langle{#1}\right|}
\def\vev#1#2{\langle{#1}|{#2}\rangle}

\def\VEV#1{\left\langle{#1}\right\rangle}
\def\nop#1{\mbox{:$#1$:}}
\def\w{{\cal W}} 
\def\del{\partial}

\def\ds{\displaystyle}
\def\ts{\textstyle}
\def\fr#1#2{{\ts\frac{#1}{#2}}} 

\def\pp{\phantom{00}}

\newcommand{\BN}{{\mathbb{N}}}

\newcommand{\BR}{{\mathbb{R}}}
\newcommand{\BZ}{{\mathbb{Z}}}
\newcommand{\Bid}{1\!{\rm l}}
\def\euf#1{\mathfrak{#1}}
\def\EH{{\euf H}}
\def\EF{{\euf F}}
 \newcommand\epl[3]   {{{\it Europhys.\ Lett.\ }{\bf C#1} (#2) #3}}
 \newcommand\jpha[3]  {{{\it J.Phys.\ }{\bf A #1} (#2) #3}}
 \newcommand\grg[3]   {{{\it Gen. Rel. Grav.\ }{\bf #1} (#2) #3}}
 \newcommand\faa[3]   {{{\it Funkt.\ Anal.\ Appl.\ }{\bf #1} (#2) #3}}
 \newcommand{\mathQA}[1]{{\tt math.QA/#1}}
\newlength\mylength
\begin{document}

\section{Introduction}\setlength\parskip\mylength

  These are notes of my lectures held at the first
  {\it School \& Workshop on Logarithmic Conformal Field Theory and
  its Applications}, which took place at the IPM (Institute for Studies in
  Theoretical Physics and Mathematics) in Tehran, Iran,
  4.-18.\ September 2001.
 
  During the last few years,
  so-called logarithmic conformal field
  theory (LCFT) established itself as a well-defined new animal in
  the zoo of conformal field theories in two dimensions. These are conformal
  field theories where, despite scaling invariance, correlation function
  might exhibit logarithmic divergences.
  To our knowledge, such
  logarithmic singularities in correlation functions were first noted by
  Knizhnik back in 1987 \cite{Knizhnik:1987}, but since LCFT had not been
  invented (or found) then, he had to discuss them away. 
  The first works we are aware of, which made a clear connection between 
  logarithms in correlation functions, indecomposability of representations
  and operator product expansions 
  containing logarithmic fields (although they were not called that way then),
  are three papers by Saleur, and then Rozansky and Saleur,
  \cite{Saleur:1991,Rozansky:1992}. But it took six years since Knizhnik's 
  publication, that the concept of a conformal field theory with logarithmic 
  divergent behavior due to logarithmic operators was considered in its own
  right by Gurarie \cite{Gurarie:1993}, who got interested in this
  matter by discussions with A.B.\ Zamolodchikov. From then one,
  there has been a considerable amount of work on analyzing the general
  structure of LCFTs, which by now has generalized almost all of the
  basic notions and tools of (rational) conformal field theories, such
  as null vectors, characters, partition functions, fusion rules,
  modular invariance etc.,
  to the logarithmic case. A complete list of references is already
  too long even for lectures notes, but see for example
  \cite{Flohr:1996a,
    Eholzer:1997,
    Gaberdiel:1996a,
    Ghezelbash:1997,
    Giribet:2001,
    Kausch:1995,
    Khorrami:1998a,
    Kogan:1998a,
    Mavromatos:1998,
    MoghimiAraghi:2000a,
    RahimiTabar:1997b,
    RahimiTabar:1998a,
    Rohsiepe:1996} and
  references therein. Besides the best understood main example of the
  logarithmic theory with central charge $c=-2$, as well as its $c_{p,1}$ 
  relatives, other specific
  models were considered such as WZW models \cite{Bernard:1997,
    Gaberdiel:2001,Kogan:1997,Nichols:2001a,Nichols:2001b} and LCFTs related
  to supergroups and supersymmetry \cite{Bhaseen:2001,Caux:1997,
    Kheirandish:2001a,Khorrami:1998b,Kogan:2001b,Ludwig:2000,
    Read:2001,Rozansky:1992}. Strikingly, Rozansky and Saleur did note
  that indecomposable representations should play a r\^ole in CFT severely
  influencing the behavior of, for example, the modular $S$- and $T$-matrices,
  before Gurarie published his work in 1993. The only concept they did not 
  explicitly introduce was that of a Jordan cell structure with respect to
  $L_0$ or other generators in the chiral symmetry algebra.
 
  Also, quite a number of applications have already been pursued, and
  LCFTs have emerged in many different areas by now. We will hear about
  some of them in the course of this school. Hence, I mention only some
  of them, which I found particularly exciting.
  Sometimes, longstanding puzzles in the description of certain
  theoretical models could be resolved, e.g.\ the enigmatic degeneracy of
  the ground state in the Haldane-Rezayi 
  fractional quantum Hall effect with filling factor $\nu=5/2$, where 
  conformal field theory descriptions of the bulk theory proved 
  difficult \cite{Cappelli:1998,Gurarie:1997,
  Read:1999},
  multi-fractality in disordered Dirac fermions, where the spectra did not
  add up correctly as long as logarithmic fields in internal channels
  were neglected \cite{Caux:1998b}, or two-dimensional conformal turbulence,
  where Polyakov's proposal of a conformal field theory solution did
  contradict phenomenological expectations on the energy spectrum 
  \cite{Flohr:1996c,RahimiTabar:1997a,Skoulakis:1998}.
  Other applications worth mentioning are
  gravitational dressing \cite{Bilal:1994a},
  polymers and Abelian sandpiles \cite{Cardy:1999,Ivashkevich:1998,Mahieu:2001,
    Saleur:1991}, the (fractional) quantum Hall effect \cite{Flohr:1996b,
    Ino:1997,Kogan:2000a}, and -- perhaps most importantly -- disorder
  \cite{Bhaseen:2000a,Bhaseen:2000b,Caux:1996,Caux:1998a,
    Gurarie:1998,Gurarie:1999,Kogan:1996b,Maassarani:1997,RahimiTabar:1998b}.
  Finally, there are even applications in string theory \cite{Kogan:1996a},
  especially in
  $D$-brane recoil \cite{CampbellSmith:2000a,Ellis:1996,Ellis:1999,
    Gravanis:2001,Kogan:1996c,Leontaris:1999,Lewis:2000a,Mavromatos:1999a},
  AdS/CFT correspondence \cite{Ghezelbash:1999,Kaviani:1999,Kim:1998,
    Kogan:1999,Kogan:2000c,MoghimiAraghi:2001b,Myung:1999,Sanjay:2000},
  and also in Seiberg-Witten solutions to supersymmetric Yang-Mills theories,
  e.g.\ \cite{Cappelli,Flohr:1998b,Lerche:1998},
  Last, but not least, a recent focus of research on LCFTs is in its
  boundary conformal field theory aspects \cite{Ishimoto:2001,Kawai:2001,
    Kogan:2000b,Lewis:2000b,MoghimiAraghi:2000a}.

  In these note, we will not cover any of the applications, and we will only
  discuss some of the general issues in LCFT. We will focus mainly on
  two issues in particular. Firstly, we discuss 
  so-called null states, and how these can help to compute correlation
  functions in LCFTs. Secondly, we look at modular invariance, whether and
  how it can be ensured in LCFTs, and what consequences it has on the
  operator algebra. More precisely, we discuss the problem of the
  generalization of characters in the case of indecomposable representations,
  a proposal for a
  Verlinde formula for fusion rules and identities relating the partition
  functions of logarithmic conformal field theories to such of well known
  ordinary conformal field theories.

  As already said, these notes cover only selected parts of the by now 
  quite extensive knowledge on logarithmic conformal field theories.
  On the other hand, 
  we have tried to make these notes rather self-contained,
  which means that some parts may overlap with other lecture notes for
  this school, and are included here for convenience. In particular, we
  did not assume any deeper knowledge of generic common conformal field
  theory. 

  \bq
  Some parts are set in smaller type, like the paragraph you are just 
  reading. They mostly contain more
  advanced material and further details which may be skipped upon first
  reading. Some of these parts, however, contain additional explanations
  addressed to a reader who is a novice to the vast theme of CFT in general,
  and may be skipped by readers already familiar with basic conformal field
  theory techniques.
  \eq
 
  \noindent For 
  those readers completely unfamiliar with CFT in general, we provide
  a (very) short list of introductory material, for their convenience which,
  however, is by no means complete.
  The reviews on string theory which we included in the list contain, 
  in our opinion, quite suitable introductions to certain aspects of conformal
  field theory.
  
  \begin{list}{}{\setlength{\parsep}{0pt}\setlength{\itemsep}{0pt}}
  \item[(1)] L.\ Alvarez-Gaum\'e, 
    {\it Helv.\ Phys.\ Acta\ }{\bf 61} (1991) 359-526.
  \item[(2)] J.\ Cardy, 
    in {\it Les Houches 1988 Summer School}, E.\ Br\'ezin and 
    J.\ Zinn-Justin, eds.\ (1989) Elsevier, Amsterdam.
  \item[(3)] Ph.\ Di~Francesco, P.\ Mathieu, D.\ S\'en\'echal, 
    {\em Conformal Field Theory},
    Graduate Texts in Contemporary Physics (1997) Springer.
  \item[(4)] R.\ Dijkgraaf, 
    {\em Les Houches Lectures on Fields, Strings and Duality},
    to appear [\hepth{9703136}].
  \item[(5)] J.\ Fuchs, 
    {\em Lectures on conformal field theory and Kac-Moody algebras}, 
    to appear in {\it Lecture Notes in Physics}, Springer [\hepth{9702194}].
  \item[(6)] M.\ Gaberdiel, 
    {\it Rept.\ Prog.\ Phys.\ }{\bf 63} (2000) 607-667
    [\hepth{9910156}].
  \item[(7)] P.\ Ginsparg, 
    in {\it Les Houches 1988 Summer School}, E.\ Br\'ezin and 
    J.\ Zinn-Justin, eds.\ (1989) Elsevier, Amsterdam
    [http://xxx.lanl.gov/hypertex/hyperlh88.tar.gz].
  \item[(8)] C.\ Gomez, M.\ Ruiz-Altaba, 
    {\it Rivista Del Nuovo Cimento\ }{\bf 16} (1993) 1--124.
  \item[(9)] M.\ Green, J.\ Schwarz, E.\ Witten, 
    {\em String Theory}, vols.\ 1,2 (1986) Cambridge University Press.
  \item[(10)] M.\ Kaku, 
    {\em String Theory\/} (1988) Springer.
  \item[(11)] S.V.\ Ketov, 
    {\em Conformal Field Theory\/} (1995) World Scientific.
  \item[(12)] D.\ L\"ust, S.\ Theisen, 
    {\em Lectures on String Theory}, Lecture Notes in Physics (1989) Springer.
  \item[(13)] A.N.\ Schellekens 
    {\em Conformal Field Theory}, 
    Saalburg Summer School lectures (1995)
 [http://www.itp.uni-hannover.de/\~{}flohr/lectures/schellekens.cft-lectures.ps.gz].
  \item[(14)]
  C.\ Schweigert, J.\ Fuchs, J.\ Walcher, {\em Conformal field theory, 
    boundary conditions and applications to string theory}
    [\hepth{0011109}].
  \item[(15)]
  A.B.\ Zamolodchikov, Al.B.\ Zamolodchikov, {\em Conformal Field Theory and 
    Critical Phenomena in Two-Dimensional Systems}, Soviet Scientific
    Reviews/Sec.\ A/Phys.\ Reviews (1989) Harwood Academic Publishers.
\end{list}

\section{CFT proper}

In these notes, we will detach ourselves from any 
string theoretic or condensed matter application motivations and
consider CFT solely on its own. This section is a very rudimentary summary of
some CFT basics. As mentioned in the basic CFT lectures,
it is customary to work on the
complex plane (or Riemann sphere) with the holomorphic coordinate $z$ and
the holomorphic differential or one-form ${\rm d}z$. A field $\Phi(z)$ is
called a {\em conformal\/} or {\em primary\/} field of {\em weight\/} $h$, if
it transforms under holomorphic mappings $z \mapsto z'(z)$ of the coordinate
as
\be
  \Phi_h(z)({\rm d}z)^h \mapsto \Phi_h(z')({\rm d}z')^h 
  = \Phi_h(z)({\rm d}z)^h\,.
\ee
In case that the conformal weight $h$ is not a (half-)integer, it is
better to write this as
\be
  \Phi_h(z) \mapsto \Phi_h(z') = \Phi_h(z)
  \left(\frac{\partial z'(z)}{\partial z}\right)^{-h}\,.
\ee
One should keep in mind that all formul\ae\ here have an anti-holomorphic
counterpart. Since a primary field factorizes into
holomorphic and anti-holomorphic parts, $\Phi_{h,\bar h}(z,\bar z) =
\Phi_h(z)\Phi_{\bar h}(\bar z)$, in most cases, we can skip half of the
story.
Infinitesimally, if $z'(z) = z + \varepsilon(z)$ with $\bar{\partial}
\varepsilon=0$, the transformation of the field is
\be
  \Phi_h(z')({\rm d}z')^h = \left(\Phi_h(z) + \varepsilon(z)\partial_z
    \Phi_h(z) + \ldots\right)({\rm d}z)^h\left(1+\partial_z
    \varepsilon(z)\right)^h\,.
\ee
Therefore, the variation of the field with respect to a holomorphic
coordinate transformation is
\be
  \delta\Phi_h(z) = \left(\varepsilon(z)\partial + h(\partial\varepsilon(z))
                    \right)\Phi_h(z)\,.
\ee
Since this transformation is supposed to be holomorphic in $\mathbb{C}^*$,
it can be expanded as a Laurent series,
\be
  \varepsilon(z) = \sum_{n\in\mathbb{Z}}\varepsilon_n z^{n+1}\,.
\ee
This suggests to take the set of infinitesimal transformations
$z \mapsto z'=z+\varepsilon_n z^{n+1}$ as a basis from which we find the
generators of this reparametrization symmetry by considering 
$\Phi_h \mapsto \Phi_h + \delta_n\Phi_h$ with
\be\label{eq:primary-trf}
  \delta_n\Phi_h(z) = \left(z^{n+1}\partial + h(n+1)z^n\right)\Phi_h(z)\,.
\ee
The generators are thus the generators of the already encountered
Witt-algebra $[\ell_n,\ell_m]=(n-m)\ell_{n+m}$, namely 
$\ell_n = -z^{n1+}\partial$. 

We are interested in a quantized theory such that conformal fields become
operator valued distributions in some Hilbert space ${\cal H}$. We therefore
seek a representation of $\ell_n\in{\it Diff}(S^1)$ by some operators
$L_n\in{\cal H}$ such that
\be
  \delta_n\Phi_h(z) = [L_n,\Phi_h(z)]\,.
\ee
We have learned this in the basic CFT lectures, where we discovered the
Virasoro algebra 
\be\label{eq:vir}
  {}[L_n,L_m] = (n-m)L_{n+m} + 
  \frac{\hat c}{12}(n^3-n)\delta_{n+m,0}\,.
\ee
We remark that
$\mathfrak{sl}(2)$ is a sub-algebra of ${\it Diff}(S^1)$ which is independent 
of the central charge $c$. So, we start with considering the consequences
of just $SL(2,\mathbb{C})$ invariance on correlation functions of primary
conformal fields of the form
\be
  G(z_1,\ldots,z_N) = \bra 0|\Phi_{h_N}(z_N)\ldots\Phi_{h_1}(z_1)|0\ket\,.
\ee
We immediately can read off the effect on primary fields from 
(\ref{eq:primary-trf}), which is $\delta_{-1}\Phi_h(z)=\partial\Phi_h(z)$,
$\delta_0\Phi_h(z)=(z\partial + h)\Phi_h(z)$, and $\delta_1\Phi_h(z)=
(z^2\partial+2hz)\Phi_h(z)$.

\subsection{Conformal Ward identities}

Global conformal invariance of correlation functions is equivalent to
the statement that $\delta_iG(z_1,\ldots,z_N)=0$ for $i\in\{-1,0,1\}$. Since
$\delta_i$ acts as a (Lie-) derivative, we find the following differential
equations for correlation functions $G(\{z_i\})$,
\be\label{eq:ward}
  \left\{\begin{array}{l}
  0=\sum_{i=1}^N\partial_{z_i}G(z_1,\ldots,z_N)\,,\\
  0=\sum_{i=1}^N(z\partial_{z_i}+h_i)G(z_1,\ldots,z_N)\,,\\
  0=\sum_{i=1}^N(z^2\partial_{z_i}+2h_iz_i)G(z_1,\ldots,z_N)\,,\\
\end{array}\right.
\ee
which are the so-called {\em conformal Ward identities}.
The general solution to these three equations is
\be\label{eq:cft-corr}
  \bra 0|\Phi_{h_N}(z_N)\ldots\Phi_{h_1}(z_1)|0\ket =
  F(\{\eta_k\})\prod_{i>j}(z_i-z_j)^{\mu_{ij}}\,,
\ee
where the exponents $\mu_{ij}=\mu_{ji}$ must satisfy the conditions
\be
  \sum_{j\neq i}\mu_{ij} = -2h_i\,,
\ee
and where $F(\{\eta_k\})$ is an arbitrary function of any set of $N-3$ 
independent harmonic ratios (a.k.a.\ crossing ratios), for example
\be
  \eta_k = \frac{(z_1-z_k)(z_{N-1}-z_N)}{(z_k-z_N)(z_1-z_{N-1})}\,,\ \ \ \
  k = 2,\ldots N-2\,.
\ee
The above choice is conventional, and maps $z_1\mapsto 0$, $z_{N-1}\mapsto 1$,
and $z_N \mapsto\infty$.
This remaining function cannot be further determined, because the harmonic
ratios are already $SL(2,\mathbb{C})$ invariant, and therefore any function
of them is too.
This confirms that $\mathfrak{sl}(2)$ invariance allows us to fix (only)
three of the variables arbitrarily.

Let us rewrite the conformal Ward identities (\ref{eq:ward}) as
\bea\label{eq:ward2}
  0 &=& \bra(\delta_i\Phi_{h_N}(z_N))\Phi_{h_{n-1}}(z_{N-1})\ldots
        \Phi_{h_1}(z_1)\ket
      + \bra(\Phi_{h_N}(z_N)(\delta_i\Phi_{h_{n-1}}(z_{N-1}))\ldots
        \Phi_{h_1}(z_1)\ket\nonumber\\
      & &\mbox{} + \ldots + \bra(\Phi_{h_N}(z_N)\Phi_{h_{n-1}}(z_{N-1})
                            (\delta_i\Phi_{h_1}(z_1))\ket\,,
\eea
where $\delta_i\Phi_h(z)=[L_i,\Phi_h(z)]$ for $i\in\{-1,0,1\}$. We assume
that the in-vacuum is $SL(2,\mathbb{C})$ invariant, i.e.\ that
$L_i|0\ket=0$ for $i\in\{-1,0,1\}$. Then (\ref{eq:ward2}) is nothing else
than $\bra 0|L_i\left(\Phi_{h_N}(z_N)\ldots\Phi_{h_1}(z_1)\right)|0\ket$ from
which it follows that $\bra 0|L_i$ must be states orthogonal to (and hence
decoupled from) any other state in the theory for $i\in\{-1,0,1\}$.

In a well-defined quantum field theory, we have an isomorphism between the
fields in the theory and states in the Hilbert space ${\cal H}$. This
isomorphism is particularly simple in CFT and induced by
\be
  \lim_{z\rightarrow 0}\Phi_h(z)|0\ket = |h\ket\,,
\ee
where $|h\ket$ is a highest-weight state of the Virasoro algebra. Indeed,
since $[L_n,\Phi_h]=(z^{n+1}\partial + h(n+1)z^n)\Phi_h$, we find with
the highest-weight property of the vacuum $|0\ket$, i.e.\ that
$L_n|0\ket = 0$ for all $n\geq -1$,
that for all $n>0$
\be
  L_n|h\ket = \lim_{z\rightarrow 0}L_n\Phi_h(z)|0\ket =
  \lim_{z\rightarrow 0}[L_n,\!\Phi_h(z)]|0\ket = 
  \lim_{z\rightarrow 0}\left(z^{n+1}\partial + 
  (n\!+\!1)hz^n\right)\!\Phi_h(z)|0\ket
  = 0\,.
\ee
Furthermore, $L_0|h\ket = h|h\ket$ by the same consideration.
Thus, primary fields correspond to highest-weight states. 

\bq
A nice exercise is to apply the conformal Ward identities to a two-point
function $G=\bra\Phi_h(z)\Phi_{h'}(w)\ket$. The constraint from $L_{-1}$ is
that $(\partial_z+\partial_w)G=0$, meaning that $G=f(z-w)$ is a function
of the distance only. The $L_0$ constraint then yields a linear ordinary
differential equation, $((z-w)\partial_{z-w}+(h+h'))f(z-w)=0$, which is
solved by ${\it const}\cdot (z-w)^{-h-h'}$. 

Finally, the $L_1$ constraint yields the condition $h=h'$. However,
we should be careful here, since this does not necessarily imply that
the two fields have to be identical. Only their conformal weights
have to coincide. In fact, we will encounter examples where the propagator
$\bra h|h'\ket = \lim_{z\rightarrow\infty}\bra 0|z^{2h}\Phi_h(z)
\Phi_{h'}(0)|0\ket$ is not diagonal.
Therefore, if more than one field of conformal weight $h$ exists, the
two-point functions aquire the form $\bra\Phi^{(i)}_{h^{}}(z)\Phi^{(j)}_{h'}(w)
\ket = (z-w)^{-2h}\delta_{h,h'}D_{ij}$ with $D_{ij}=\bra h;i|h;j\ket$ the 
propagator matrix.
The matrix $D_{ij}$ then induces a metric on the space of fields. In the
following, we will assume that $D_{ij}=\delta_{ij}$ except otherwise
stated.
\eq

It is worth noting that the conformal Ward identities (\ref{eq:ward}) allow
us to fix the two- and three-point functions completely upto constants.
In fact, the two-point functions are simply given by
\be\label{eq:2pt}
  \bra\Phi_{h^{}}(z)\Phi_{h'}(w)\ket = \frac{\delta_{h^{},h'}}{(z-w)^{2h}}\,,
\ee
where we have taken the freedom to fix the normalization of our primary 
fields. The three-point functions turn out to be
\be\label{eq:3pt}
  \bra\Phi_{h_i}(z_i)\Phi_{h_j}(z_j)\Phi_{h_k}(z_k)\ket =
  \frac{C_{ijk}}{(z_{ij})^{h_i+h_j-h_k}(z_{ik})^{h_i+h_k-h_j}
  (z_{jk})^{h_j+h_k-h_i}}\,,
\ee
where we again used the abbreviation $z_{ij}=z_i-z_j$. The constants
$C_{ijk}$ are not fixed by $SL(2,\mathbb{C})$ invariance and are called
the {\em structure constants\/} of the CFT. Finally, the four-point function
is determined upto an arbitrary function of one crossing ratio, usually
chosen as $\eta=(z_{12}z_{34})/(z_{24}z_{13})$. The solution for $\mu_{ij}$
is no longer unique for $N\geq 4$, and the customary one for $N=4$ is
$\mu_{ij}=H/3-h_i-h_j$ with $H=\sum_{i=1}^4h_i$, such that the four-point
functions reads
\be\label{eq:4pt}
  \bra\Phi_{h_4}(z_4)\Phi_{h_3}(z_3)\Phi_{h_2}(z_2)\Phi_{h_1}(z_1)\ket =
  \prod_{i>j}(z_{ij})^{H/3-h_i-h_j}F(\frac{z_{12}z_{34}}{z_{24}z_{13}})\,.
\ee
Note again that $SL(2,\mathbb{C})$ invariance cannot tell us anything
about the function $F(\eta)$, since $\eta$ is invariant under M\"obius
transformations.

\subsection{Virasoro representation theory: Verma modules}

We already encountered highest-weight states, which are the states corresponding
to primary fields. On each such highest-weight state we can construct a
{\em Verma module\/} $V_{h,c}$ with respect to the Virasoro algebra
${\it Vir}$ by applying the negative modes $L_n$, $n<0$ to it. Such states
are called {\em descendant\/} states.
In this way
our Hilbert space decomposes as
\be\begin{array}{rcl}
  {\cal H}&=&\bigoplus_{h,\bar h} V_{h,c}\otimes V_{\bar h,c}\,,\\
  V_{h,c}&=&{\rm span}\left\{(\prod_{i\in I} L_{-n_i}|h\ket
             : \mathbb{N}\supset I=\{n_1,\ldots n_k\}, n_{i+1}\geq n_i\right\}
  \,,
  \end{array}
\ee
where we momentarily have sketched the fact that the full CFT has a 
holomorphic and an anti-holomorphic part. Note also, that we indicate the
value for the central charge in the Verma modules. We have so far chosen
the anti-holomorphic part of the CFT to be simply a copy of the
holomorphic part, which guarantees the full theory to be local. However,
this is not the only consistent choice, and heterotic strings are an
example where left and right chiral CFT definitely are very much different
from each other.

A way of counting the number of states in $V_{h,c}$ is to introduce
the {\em character\/} of the Virasoro algebra, which is a formal power
series
\be\label{eq:vir-char}
  \chi_{h,c}(q) = {\rm tr}_{V_{h,c}} q^{L_0 -c/24}\,.
\ee
For the moment, we consider $q$ to be a formal variable, but we will later
interpret it in physical terms, where it will be defined by $q={\rm e}^{2\pi
{\rm i}\tau}$ with a complex parameter $\tau$ living in the upper half
plane, i.e.\ $\Im{\rm m}\,\tau>0$. The meaning of the constant term $-c/24$ will
also become clear further ahead.

The Verma module possesses a natural gradation in terms of the eigen value
of $L_0$, which for any descendant state $L_{-\vec{n}}|h\ket\equiv 
L_{-n_1}\ldots L_{-n_k}
|h\ket$ is given by $L_0L_{-\vec{n}}|h\ket=(h+|\vec{n}|)|h\ket\equiv
(h+n_1+\ldots+n_k)|h\ket$. One calls $|\vec{n}|$ the level of the descendant
$L_{-\vec{n}}|h\ket$.
The first descendant states in $V_{h,c}$ are easily found. At level zero,
there exists of course only the highest-weight state itself, $|h\ket$.
At level one, we only have one state, $L_{-1}|h\ket$. At level two, we
find two states, $L_{-1}^2|h\ket$ and $L_{-2}|h\ket$. In general, we have
\be\begin{array}{rcl}
  V_{h,c}&=&\bigoplus_N V_{h,c}^{(N)}\,,\\
  V_{h,c}^{(N)}&=&{\rm span}\left\{L_{-\vec{n}}|h\ket : |\vec{n}|=N\right\}\,,
  \end{array}
\ee
i.e.\ at each level $N$ we generically have $p(N)$ linearly independent
descendants, where $p(N)$ denotes the number of partitions of $N$ into
positive integers. If all these states are physical, i.e.\ do not decouple
from the spectrum, we easily can write down the character of this
highest-weight representation,
\be\label{eq:part-gen}
  \chi_{h,c}(q) = q^{h-c/24}\prod_{n\geq 1}\frac{1}{1-q^n}\,.
\ee
To see this, the reader should make herself clear that we may act on $|h\ket$
with any power of $L_{-m^{}}$ independently of the powers of any other mode
$L_{-m'}$, quite similar to a Fock space of harmonic oscillators.
A closer look reveals that (\ref{eq:vir-char}) is
indeed formally equivalent to the partition function of an infinite number
of oscillators with energies $E_n=n$. The expression (\ref{eq:part-gen})
contains the generating function for the numbers of partitions, since
expanding it in a power series yields
\bea
  & & \prod_{n\geq 1}(1-q^n)^{-1} = \sum_{N\geq 0} p(N)q^N \\
  &=& 1 + q + 2q^2 + 3q^3 + 5q^4 +7q^5 + 11q^6 + 15q^7 + 22q^8
     + 30q^9 + 42q^{10} + \ldots\nonumber\,.
\eea

\subsection{Virasoro representation theory: Null vectors}

The above considerations are true in the generic case. But if we start to
fix our CFT by a choice of the central charge $c$, we have to be careful
about the question whether all the states are really linearly independent.
In other words: May it happen that for a given level $N$ a particular linear 
combination
\be
  |\chi^{(N)}_{h,c}\ket = \sum_{|\vec{n}|=N}\beta^{\vec{n}}L_{-\vec{n}}|h\ket 
  \equiv 0\,?
\ee
With this we mean that $\bra\psi|\chi^{(N)}_{h,c}\ket=0$ for all $|\psi\ket\in
{\cal H}$. To be precise, this statement assumes that our space of states
admits a sesqui-linear form $\bra.|.\ket$. In most CFTs, this is the case,
since we can define asymptotic out-states by
\be\label{eq:out-state}
  \bra h| \equiv \lim_{z\rightarrow\infty} \bra 0|\Phi_h(z)z^{2h}\,.
\ee
This definition is forced by the requirement to be compatible with 
$SL(2,\mathbb{C})$ invariance of the two-point function (\ref{eq:2pt}).
We then have $\bra h'|h\ket = \delta_{h',h^{}}$. The exponent $z^{2h}$ arises
due to the conformal transformation $z\mapsto z'=1/z$ we implicitly have
used. We further assume the hermiticity condition $L_{-n}^{\dag}=L_{n}$ to
hold.

\bq
The hermiticity condition is certainly fulfilled for unitary theories.
We already know from the calculation of the two-point function of the
stress-energy tensor,
$\bra T(z)T(w)\ket = \frac12c(z-w)^{-4}$, 
that necessarily $c\geq 0$ for unitary theories.
Otherwise, $\|L_{-n}|0\ket\|^2 = \bra 0|L_nL_{-n}|0\ket =
\bra 0|[L_n,L_{-n}]|0\ket = \frac{1}{12}c(n^3-n)\bra 0|0\ket$ would be
negative for $n \geq 2$. Moreover, redoing the same calculation for the
highest-weight state $|h\ket$ instead of $|0\ket$, we find
$\|L_{-n}|h\ket\|^2 = \left(\frac{1}{12}c(n^3-n) + 2nh\right)\bra h|h\ket$.
The first term dominates for large $n$ such that again $c$ must be non-negative,
if this norm should be positive definite. The second term dominates for $n=1$,
from which we learn that $h$ must be non-negative, too. To summarize,
unitary CFTs necessarily require $c\geq 0$ and $h\geq 0$, where the theory
is trivial for $c=0$ and where $h=0$ implies that $|h=0\ket=|0\ket$ is the
(unique) vacuum.
\eq

To answer the above question, we consider the $p(N)\times p(N)$ matrix 
$K^{(N)}$ of all possible scalar products $K^{(N)}_{\vec{n}',\vec{n}^{}}=
\bra h|L_{\vec{n}'}L_{-\vec{n}^{}}|h\ket$. This matrix is hermitian by
definition. If this matrix has a vanishing or negative determinant, then
it must possess an eigen vector (i.e.\ a linear combination of level
$N$ descendants) with zero or negative norm, respectively.
The converse is not necessarily true, such that a positive determinant
could still mean the presence of an even number of negative eigen values.
For $N=1$, this reduces to the simple statement ${\rm det}K^{(1)} =
\bra h|L_1L_{-1}|h\ket = \|L_{-1}|h\ket\|^2 = \bra h|2L_0|h\ket = 
2h\bra h|h\ket = 2h$, where we used the Virasoro algebra (\ref{eq:vir}). 
Thus, there exists a null vector at level $N=1$ only
for the vacuum highest-weight representation $h=0$.

We note a view points concerning the general case. Firstly, due to the
assumption that all highest-weight states are unique (i.e.\ $\bra h'|h\ket
=\delta_{h',h^{}}$), it follows that it suffices to analyze the matrix
$K^{(N)}$ in order to find conditions for the presence of null states.
Note that scalar products $\bra h|L_{\vec{n}'}L_{-\vec{n}^{}}|h\ket$ are
automatically zero for $|\vec{n}'|-|\vec{n}|\neq 0$ due to the highest-weight
property. Secondly, using the Virasoro algebra, each
matrix element can be reduced to a polynomial function of $h$ and $c$. This
must be so, since the total level of the descendant 
$L_{\vec{n}'}L_{-\vec{n}^{}}|h\ket$ is zero such that use of the 
Virasoro algebra
allows to reduce it to a polynomial $p_{\vec{n}',\vec{n}^{}}(L_0,\hat c)|h\ket$.
It follows that $K^{(N)}_{\vec{n}',\vec{n}^{}}=p_{\vec{n}',\vec{n}^{}}(h,c)$.

\bq
It is an extremely useful exercise to work out the level $N=2$ case by hand.
Since $p(2)=2$, The matrix $K^{(2)}$ is the $2\times 2$ matrix
\be
  K^{(2)} = \left(\begin{array}{cc}
    \bra h|L_2L_{-2}|h\ket    & \bra h|L_2L_{-1}L_{-1}|h\ket \\
    \bra h|L_1L_1L_{-2}|h\ket & \bra h|L_1L_1L_{-1}L_{-1}|h\ket
  \end{array}\right)\,.
\ee
The Virasoro algebra reduces all the four elements to expressions in $h$ and
$c$. For example, we evaluate $L_1L_1L_{-2}|h\ket = L_1[L_1,L_{-2}]|h\ket =
3L_1L_{-1}|h\ket = 6L_0|h\ket$ etc., such that we arrive at
\be
  K^{(2)} = \left(\begin{array}{cc}
    4h+\frac12c & 6h \\
    6h          & 4h+8h^2
  \end{array}\right)\bra h|h\ket\,.
\ee
For $c,h\gg 1$, the diagonal dominates and the eigen values are hence both
positive. The determinant is 
\be
  {\rm det}K^{(2)} = 2h\left(16h^2+2(c-5)h + c\right)\bra h|h\ket^2\,.
\ee
\eq

\noindent At level $N=2$, there are three values of the highest weight $h$,
\be\label{eq:level2-h}
  h \in \left\{0,{\textstyle\frac{1}{16}}(5-c\pm\sqrt{(c-1)(c-25)})
        \right\}\,,
\ee
where the matrix $K^{(2)}$
develops a zero eigen value. Note that one finds two values $h_{\pm}$ for
each given central charge $c$, besides the value $h=0$ which is a remnant
of the level one null state. The corresponding eigen vector is easily found
and reads
\be\label{eq:level2}
  |\chi^{(2)}_{h_{\pm},c}\ket = \left({\textstyle\frac23}(2h_{\pm}+1)L_{-2}-
  L_{-1}^2\right)|h_{\pm}\ket\,.
\ee

\bq
This can be generalized. The reader might occupy herself some time with
calculating the null states for the next few levels. Luckily, there exist
at least general formul\ae\ for the zeroes of the so-called Kac determinant
${\rm det}K^{(N)}$, which are curves in the $(h,c)$ plane. Reparametrizing
with some hind-sight 
\be\label{eq:cseries}
  c = c(m) = 1 - 6\frac{1}{m(m+1)}\,,\ \ \ {\rm i.e.}\ \ \ 
      m = -{\textstyle\frac12}\left(1\pm\sqrt{\frac{c-25}{c-1}}\right)\,,
\ee
one can show that the vanishing lines are given by
\bea\label{eq:hseries}
  h_{p,q}(c) &=& \frac{\left((m+1)p-mq\right)^2 - 1}{4m(m+1)}\\
             &=& -{\textstyle\frac12}pq+{\textstyle\frac{1}{24}}(c-1)
                 + {\textstyle\frac{1}{48}}\left(
                  (13 - c \mp \sqrt{(c-1)(c-25)})p^2
                 +(13 - c \pm \sqrt{(c-1)(c-25)})q^2\right)\,.\nonumber
\eea 
Note that the two solutions for $m$ lead to the same set of $h$-values,
since $h_{p,q}(m_+(c)) = h_{q,p}(m_-(c))$.
With this notation for the zeroes, the Kac determinant can be written upto
a constant $\alpha_N$ of combinatorial origin as
\be
  {\rm det}K^{(N)} = \alpha_N\prod_{pq\leq N}(h-h_{p,q}(c))^{p(n-pq)}
                   \propto  {\rm det}K^{(N-1)}\prod_{pq=N}(h-h_{p,q}(c))\,,
\ee
where we have set $\bra h|h\ket = 1$, and where $p(n)$ denotes again the
number of partitions of $n$ into positive integers. 

A deeper analysis not only reveals null states, where the scalar product
would be positive semi-definite, but also regions of the $(h,c)$ plane
where negative norm states are present. A physical sensible string theory
should possess a Hilbert space of states, i.e.\ the scalar product should
be positive definite. Therefore, an analysis which regions of the $(h,c)$
plane are free of negative-norm states is a very important issue in 
string theory. As a result, for $0\leq c<1$, only the discrete set of points
given by the values $c(m)$ with $m\in\mathbb{N}$ in (\ref{eq:cseries}) and 
the corresponding values $h_{p,q}(c)$ with $1\leq p<m$ and $1\leq q<m+1$ 
in (\ref{eq:hseries}) turns out to be free of negative-norm states. 
In string theory, one learns that the region
$c\geq 25$ is particularly interesting, and that indeed $c=26$ admits
a positive definite Hilbert space.
\eq

\EPSFIGURE{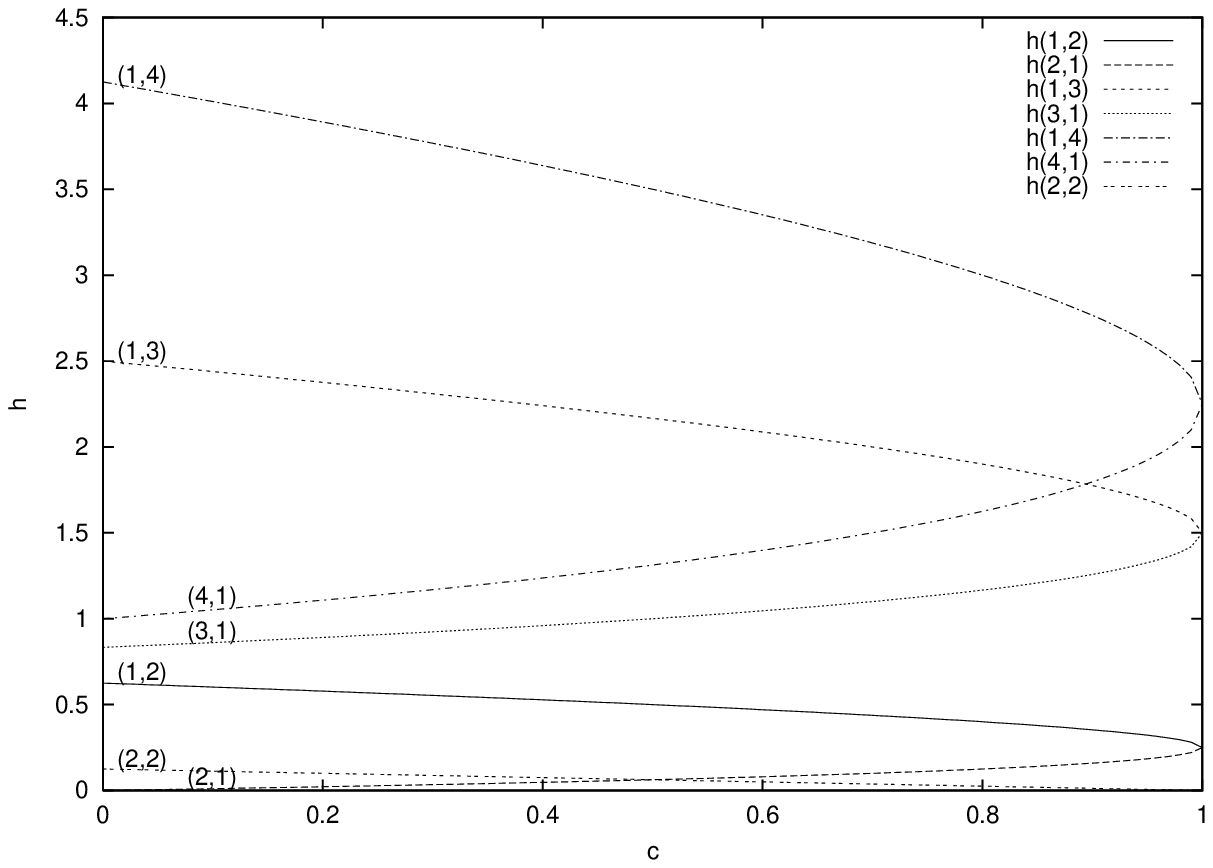,width=10cm}{The first few of the lines $h_{p,q}(c)$
where null states exist. They are also the lines where the Kac determinant
has a zero, indicating a sign change of an eigenvalue.}

To complete our brief discussion of Virasoro representation theory, we
note the following: If null states are present in a given Verma module
$V_{h,c}$, they are states which are orthogonal to all other states. It
follows, that they, and all their descendants, decouple from the other
states in the Verma module. Hence, the correct representation module is
the irreducible sub-module with the ideal generated by the null
state divided out, or more precisely, with the maximal proper sub-module
divided out, i.e.
\be
  V_{h_{p,q}(c),c} \longrightarrow M_{h_{p,q}(c),c} = V_{h_{p,q}(c),c}
  \left/{\rm span}\{|\chi_{h_{p,q}(c),c}^{(N)}\ket\equiv 0\}\right.\,,
\ee
or mathematically more rigorously, $M_{h_{p,q}(c),c}$ is the unique 
sub-module such that 
\be
  V^{}_{h_{p,q}(c),c}\longrightarrow M'_{h_{p,q}(c),c}
  \longrightarrow M^{}_{h_{p,q}(c),c}
\ee
is exact for all $M'$. Due to the state-field isomorphism, it is clear 
that this decoupling of states must reflect itself in partial differential 
equations for correlation functions, since descendants of primary fields
are made by acting with modes of the stress energy tensor on them. These
modes, as we have seen, are represented as differential operators. The
precise relationship will be worked out further below. 
Thus, null states provide a very
powerful tool to find further conditions for expectation values. They allow
us to exploit the infinity of local conformal symmetries as well, and under
special circumstances enable us -- at least in principle -- to compute {\em 
all\/} observables of the theory.

\subsection{Descendant fields and operator product expansion}

As we associated to each highest-weight state a primary field, we may
associate to each descendant state a descendant field in the following way:
A descendant is a linear combination of monomials $L_{-n_1}\ldots L_{-n_k}
|h\ket$. We heard in the basic CFT lectures that the modes $L_n$ are 
extracted from the stress-energy tensor via
a contour integration. This suggests to create the descendant
field $\Phi_h^{(-n_1,\ldots,-n_k)}(z)$ by a successive application of
contour integrations
\bea\label{eq:desc}
  & &\Phi_h^{(-n_1,\ldots,-n_k)}(z) = \\    & &
    \oint_{C_1}\frac{{\rm d}w_1}{(w_1-z)^{n_1-1}}T(w_1)
    \oint_{C_2}\frac{{\rm d}w_2}{(w_2-z)^{n_2-1}}T(w_2)\ldots
    \oint_{C_k}\frac{{\rm d}w_k}{(w_k-z)^{n_k-1}}T(w_k)\Phi_h(z)\,,\nonumber
\eea
where from now on we include the prefactors $\frac{1}{2\pi{\rm i}}$ into the
definition of $\oint{\rm d}z$. The contours $C_i$ all encircle $z$ and
$C_{i}$ completely encircles $C_{i+1}$, in short $C_i \succ C_{i+1}$.

\EPSFIGURE{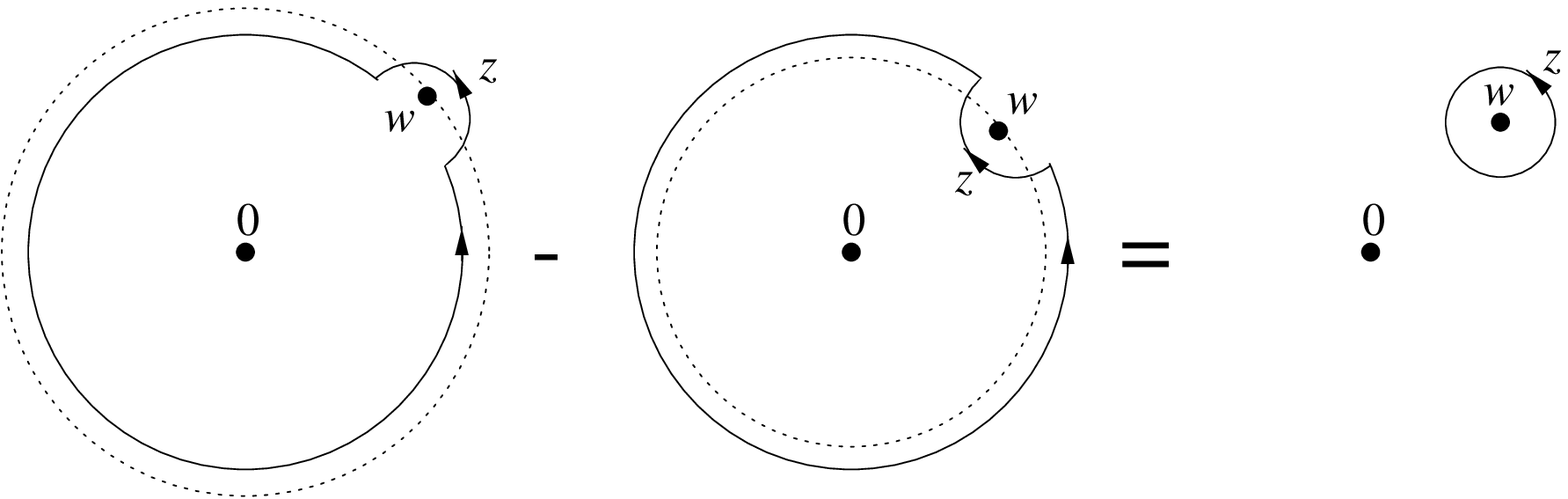,width=6cm}{Typical contour deformation for OPE 
calculations.}
There is only one problem with this definition, namely that it involves
products of operators. In quantum field theory, this is a notoriously
difficult issue. Firstly, operators may not commute, secondly, and more
seriously, products of operators at equal points are not well-defined
unless normal ordered. As we defined (\ref{eq:desc}), we took care to
respect ``time'' ordering, i.e.\ radial ordering on the complex plane.
In order to evaluate equal-time commutators, we define for operators
$A,B$ and arbitrary functions $f,g$ the densities
\be
  A_f = \oint_0{\rm d}z f(z)A(z)\,,\ \ \ \ B_g=\oint_0{\rm d}w g(w)B(w)\,,
\ee
where the contours are circles around the origin with radii $|z|=|w|=1$. Then,
the equal-time commutator of these objects is
\be
  [A_f,B_g]_{{\rm e.t.}} = 
  \oint_{C_1}{\rm d}z f(z)A(z)\oint_{C_2}{\rm d}w g(w)B(w) -
  \oint_{C_2}{\rm d}w g(w)B(w)\oint_{C_1}{\rm d}z f(z)A(z)\,,
\ee
where we took the freedom to deform the contours in a homologous way such that
radial ordering is kept in both terms. As indicated in the figure five, 
both terms together result in the following expression,
\be\label{eq:et-comm}
  [A_f,B_g]_{{\rm e.t.}} = \oint_0{\rm d}w g(w)\oint_w{\rm d}z f(z) A(z)B(w)
\ee
with the contour around $w$ as small as we wish. The inner integration is
thus given by the singularities of the operator product expansion (OPE) of
$A(z)B(w)$. We suppose that products of operators have an asymptotic expansion
for short distances of their arguments. The singular part of this
short-distance expansion determines via contour integration the corresponding
equal-time commutators. For example, with
\be
  T_{\varepsilon} = \oint_0{\rm d}z \varepsilon(z)T(z)\,,
\ee
we recognize immediately $\delta_{\varepsilon}\Phi_h(w) =
(\varepsilon\partial_w + h(\partial_w\varepsilon))\Phi_h(w) =
[T_{\varepsilon},\Phi_h(w)]$. 
Note that this is simply the general version of the common definition of the
Virasoro modes $L_n = \frac{1}{2\pi{\rm i}}\oint_{0}z^{n+1}T(z)$
for $\varepsilon(z)=z^{n+1}$.
If this is to be reproduced by an OPE, it must
be of the form
\be\label{eq:opeTprim}
  T(z)\Phi_h(w) = \frac{h}{(z-w)^2}\Phi_h(w) + \frac{1}{(z-w)}
  \partial_w\Phi_h(w) + {\it regular\ terms}\,.
\ee
To see this, one essentially has to apply Cauchy's integral formula
$\oint{\rm d}zf(z)(z-w)^{-n} = \frac{1}{(n-1)!}\partial^{n-1}f(w)$.
Of course, we may also attempt to find the OPE of the stress-energy tensor
with itself from the Virasoro algebra in the same way,
which yields
\be\label{eq:opeTT}
  T(z)T(w) = \frac{c/2}{(z-w)^4}\mbox{$1\!{\rm l}$} + \frac{2}{(z-w)^2}T(w)
  + \frac{1}{(z-w)}\partial_wT(w) + {\it regular\ terms}\,.
\ee
The reader is encouraged to verify that the above OPE does indeed yield
the Virasoro algebra, if substituted into (\ref{eq:et-comm}).

Note that $T(z)$ is not a proper primary field of weight two due to the
term involving the central charge. Since $T(z)$ behaves as a primary field
under $L_i$, $i\in\{-1,0,1\}$ meaning that it is a weight two tensor with
respect to $SL(2,\mathbb{C})$, it is called quasi-primary. 
One important consequence of this is that the stress-energy tensor on
the complex plane and the original stress energy tensor on the cylinder
differ by a constant term. Indeed, remembering that the transfer from
the complexified cylinder coordinate $w$ to the complex plane coordinate
$z$ was given by the conformal map $z={\rm e}^w$, one obtains
\be
  T_{{\rm cyl}}(w) = z^2T(z) - \frac{c}{24}\mbox{$1\!{\rm l}$}
  \,,\ \ \ {\rm i.e.}\ \ \
  (L_n)_{{\rm cyl}} = L_n - \frac{c}{24}\delta_{n,0}\,.
\ee
This explains the appearance of the factor $-c/24$ in the definition
(\ref{eq:vir-char}) of the Virasoro characters.

The structure of OPEs in CFT is fixed to some degree by two requirements.
Firstly, the OPE is not a commutative product, but it should be associative,
i.e.\ $(A(x)B(y))C(z) = A(x)(B(y)C(z))$.
The motivation for this presumption comes from the duality properties of
string amplitudes. Duality is crossing symmetry in CFT correlation functions,
which can be seen to be equivalent to associativity of the OPE. For
example, one may evaluate a four-point function in several regions, where
different pairs of coordinates are taken close together such that OPEs
can be applied. 
Secondly, the OPE must be consistent with global conformal invariance, i.e.\
it must respect (\ref{eq:2pt}), (\ref{eq:3pt}), and (\ref{eq:4pt}). This
fixes the OPE to be of the following generic form,
\be\label{eq:ope}
   \Phi_{h_i}(z)\Phi_{h_j}(w) = \sum_{k}\frac{C_{ij}^k}{(z-w)^{h_i+h_j-h_k}}
   \Phi_{h_k}(w) + \ldots\,,
\ee
where the structure constants are identical to the structure constants
which appeared in the three-point functions (\ref{eq:3pt}). Note that due
to our normalization of the propagators (two-point functions), raising
and lowering of indices is trivial (unless the two-point functions are
non-trivial, i.e.\ $D_{ij}\neq\delta_{ij}$). 

\bq
We can divide all fields in a CFT into a few classes. First, there are
the primary fields $\Phi_h$ corresponding to highest-weight states
$|h\ket$ and second, there are all their Virasoro descendant fields 
$\Phi_h^{(-\vec{n})}$ corresponding to the descendant states 
$L_{-\vec{n}}|h\ket$
given by (\ref{eq:desc}). For instance, the stress energy tensor itself
is a descendant of the identity, $T(z)=\mbox{$1\!{\rm l}$}^{(-2)}$. 
We further divide descendant fields into two sub-classes, namely fields
which are quasi-primary, and fields which are not. Quasi-primary fields
transform conformally covariant for $SL(2,\mathbb{C})$ transformations
only.

General local conformal transformations are implemented in a correlation 
function by simply inserting the Noether charge, which yields
\be\label{eq:conf-transf}
  \delta_{\varepsilon}\bra 0|\Phi_{h_N}(z_N)\ldots\Phi_{h_1}(z_1)|0\ket
  = \bra 0|\oint{\rm d}z\varepsilon(z)T(z)\Phi_{h_N}(z_N)\ldots\Phi_{h_1}(z_1)
  |0\ket\,,
\ee
where the contour encircles all the coordinates $z_i$, $i=1,\ldots,N$. This
contour can be deformed into the sum of $N$ small contours, each encircling
just one of the coordinates, which is a standard technique in complex
analysis. That is equivalent to rewriting (\ref{eq:conf-transf}) as
\be\label{eq:conf-transf2}
  \sum_i\bra 0|\Phi_{h_N}(z_N)\ldots(\delta_{\varepsilon}\Phi_{h_i}(z_i))\ldots
  \Phi_{h_1}(z_1)|0\ket = \sum_i\bra 0|\Phi_{h_N}(z_N)\ldots\left(
  \oint_{z_i}{\rm d}z\varepsilon(z)T(z)\Phi_{h_i}(z_i)\right)\ldots
  \Phi_{h_1}(z_1)|0\ket\,.
\ee
Since this holds for any $\varepsilon(z)$, we can proceed to a local
version of the equality between the right hand sides of (\ref{eq:conf-transf})
and (\ref{eq:conf-transf2}), yielding 
\be\label{eq:Tward}
  \bra 0|T(z)\Phi_{h_N}(z_N)\ldots\Phi_{h_1}(z_1)|0\ket =
  \sum_i\left(\frac{h_i}{(z-z_i)^2} + \frac{1}{(z-z_i)}\partial_{z_i}\right)
  \bra 0|\Phi_{h_N}(z_N)\ldots\Phi_{h_1}(z_1)|0\ket\,.
\ee

This identity is extremely useful, since it allows us to compute any
correlation function involving descendant fields in terms of the
corresponding correlation function of primary fields.
For the sake of simplicity, let us consider the correlator
$\bra 0|\Phi_{h_N}(z_N)\ldots\Phi_{h_1}(z_1)\Phi_h^{(-k)}(z)|0\ket$ with only
one descendant field involved. Inserting the definition (\ref{eq:desc}) and
using the conformal Ward identity (\ref{eq:Tward}), this gives
\bea
  & & \oint\frac{{\rm d}w}{(w-z)^{k-1}}\\
  &\times& \left[
    \bra 0|T(z)\Phi_{h_N}(z_N)\ldots\Phi_{h_1}(z_1)\Phi_h(z)|0\ket
  - \sum_i\left(\frac{h_i}{(w-z_i)^2} + \frac{1}{(w-z_i)}\partial_{z_i}\right)
    \bra 0|\Phi_{h_N}(z_N)\ldots\Phi_{h_1}(z_1)\Phi_h(z)|0\ket\right]\,.
    \nonumber
\eea
The contour integration in the first term encircles all the coordinates
$z$ and $z_i$, $i=1,\ldots,N$. Since there are no other sources of poles,
we can deform the contour to a circle around infinity by pulling it over 
the Riemann sphere accordingly. The highest-weight property
$\bra 0|L_{k} = 0$ for $k\leq 1$ ensures that the integral around $w=\infty$
vanishes. The other terms are evaluated with the help of Cauchy's formula to
\be
  {\cal L}_{-k}^i \equiv -\oint_{z_i}\frac{{\rm d}w}{(w-z)^{k-1}}\left(
  \frac{h_i}{(w-z_i)^2} + \frac{1}{(w-z_i)}\partial_{z_i}\right)
  = \frac{(k-1)h_i}{(z_i-z)^k} + \frac{1}{(z_i-z)^{k-1}}\partial_{z_i}\,.
\ee
\eq

\noindent Going through the above small-print shows that a correlation
function involving descendant fields can be expressed in terms of the
correlation function of the corresponding primary fields only, on which
explicitly computable partial differential operators act. 
Collecting ${\cal L}_{-k} = \sum_i{\cal L}_{-k}^i$ yields a partial differential
operator (which implicitly depends on $z$) such that
\be\label{eq:pde}
  \bra 0|\Phi_{h_N}(z_N)\ldots\Phi_{h_1}(z_1)\Phi_h^{(-k)}(z)|0\ket
  = {\cal L}_{-k}\bra 0|\Phi_{h_N}(z_N)\ldots\Phi_{h_1}(z_1)\Phi_h(z)|0\ket\,,
\ee
where this operator ${\cal L}_{-k}$ has the explicit form
\be
  {\cal L}_{-k} = \sum_{i=1}^N\left(\frac{(k-1)h_i}{(z_i-z)^k} 
                  + \frac{1}{(z_i-z)^{k-1}}\partial_{z_i}\right)
\ee
for $k > 1$. Due to the global conformal Ward identities, the case
$k=1$ is much simpler, being just the derivative of the primary field,
i.e.\ ${\cal L}_{-1} = \partial_z$.
Thus, correlators involving descendant fields are entirely expressed in
terms of correlators of primary fields only. Once we know the latter, we
can compute all correlation functions of the CFT. 

On the other hand, if we use a descendant, which is a null field, i.e. 
\be
  \chi_{h,c}^{(N)}(z) = \sum_{|\vec{n}|=N} \beta^{\vec{n}}\Phi_h^{(-\vec{n})}(z)
\ee
with $|\chi_{h,c}^{(N)}\ket$ orthogonal to all other states,
we know that it completely decouples from the physical states. Hence, every
correlation function involving $\chi_{h,c}^{(N)}(z)$ must vanish. Hence,
we can turn things around and use this knowledge to find partial differential
equations, which must be satisfied by the correlation function involving
the primary $\Phi_h(z)$ instead. For example, the level $N=2$ null field
yields according to (\ref{eq:level2}) the equation
\be\label{eq:lev2-nullfield}
  \left({\textstyle\frac23}(2h_{\pm}+1){\cal L}_{-2}-\partial_z^2\right)
  \bra 0|\Phi_{h_N}(z_N)\ldots\Phi_{h_1}(z_1)\Phi_{h_{\pm}}(z)|0\ket = 0
\ee
with $h_{\pm}$ given by the non-trivial values in (\ref{eq:level2-h}).

A particular interesting case is the four-point function. The three
global conformal Ward identities (\ref{eq:ward}) then allow us to
express derivatives with respect to $z_1,z_2,z_3$ in terms of derivatives
with respect to $z$. Every new-comer to CFT should once in her life go
through this computation for the level two null field: 
If the field $\Phi_h(z)$ is degenerate of level two, i.e.\ possesses a
null field at level two, we can reduce the partial differential equation
(\ref{eq:lev2-nullfield}) for $G_4 = 
\bra\Phi_{h_3}(z_3)\Phi_{h_2}(z_2)\Phi_{h_1}(z_1)\Phi_{h}(z)\ket$
to an ordinary Riemann differential equation
\bea
  0 &=& \left(\frac{3}{2(2h+1)}\partial_z^2 - \sum_{i=1}^3\left(\frac{h_i}{
  (z-z_i)^2} + \frac{1}{z-z_i}\partial_{z_i}\right)\right)G_4 \\
  &=& \left(\frac{3}{2(2h+1)}\partial_z^2 + \sum_{i=1}^3\left(
  \frac{1}{z-z_i}\partial_z - \frac{h_i}{(z-z_i)^2}\right)
  + \sum_{i<j}\frac{h+h_i+h_j-\varepsilon_{ij}^kh_k}{(z-z_i)(z-z_j)}\right)
  G_4\,.\nonumber
\eea
This can be brought into the well-known form of the Gauss hypergeometric
equation by extracting a suitable factor $x^p(1-x)^q$ from $G_4$ with
$x$ the crossing ratio $x=\frac{(z-z_1)(z_2-z_3)}{(z-z_3)(z_2-z_1)}$.
Using the general ansatz (\ref{eq:4pt}), 
we first rewrite the four-point function
for the particular choice of coordinates $z_3=\infty$, $z_2=1$, and $z_1=0$ 
(i.e.\ $z\equiv x$) in the following form, where we renamed $h=h_0$
to allow consistent labeling:
\bea\label{eq:4pt-level2}
   \bra\Phi_{h_3}(\infty)\Phi_{h_2}(1)\Phi_{h_1}(0)\Phi_{h_0}(z)\ket
   &=& z^{p+\mu_{01}}(1-z)^{q+\mu_{20}}F(z)\,,\\
   \mu_{ij} &=& (h_0+h_1+h_2+h_3)/3-h_i-h_j\,,\nonumber\\
   p &=& {\textstyle\frac16 - \frac23h_0-\mu_{01}-\frac16\sqrt{r_1}}
         \,,\nonumber\\
   q &=& {\textstyle \frac16 - \frac23h_0-\mu_{01}-\frac16\sqrt{r_2}}
         \,,\nonumber\\
   r_i &=& {\textstyle 1 - 8h_0 + 16h_0^3 + 48h_ih_0 + 24h_i}\,.\nonumber
\eea
The remaining function $F(z)$ then is a solution of the hypergeometric 
system ${}_2F_1(a,b;c;z)$ given by
\bea\label{eq:hypergeom}
  0 &=& \left(z(1-z)\partial_z^2 + [c - (a+b+1)z]\partial_z - ab\right)F(z)\,,\\
  a &=& {\textstyle\frac12-\frac16\sqrt{r_1}-\frac16\sqrt{r_2}
        -\frac16\sqrt{r_3}}\,,\nonumber\\
  b &=& {\textstyle\frac12-\frac16\sqrt{r_1}-\frac16\sqrt{r_2}
        +\frac16\sqrt{r_3}}\,,\nonumber\\
  c &=& {\textstyle 1 - \frac13\sqrt{r_1}}\,.\nonumber
\eea
The general solution is then a linear combination of the two
linearly independent solutions ${}_2F_1(a,b;c;z)$ and
$z^{1-c}{}_2F_1(a-c+1,b-c+1;2-c;z)$. Which linear combination one has to take
is determined by the requirement that the full four-point function 
involving holomorphic and anti-holomorphic dependencies must be
single-valued to represent a physical observable quantity.
For $|z|<1$, the hypergeometric function enjoys a convergent power series
expansion
\be
  {}_2F_1(a,b;c;z) = 
  \sum_{n=0}^{\infty}\frac{(a)_n(b)_n}{(c)_n}\frac{z^n}{n!}\,,\ \ \ \
  (x)_n = \Gamma(x+n)/\Gamma(x)\,,
\ee
but it is a  quite interesting point to note that the integral representation
has a remarkably similarity to expressions of dual string-amplitudes
encountered in string theory, namely
\be
  {}_2F_1(a,b;c;z) = \frac{\Gamma(c)}{\Gamma(b)\Gamma(c-b)}\int_0^1{\rm d}t\,
  t^{b-1}(1-t)^{c-b-1}(1-zt)^{-a}\,,
\ee
which, of course, is no accident. However, we must leave this issue to
the curiosity of the reader, who might browse through the literature 
looking for the keyword {\em free field construction}.

\bq
A further consequence of the fact, that descendants are entirely determined
by their corresponding primaries is that we can refine the structure of OPEs.
Let us assume we want to compute the OPE of two primary fields. The right
hand side will possibly involve both, primary and descendant fields. Since
the coefficients for the descendant fields are fixed by local conformal 
covariance, we may rewrite (\ref{eq:ope}) as
\be
  \Phi_{h_i}(z)\Phi_{h_j}(w) = \sum_{k,\vec{n}}
  {\cal C}_{ij}^k\beta_{ij}^{k,\vec{n}}\,
  (z-w)^{h_k+|\vec{n}|-h_i-h_j}\Phi_{h_k}^{(-\vec{n})}(w)\,,
\ee
where the coefficients $\beta$ are determined by conformal covariance.
Note that we have skipped the anti-holomorphic part, although an OPE
is in general only well-defined for fields of the full theory, i.e.\
for fields $\Phi_{h,\bar h}(z,\bar z)$. An exception is the case where
all conformal weights satisfy $2h\in\mathbb{Z}$, since then holomorphic
fields are already local.

Finally, we can explain how associativity of the OPE and crossing
symmetry are related. Let us consider a four-point function
$G_{ijkl}(z,\bar z) = \bra 0|\phi_l(\infty,\infty)\phi_k(1,1)\phi_j(z,\bar z)
\phi_i(0,0)|0\ket$. There are three different regions for the free
coordinate $z$, for which an OPE makes sense, corresponding to the
contractions $z\rightarrow 0:(i,j)(k,l)$, $z\rightarrow 1:(k,j)(i,l)$, and
$z\rightarrow\infty:(l,j)(k,i)$. In fact, these three regions correspond
to the $s$, $t$, and $u$ channels. Duality states, that the evaluation of
the four-point function should not depend on this choice. Absorbing all
descendant contributions into functions ${\cal F}$ called {\em conformal
blocks}, duality imposes the conditions
\bea\label{eq:crossing-symm}
  G_{ijkl}(z,\bar z) &=& \sum_m C_{ij}^mC_{mkl}{\cal F}_{ijkl}(z|m)
                                         \bar{{\cal F}}_{ijkl}(\bar z|m)\\
  &=& \sum_m C_{jk}^mC_{mli}{\cal F}_{ijkl}(1-z|m)
                      \bar{{\cal F}}_{ijkl}(1-\bar z|m)\nonumber\\
  &=& \sum_m C_{jl}^mC_{mki}z^{-2h_j}{\cal F}_{ijkl}(
                 {\displaystyle\frac{1}{z}}|m)
                 \bar z^{-2\bar h_j}\bar{{\cal F}}_{ijkl}(
                 {\displaystyle\frac{1}{z}}|m)\nonumber\,,
\eea
where $m$ runs over all primary fields which appear on the right hand side
of the corresponding OPEs. The careful reader will have noted that these
last equations were written down in terms of the full fields in the
so-called {\em diagonal\/} theory, i.e.\ where $\bar h=h$ for all fields.
This is one possible solution to the physical requirement that the full 
correlator be a single-valued analytic function. Under certain circumstances,
other solutions, so-called non-diagonal theories, do exist. 
\eq

\EPSFIGURE{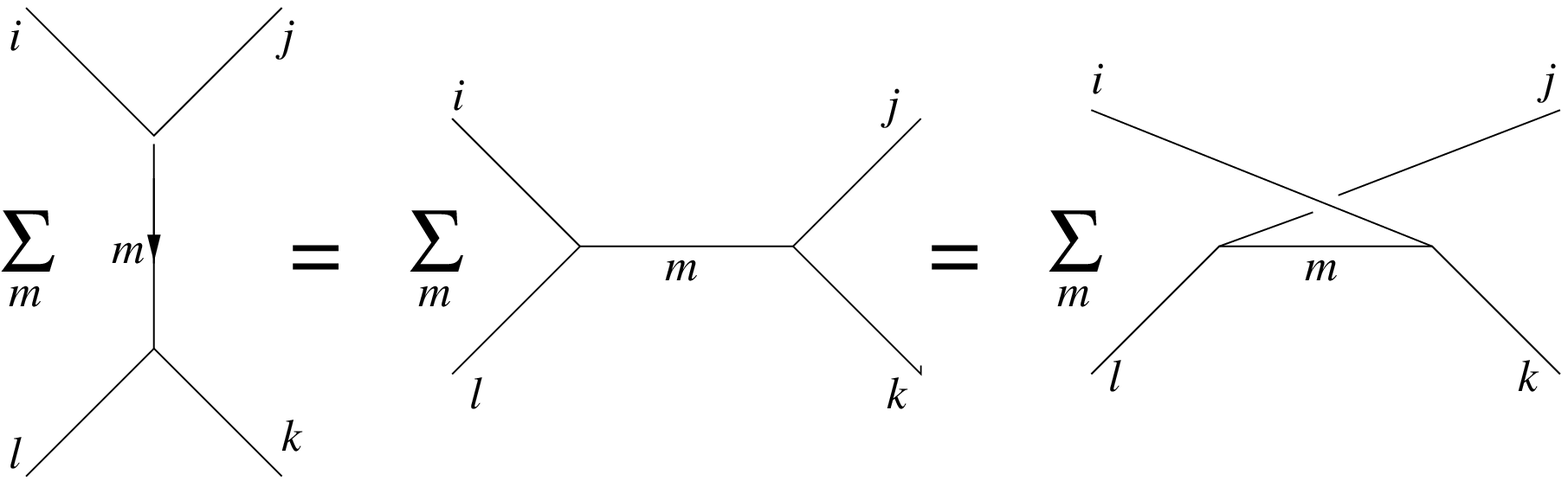,width=10cm}{The three different ways to evaluate
a four-point amplitude, i.e.\ $s$- $t$- and $u$-channels.}

\bq
In the full theory, with left- and right-chiral parts combined, the OPE
has the following structure, where the contributions from descendants
have been made explicit:
\be
  \Phi_{h_i,\bar h_i}(z,\bar z)\Phi_{h_j,\bar h_j}(w,\bar w) = \sum_{k,\vec{n}}
  \sum_{\bar k,\bar{\vec{n}}}
  {\cal C}_{ij}^k\beta_{ij}^{k,\vec{n}}
  {\cal C}_{\bar\imath\bar\jmath}^{\bar k}\beta_{
  \bar\imath\bar\jmath}^{\bar k,\bar{\vec{n}}}\,
  (z-w)^{h_k+|\vec{n}|-h_i-h_j}
  (\bar z-\bar w)^{\bar h_k+|\bar{\vec{n}}|-\bar h_i-\bar h_j}
  \Phi_{h_k,\bar h_k}^{(-\vec{n},-\bar{\vec{n}})}(w,\bar w)\,.
\ee
Correlation functions in the full CFT should be single valued in order to
represent observables, i.e.\ physical measurable quantities. This imposes
further restrictions on the particular linear combinations of the
conformal blocks ${\cal F}_{ijkl}(z|m)$ in (\ref{eq:crossing-symm}).
In most CFTs, the diagonal combination $\bar h=h$ is a solution, but it is
easy to see, that the monodromy of a field $\Phi_{h,\bar h}(z,\bar z)$ under
$z\mapsto{\rm e}^{2\pi{\rm i}}z$ yields the less restrictive condition
$h-\bar h\in\mathbb{Z}$, such that off-diagonal solutions can be possible.

The success story of CFT is much rooted in the following observation first
made by Belavin, Polyakov and Zamolodchikov \cite{BPZ}: If an OPE of
two primary fields $\Phi_i(z)\Phi_j(w)$ is considered, which both are 
degenerated at levels $N_i$ and $N_j$ respectively, 
then the right hand side will
only involve contributions from primary fields, which {\em all\/} are
degenerate at a certain levels $N_k \leq N_i+N_j$. In particular, the sum over
conformal families $k$ on the right hand side is then always finite, and so
is the set of conformal blocks one has to know. In particular, the set
of degenerate primary fields (and their descendants) forms a closed operator 
algebra. For example, considering
a four-point function where all four fields are degenerate at level two,
we find only two conformal blocks for each channel, which precisely are the 
hypergeometric functions computed above and their analytic continuations.
Even more remarkably, for the special values $c(m)$ in (\ref{eq:cseries})
with $m\in\mathbb{N}$, there are only {\em finitely\/} many primary fields
with conformal weights $h_{p,q}(c)$ with $1\leq p<m$ and $1\leq q<m+1$
given by(\ref{eq:hseries}). All other degenerate primary fields
with weights $h_{p,q}(c)$ where $p$ or $q$ lie outside this range turn out
to be null fields within the Verma modules of the descendants of these former
primary fields. Hence, such CFTs have a finite field content and are
actually the ``smallest'' CFTs. This is why they are called {\em minimal
models}. Unfortunately, they are not very useful for string theory, but
turn up in many applications of statistical physics \cite{ISZ}.
\eq 

\section{Logarithmic null vectors}

We have learned in the basic introductionary lectures that logarithmic 
conformal field theory (LCFT) arises due to the existence of indecomposable
representations. Thus, instead of a unique highest weight state, on which
the representation module is built, we have to deal with a Jordan cell
of states which are linked by the action of some operator which cannot
be diagonalized. In most cases, this will be the action of the stress-energy
tensor, but in general Jordan cells might occur due to the action of any 
generator of the (extended) chiral symmetry algebra. To keep things simple,
we will confine ourselves to the Virasoro case within these notes. We will
see other examples in the lectures by Matthias Gaberdiel.

Let us briefly recall what we mean by
Jordan cell structure. Suppose we have two operators $\Phi(z),\Psi(z)$ with
the same conformal weight $h$, or more precisely, with an equivalent set
of quantum numbers with respect to the maximally extended chiral symmetry
algebra. As was first realized in \cite{Gurarie:1993}, this
situation leads to logarithmic correlation functions and to the fact that
$L_0$, the zero mode of the Virasoro algebra, can no longer be diagonalized:
  \bea
  L_0\vac{\Phi} & = & h\vac{\Phi}\,,\nonumber\\
  L_0\vac{\Psi} & = & h\vac{\Psi} + \vac{\Phi}\,,
  \eea
where we worked with states instead of the fields themselves. The field
$\Phi(z)$ is then an ordinary primary field, whereas the field $\Psi(z)$
gives rise to logarithmic correlation functions and is therefore called
a {\em logarithmic partner\/} of the primary field $\Phi(z)$. We would like
to note once more that two fields of the same conformal dimension {\em do
not automatically\/} lead to LCFTs with respect to the Virasoro algebra.
Either, they differ in some other quantum numbers (for examples of such CFTs
see \cite{Flo93}), or they form a Jordan cell structure with respect to
an extended chiral symmetry only (see \cite{Kogan:1998a} for a description of
the different possible cases).

We remember that a singular or null vector
$\vac{\chi}$ is a state which is orthogonal to all states,
  \be
  \vev{\psi}{\chi} = 0\ \forall \vac{\psi}\,,
  \ee
where the scalar product is given by the Shapovalov form. Such
states can be considered to be identically zero.

A pair of fields $\Phi(z),\Psi(z)$ forming a Jordan cell structure brings
the problem of off-diagonal terms produced by the action of the Virasoro
field, such that the corresponding representation is indecomposable.
Therefore, if $\Vac{\chi^{}_{\Phi}}$ is a null vector in the Verma
module on the highest weight state $\vac{\Phi}$ of the primary field, we
cannot just replace $\vac{\Phi}$ by $\vac{\Psi}$ and obtain another null
vector.

Before we define general null vectors for Jordan cell structures, we
present a formalism which might be useful in the future for all kinds of
explicit calculations in the LCFT setting. This formalism, 
has the advantage that the Virasoro modes are still
represented as linear differential operators, and that it is compact and
elegant allowing for arbitrary rank Jordan cell structures. Moreover,
the connection between LCFTs and supersymmetric CFTs, which one could glimpse
here and there \cite{Caux:1997,Flohr:1996a,Rozansky:1992,Saleur:1991} 
(see also \cite{EhGa96}),
seems to be a quite fundamental one. 

\subsection{Jordan cells and nilpotent variable formalism}

LCFTs are characterized by the fact that some of their highest weight
representations are indecomposable. This is usually described by saying that
two (or more) highest weight states with the same highest weight span
a non-trivial Jordan cell. In the following we call the dimension of
such a Jordan cell the {\em rank\/} of the indecomposable representation.

Therefore, let us assume that a given LCFT has an indecomposable representation
of rank $r$ with respect to its maximally extended chiral symmetry algebra
${\cal W}$. This Jordan cell is spanned by $r$ states $\vac{w_0,w_1,\ldots;n}$,
$n=0,\ldots,r-1$ such that the modes of the generators of the chiral
symmetry algebra act as
  \bea
  \label{eq:jordan}
  \Phi^{(i)}_0\vac{w_0,w_1,\ldots;n} &=& w_i\vac{w_0,w_1,\ldots;n}
  + \sum_{k=0}^{n-1}a_{i,k}\vac{w_0,w_1,\ldots;k}\,,\\
  \Phi^{(i)}_m\vac{w_0,w_1,\ldots;n} &=& 0\ {\rm for}\ m>0\,,
  \eea
where usually $\Phi^{(0)}(z) = T(z)$ is the stress energy tensor which gives
rise to the Virasoro field, i.e.\ $\Phi^{(0)}_0 = L_0$, and $w_0 = h$ is the
conformal weight. For the sake of simplicity, we concentrate in these notes
on the representation theory of LCFTs with respect to the pure Virasoro
algebra such that (\ref{eq:jordan}) reduces to
  \bea
  \label{eq:L0}
  L_0\vac{h;n} &=& h\vac{h;n} + (1 - \delta_{n,0})\vac{h;n-1}\,,\\
  L_m\vac{h;n} &=& 0\ {\rm for}\ m>0\,,
  \eea
where we have normalized the off-diagonal contribution to 1. As in ordinary
CFTs, we have an isomorphism between states and fields. Thus, the state
$\vac{h;0}$, which is the highest weight state of the irreducible
sub-representation contained in every Jordan cell, corresponds to an ordinary
primary field $\Psi_{(h;0)}(z)\equiv\Phi_{h}(z)$, whereas states $\vac{h;n}$
with $n>0$ correspond to the so-called logarithmic partners $\Psi_{(h;n)}(z)$
of the primary field. The action of the modes of the Virasoro field on these
primary fields and their logarithmic partners is given by
  \bea
  \label{eq:virl}
  \lefteqn{{\cal L}_{-k}(z)\Psi_{(h;n)}(w) = }\\
  & & \frac{(1-k)h}{(z-w)^k}\Psi_{(h;n)}(w)
  - \frac{1}{(z-w)^{k-1}}\frac{\del}{\del w}\Psi_{(h;n)}(w)
  - (1 - \delta_{n,0})\frac{\lambda(1-k)}{(z-w)^k}\Psi_{(h;n-1)}(w)\,,
  \nonumber
  \eea
with $\lambda$ normalized to 1 in the following.\footnote{The reader should
recall from linear algebra that it is always possible to normalize
the off-diagonal entries in a Jordan block to one.} 
As it stands, the off-diagonal
term spoils writing the modes ${\cal L}_{-k}(z)$ as linear differential
operators.

\bq
There is one subtlety here. In these notes we {\it assume\/} that the
logarithmic partner fields of a primary field are all quasi-primary in the
sense that the corresponding states $\vac{h;n}$ are all annihilated by the 
action of modes $L_m$, $m>0$. This is not necessarily the case, and
there are examples of LCFTs where Jordan blocks occur, where the
logarithmic partner is not quasi-primary.\footnote{The author thanks
Matthias Gaberdiel to pointing this out.} For instance, the Jordan
block of $h=1$ fields in the $c=-2$ LCFT is made up of a primary field
with highest weight state $\vac{\phi}$ and a logarithmic partner $\vac{\psi}$
such that 
$$ 
  L_0\vac{\phi}=\vac{\phi}\,,\ \ \ \ 
  L_0\vac{\psi}=\vac{\psi}+\vac{\phi}\,,\ \ \ \
  L_1\vac{\phi}=0\,,\ \ \ \
  L_1\vac{\psi}=\vac{\xi}\,,
$$
where $\vac{\xi}$, a state corresponding to a field of zero conformal
weight, is related to the primary field via $L_{-1}\vac{\xi}=\vac{\phi}$.
Note that in this particular example, the primary field corresponding
to $\vac{\phi}$ is a current, and a descendant of the field corresponding
to $\vac{\xi}$.
However, there are indications that such indecomposable representations
with non-quasi-primary states of weight $h$ only occur together with a 
corresponding indecomposable representation of only quasi-primary states
of weight $h-k$, $k\in\mathbb{Z}_+$. We are not
going to investigate this issue further, but note that all so far 
explicitly known LCFTs possess at least one indecomposable representation
where all states of the basic Jordan block are quasi-primary. Since it
is a very difficult task to construct null vectors on non-quasi-primary
states, we will not consider such indecomposable representations here.
For more details on the issue of Jordan cells with non-quasi-primary fields
see the last reference in \cite{Flohr:1996a}.
\eq

Our first aim is simply to prepare a formalism in which the
Virasoro modes are expressed as linear differential operators. To this end, we
introduce a new -- up to now purely formal -- variable $\theta$ with the
property $\theta^r = 0$. We may then view an arbitrary state in the Jordan
cell, i.e.\ a particular linear combination
  \be
  \Psi_{h}(\vec{a})(z) = \sum_{n=0}^{r-1}a_n\Psi_{(h;n)}(z)\,,
  \ee
as a formal series expansion describing an arbitrary function $a(\theta)$
in $\theta$, namely
  \be
  \label{eq:expand}
  \Psi_h(a(\theta))(z) =
  \sum_{n}a_n\frac{\theta^n}{n!}\Psi_h(z)\,.
  \ee
This means that the space of all states in a Jordan cell can be described
by tensoring the primary state with the space of power series in $\theta$,
i.e. $\Theta_r(\Psi_h)\equiv\Psi_h(z)\otimes\mathbb{C}[\![\theta]\!]/{\cal I}$,
where we divided out the ideal generated by the relation ${\cal I}=\langle
\theta^r\!=\!0\rangle$.
In fact, the action of the Virasoro algebra is now simply given by
  \be
  \label{eq:vir2}
  {\cal L}_{-k}(z)\Psi_h(a(\theta))(w) = \left(\frac{(1-k)h}{(z-w)^k}
  - \frac{1}{(z-w)^{k-1}}\frac{\del}{\del w}
  - \frac{\lambda(1-k)}{(z-w)^k}\frac{\del}{\del\theta}\right)
  \Psi_h(a(\theta))(w)\,.
  \ee
Clearly, $\Psi_{(h;n)}(z) = \Psi_h(\theta^n/n!)(z)$, but we will often simplify
notation and just write $\Psi_h(\theta)(z)$ for a generic element in
$\Theta_r(\Psi_h)$. However, the context should always make it clear, whether
we mean a generic element or really $\Psi_{(h;1)}(z)$. The corresponding
states are denoted by $\vac{h;a(\theta)}$ or simply $\vac{h;\theta}$.
To project onto the $k^{{\rm th}}$ highest weight state\footnote{
More precisely, only $\vac{h;0}$ is a proper highest weight state, so 
calling $\vac{h;n}$ for $n>0$ highest weight states is a sloppy abuse
of language.} of the Jordan cell, we just use
$a_k\vac{h;k} = \left.\del_{\theta}^k\vac{h;a(\theta)}\right|_{\theta=0}$.
In order to avoid confusion with $\vac{h;1}$ we write $\vac{h;\mathbb{I}}$ if
the function $a(\theta)\equiv 1$.

It has become apparent by now that LCFTs are somehow closely linked to
supersymmetric CFTs \cite{Caux:1997,Flohr:1996a,Rozansky:1992,Saleur:1991} 
(see also \cite{EhGa96}).
We suggestively denoted our formal variable by $\theta$,
since it can easily be constructed with the help of Grassmannian variables
as they appear in supersymmetry. Taking $N\!=\!r-1$ supersymmetry with
Grassmann variables $\theta_i$ subject to $\theta_i^2=0$, we may define
$\theta = \sum_{i=1}^{r-1}\theta_i$. More generally, $\theta$ and its powers
constitute a basis of the totally symmetric, homogenous polynomials in the
Grassmannians $\theta_i$.
 
Finally, we remark that the $\theta$ variables are associated {\em not\/} with
the coordinates the fields are localized in coordinate space, but with the
positions the fields are localized in $h$-space (the Jordan cells). Therefore,
the $\theta$ variables will be labeled by the conformal weight they refer to,
whenever the context makes it necessary.

\subsection{Logarithmic null vectors}

Next, we derive the consequences of our formalism. An arbitrary state in a
LCFT of level $n$ is a linear combination of descendants of the form
  \be
  \vac{\psi(\theta)} = \sum_k\sum_{\{n_1+n_2+\ldots+n_m=n\}}
  b^{\{n_1,n_2,\ldots,n_m\}}_k
  L_{-n_m}\ldots L_{-n_2}L_{-n_1}\vac{h;k}
  \ee
which we often abbreviate as
  \be
  \label{eq:descendant}
  \vac{\psi(\theta)} = \sum_{|\vec{n}|=n}
  L_{-\vec{n}}b^{\vec{n}}(\theta)\vac{h}\,.
  \ee
We will mainly be concerned with calculating Shapovalov forms
$\vev{\psi'(\theta')}{\psi(\theta)}$ which ultimately cook down (by
commuting Virasoro modes through) to expressions of the form
  \be
  \label{eq:shapovalov}
  \vev{\psi'(\theta')}{\psi(\theta)} =
  \avac{h';a'(\theta')}\sum_m f_m(c)(L_0)^m\vac{h;a(\theta)}\,,
  \ee
where we explicitly noted the dependence of the coefficients on the central
charge $c$.
Combining (\ref{eq:shapovalov}) with (\ref{eq:descendant}) we write
$\vev{\psi'(\theta')}{\psi(\theta)}=
\avac{h';a'(\theta')}f_{\vec{n}',\vec{n}}(L_0,C)\vac{h;a(\theta)}$
for the Shapovalov form between two {\em monomial\/} descendants, i.e.\
  \be
  \avac{h';a'(\theta')}f_{\vec{n}',\vec{n}}(L_0,C)\vac{h;a(\theta)} =
  \avac{h';a'(\theta')}L_{n'_1}L_{n'_2}\ldots L_{-n_2}L_{-n_1}\vac{h;a(\theta)}
  \,.
  \ee
More generally, since $L_0\vac{h;a(\theta)} = (h + \del_{\theta})
\vac{h;a(\theta)}$, it is easy to
see that an arbitrary function $f(L_0,C)\in\mathbb{C}[\![L_0,C]\!]$ acts as
  \be
  \label{eq:fL0}
  f(L_0,C)\vac{h;n} = \sum_k\frac{1}{k!}\left(\frac{\del^k}{\del h^k}f(h,c)
  \right)\vac{h;n-k}\,,
  \ee
and therefore $f(L_0,C)\vac{h;a(\theta)} = \vac{h;\tilde{a}(\theta)}$, where
with $a(\theta) = \sum_n a_n\frac{\theta^n}{n!}$ we have
  \be
  \tilde{a}_n = \sum_k\frac{a_{n+k}}{k!}\frac{\del^k}{\del h^k}f(h,c)\,.
  \ee

\bq
It may be instructive to check this statement explicitly for the
simple case $f(L_0,C)=L_0^m$. Keeping in mind that $\vac{h;n}=
\vac{h;\frac{1}{n!}\theta^n}$, one then finds
\bea
  L_0^m\vac{h;n} &=& (h+\del^{}_{\theta})^m\vac{h;\frac{1}{n!}\theta^n}
  = \sum_k {m\choose k}h^{m-k}\del_{\theta}^{k}\vac{h;\frac{1}{n!}\theta^n}
  = \sum_k {m\choose k}h^{m-k}\frac{n(n-1)\ldots(n-k+1)}{n!}
    \vac{h;\theta^{n-k}}\nonumber\\ 
 &=&\sum_k \frac{m!}{k!(m-k)!}\frac{1}{(n-k)!}h^{m-k}\vac{h;\theta^{n-k}}
  = \sum_k \frac{1}{k!}m(m-1)\ldots(m-k+1)h^k\vac{h;n-k}\nonumber\\
 &=&\sum_k \frac{1}{k!}(\del_h^{k} h^m)\vac{h;n-k}
  = \sum_k \frac{1}{k!}\del_h^kf(h,c)\vac{h;n-k}\,.
\eea
Since more general functions $f(L_0,C)$ are merely linear combinations
of the above example with different $m$, the general statement should be
clear. Note, however, that sofar the central charge only enters as an
external parameter.
\eq

\noindent This
puts the convenient way of expressing the action of $L_0$ on Jordan cells
by derivatives with respect to the conformal weight $h$, which appeared earlier
in the literature, on a firm ground. Moreover, from now on we do not worry
about the range of summations, since all series automatically truncate in the
right way due to the condition $\theta^r=0$.
 
It is evident that choosing $a(\theta) = \mathbb{I}$ extracts the irreducible
sub-representation which is invariant under the action of $L_0$. All other
non-trivial choices of $a(\theta)$ yield states which are not invariant
under the action of $L_0$. The existence of null vectors of level $n$ on such
a particular state is subject to the conditions that
  \bea \lefteqn{
  \sum_{|\vec{n}|=n}f_{\vec{n}',\vec{n}}(L_0,C) b^{\vec{n}}(\theta,h,c)
  \vac{h}}\\ 
  &\equiv&
  \sum_{|\vec{n}|=n}f_{\vec{n}',\vec{n}}(L_0,C)\sum_k b^{\vec{n}}_k(h,c)
  \vac{h;k} = 0\ \ \ \ \forall\ \vec{n}':|\vec{n}'|=n\,.\nonumber
  \eea
Notice that we have the freedom that each highest weight state of the
Jordan cell comes with its own descendants. These conditions determine the
$b^{\vec{n}}_k(h,c)$ as functions in the conformal
weight and the central charge. Clearly, for $a(\theta) = \mathbb{I}$ this would
just yield the ordinary results as known since BPZ \cite{BPZ},
i.e.\ the solutions for $b^{\vec{n}}_0(h,c)$.
The question is now, under which circumstances null vectors
exist on the whole Jordan cell, i.e.\ for non-trivial choices of $a(\theta)$.
Obviously, these null vectors, which we call {\em logarithmic null vectors\/}
can only constitute a subset of the ordinary null vectors. From (\ref{eq:fL0})
we immediately learn that the conditions imply
  \be\label{eq:nully}
  \sum_{k=0}^{s-1}\sum_{|\vec{n}|=n}b^{\vec{n}}_k(h,c)\frac{1}{(s-1-k)!}
  \frac{\del^{s-1-k}}{\del h^{s-1-k}}f_{\vec{n}',\vec{n}}(h,c)
  = 0\ \ \forall\ \vec{n}':|\vec{n}'|=n\,,\ 1\leq s\leq r\,.
  \ee

\bq
To see this, simply start with $s=1$ and observe that this recovers
the well known condition for a generic null vector of a ordinary
non-logarithmic CFT, $\sum_{|\vec{n}|=n}
b_0^{\vec{n}}(h,c)f_{\vec{n}',\vec{n}}(h,c) = 0$.
Then proceed inductively. In the next step, $s=2$, one now finds
a condition which relates the coefficients $b_1^{\vec{n}}(h,c)$ and
the coefficients $b_0^{\vec{n}}(h,c)$, 
$$
  \sum_{|\vec{n}|=n}\left(
  b_1^{\vec{n}}(h,c)f_{\vec{n}',\vec{n}}(h,c)+
  b_0^{\vec{n}}(h,c)\partial_hf_{\vec{n}',\vec{n}}(h,c)\right) = 0\,,
$$
which is clear since the action of
$L_0$ on $\vac{h;1}$ will produce terms proportional to $\vac{h;0}$.
Since $L_0$ never moves up within a Jordan block, the
condition for the coefficients for $\vac{h;s-1}$ can only involve
the coefficients for states $\vac{h;s'-1}$, $0\leq s'<s$. Thus, we
arrive at the above statement.
\eq

\noindent The conditions (\ref{eq:nully})  can be satisfied if we put
  \be
  b^{\vec{n}}_k(h,c) = \frac{1}{k!}\frac{\del^k}{\del h^k}b^{\vec{n}}_0(h,c)\,.
  \ee
In fact, choosing the $b^{\vec{n}}_k(h,c)$ in this way allows one to rewrite
the conditions as total derivatives of the standard condition for
$b^{\vec{n}}_0(h,c)$. Keeping in mind that each Jordan cell module of rank $r$
has Jordan cells of ranks $r'$, $1\leq r' \leq r$, as submodules, we can find
intermediate null vector conditions, where the null vector only lies in
the rank $r'$ submodule (think of $r'=1$ as a trivial example), if we restrict
the range of $s$ in (\ref{eq:nully}) accordingly.
Of course, this determines the $b^{\vec{n}}_k(h,c)$
only up to terms of lower order in the derivatives such that the conditions
finally take the general form
  \be
  \sum_k\frac{\lambda_k}{k!}\frac{\del^k}{\del h^k}\left(
  \sum_{|\vec{n}|=n}f_{\vec{n}',\vec{n}}(h,c)b^{\vec{n}}_0(h,c)\right)
  = 0\ \forall\ \vec{n}':|\vec{n}'|=n\,,
  \ee
which, however, does not yield any different results. Moreover, the
coefficients $b^{\vec{n}}_k(h,c)$ can only be determined up to an overall
normalization. Clearly, there are $p(n)$ coefficients, where $p(n)$ denotes
the number of partitions of $n$ into positive integers. This means that only
$p(n)-1$ of the standard coefficients
$b^{\vec{n}}_0(h,c)$ are determined to be functions in $h,c$ multiplied by
the remaining coefficient, e.g.\ $b^{\{1,1,\ldots,1\}}_0$ (if this coefficient
is not predetermined to vanish). In order to be able to write the coefficients
$b^{\vec{n}}_k(h,c)$ with $k>0$ as derivatives with respect to $h$, one
needs to fix the remaining free coefficient $b^{\{1,1,\ldots,1\}}_0=h^{p(n)}$
as a function of $h$. The choice given here ensures that all coefficients
are always of sufficient high degree in $h$.\footnote{
We usually choose the least common multiple of the denominators of the
resulting rational functions in $h,c$ of the other coefficients in order to
simplify the calculations. This, however, occasionally leads to additional --
trivial -- solutions which are the price we pay for doing all calculations
with polynomials only.} Clearly, this works
only for $h\neq 0$. To find null vectors with $h=0$ needs some extra care. One
foolproof choice is to put the remaining free coefficient to $\exp(h)$.
The problem is that the Hilbert space of states is a projective space due to
the freedom of normalization, and that we used $h$ as a projective coordinate
in this space, which only works for $h\neq 0$.

It is important to
understand that the above is only a necessary condition due to the
following subtlety: The derivatives with respect to $h$ are done in a purely
formal way. But already determining the standard solution $b^{\vec{n}}_0(h,c)$
is not sufficient in itself, and the conditions for the existence of
standard null vectors yield one more constraint, namely $h=h_i(c)$ or vice
versa $c=c_i(h)$ (the index $i$ denotes possible different solutions, since
the resulting equations are higher degree polynomials $\in\mathbb{C}[h,c]$).
These constraints must be plugged in {\em after\/} performing the
derivatives and, as it will turn out, this will severely restrict the
existence of logarithmic null vectors, yielding only some {\em discrete\/}
pairs $(h,c)$ for each level $n$. Moreover, the set of solutions gets
rapidly smaller if for a given level $n$ the rank $r$ of the assumed Jordan
cell is increased. Since there are $p(n)$ linearly independent conditions
for the $b^{\vec{n}}_0(h,c)$ of a standard null vector of level $n$,
a necessary condition is $r\leq p(n)$.
As mentioned above, $h$ is not a good coordinate for $h=0$, but $c_i(h)$ still
is.\footnote{Again, this is only true as long as $c\neq 0$. The special
point $(c=0,h=0)$ unfortunately cannot be treated within our scheme, but
must be checked by direct calculations.}
Therefore, for $h=0$ we should use $c$ for normalization, meaning that for
$h=0$, the $c_i(h)$ have to be plugged in {\em before\/} doing the derivatives.

\subsection{An example}

Now we will go through a rather elaborate example to see how all this is
supposed to work. So, we are going to demonstrate what a logarithmic null vector is
and under which conditions it exists. Null vectors are of particular
importance for rational CFTs. For any CFT given by its maximally extended
symmetry algebra ${\cal W}$ and a value $c$ for the central charge we can
determine the so-called degenerate ${\cal W}$-conformal families which contain
at least one null vector. The corresponding highest weights turn out to
be parametrized by certain integer labels, yielding the so-called
Kac-table. If ${\cal W} = \{T(z)\}$ is just the Virasoro algebra, all degenerate
conformal families have highest weights labeled by two integers $r,s$,
  \be
  \label{eq:h}
  h_{r,s}(c) = \frac{1}{4}\left(
  \frac{1}{24}\left(\sqrt{(1-c)}(r+s)-\sqrt{(25-c)}(r-s)\right)^2
  - \frac{1-c}{6}\right)\,.
  \ee
The level of the (first) null vector contained in the conformal families
over the highest weight state $\vac{h_{r,s}(c)}$ is then $n=rs$.

LCFTs have the special property that there are at least two conformal
families with the same highest weight state, i.e.\ that we must have
$h = h_{r,s}(c) = h_{t,u}(c)$. This does not happen for the so-called minimal
models since their truncated conformal grid precisely excludes this.
However, LCFTs may be constructed for example for $c=c_{p,1}$, where
formally the conformal grid is empty, or by augmenting the field
content of a CFT by considering an enlarged conformal grid. However,
if we have the situation typical for a LCFT, we have two non-trivial and
{\em different\/} null vectors, one at level $n=rs$ and one at $n'=tu$ where
we assume without loss of generality $n\leq n'$.\footnote{
It follows from this reasoning that there can be no logarithmic null vector
at level 1. Thus, the only null vector at level 1 is the trivial
null vector $\vac{\chi^{(1)}_{h=0,c}}= L_{-1}\vac{0}$.} Then the null vector at
level $n$ is an ordinary null vector on the highest weight state of the
irreducible sub-representation $\vac{h;0}$ of the rank 2 Jordan cell spanned
by $\vac{h;0}$ and $\vac{h;1}$, but what about the null vector at level $n'$?

Let us consider the particular LCFT with $c=c_{3,1}=-7$. This LCFT admits
the highest weights
$h\in\{0,\frac{-1}{4},\frac{-1}{3},\frac{5}{12},1,\frac{7}{4}\}$ which yield
the two irreducible representations at $h_{1,3}=\frac{-1}{3}$ and
$h_{1,6}=\frac{5}{12}$ as well as two indecomposable representations with
so-called staggered module structure (roughly a generalization of Jordan cells
to the case that some highest weights differ by integers \cite{Gaberdiel:1996a,
Rohsiepe:1996})
constituted by the
triplets $(h_{1,1}\!=\!0,h_{1,5}\!=\!0,h_{1,7}\!=\!1)$ and
$(h_{1,2}\!=\!\frac{-1}{4},h_{1,4}\!=\!\frac{-1}{4},h_{1,8}\!=\!\frac{7}{4})$.
We note that similar to the case
of minimal models we have the identification $h_{1,s} = h_{2,9-s}$ such that
the actual level of the null vector might be reduced. In the following we
will determine the null vectors at level 2 and 4 for the rank 2 Jordan
cell with $h=\frac{-1}{4}$. First, we start with the level 2 null vector,
whose general ansatz is
  \be
  \vac{\chi^{(2)}_{h,c}} =
  \left(b^{\{1,1\}}_0L_{-1}^2 + b^{\{2\}}_0L_{-2}\right)
  \vac{h;a(\theta)} +
  \left(b^{\{1,1\}}_1L_{-1}^2 + b^{\{2\}}_1L_{-2}\right)
  \vac{h;\del_{\theta}a(\theta)}\,,
  \ee
where we explicitly made clear how we counteract the off-diagonal action of the
Virasoro null mode. 

\bq
For null vectors of level $n>1$ we make the general ansatz
  \be
  \vac{\chi^{(n)}_{h,c}} = \sum_j\sum_{|\vec{n}|=n}b^{\vec{n}}_j(h,c)
  L_{-\vec{n}}\Vac{h;\del_{\theta}^ja(\theta)}
  \ee
and define matrix elements
  \bea
  N_{k,l}^{(n)} &=& \left.\frac{\del^k}{\del\theta^k}\left(
  \sum_j\sum_{|\vec{n}|=n}b^{\vec{n}}_j(h,c)
  \left\langle{h}\left|L_{\vec{n}'_l}L_{-\vec{n}}
  \Vac{h;\del_{\theta}^ja(\theta)}\right.\right.
  \right)\right|_{\theta=0}
  \nonumber\\
          &=& \sum_{j=0}^k\sum_{|\vec{n}|=n}b^{\vec{n}}_j(h,c)
  \frac{1}{j!}\frac{\del^j}{\del h^j}
  \Avac{h}L_{\vec{n}'_l}L_{-\vec{n}}\Vac{h}\,,
  \eea
where $\vec{n}'_l$ is some enumeration of the $p(n)$ different partitions of
$n$. Since the maximal possible rank of a Jordan cell representation which
may contain a logarithmic null vector is $r\leq p(n)$, we consider $N^{(n)}$
to be
a $p(n)\times p(n)$ square matrix. Our particular ansatz is conveniently chosen
to simplify the action of the Virasoro modes on Jordan cells. Notice, that
the derivatives with respect to the conformal weight $h$ do not act on the
coefficients $b^{\vec{n}}_j(h,c)$. Of course, we assume that $a(\theta)$ has
maximal degree in $\theta$, i.e.\ ${\rm deg}(a(\theta)) = r-1$.

In our example at level 2, we have $p(2)=2$ and the matrix $N^{(2)}$ we have
to evaluate is
\be
  N^{(2)} = \left[\begin{array}{cc}
  b^{\{1,1\}}_0\avac{h}L_1^2L_{-1}^2\vac{h}
  +b^{\{2\}}_0\avac{h}L_1^2L_{-2}\vac{h} & \ \
  b^{\{1,1\}}_0\del_h\avac{h}L_1^2L_{-1}^2\vac{h}
  +b^{\{2\}}_0\del_h\avac{h}L_1^2L_{-2}\vac{h}\\
  & \ \ {}+b^{\{1,1\}}_1\avac{h}L_1^2L_{-1}^2\vac{h}     
  +b^{\{2\}}_1\avac{h}L_1^2L_{-2}\vac{h}\\ \noalign{\medskip}
  b^{\{1,1\}}_0\avac{h}L_2L_{-1}^2\vac{h}
  +b^{\{2\}}_0\avac{h}L_2L_{-2}\vac{h} & \ \
  b^{\{1,1\}}_0\del_h\avac{h}L_2L_{-1}^2\vac{h}
  +b^{\{2\}}_0\del_h\avac{h}L_2L_{-2}\vac{h}\\
  & \ \ {}+b^{\{1,1\}}_1\avac{h}L_2L_{-1}^2\vac{h}     
  +b^{\{2\}}_1\avac{h}L_2L_{-2}\vac{h}
  \end{array}\right]\,.
\ee
Doing the computations, this reads
  \be
  N^{(2)} = \left[\begin{array}{cc}
  b^{\{1,1\}}_0\left(8h^2+4h\right)+6b^{\{2\}}_0h &
  b^{\{1,1\}}_0\left(16h+4\right)+6b^{\{2\}}_0
  +b^{\{1,1\}}_1\left(8h^2+4h\right)+6b^{\{2\}}_1h\\\noalign{\medskip}
  6b^{\{1,1\}}_0h+b^{\{2\}}_0\left(4h+\fr{1}{2}c\right) &
  6b^{\{1,1\}}_0+4b^{\{2\}}_0+6b^{\{1,1\}}_1h
  +b^{\{2\}}_1\left(4h+\fr{1}{2}c\right)
  \end{array}\right]\,.
  \ee
A null vector is logarithmic of rank $k\geq 0$ if the first $k+1$ columns
of $N^{(n)}$ are zero, where $k=0$ means an ordinary null vector. As described
in
the text, one first solves for ordinary null vectors (such that the first
column vanishes up to one entry). This determines the $b^{\vec{n}}_0(h,c)$.
Then one puts $b^{\vec{n}}_k(h,c) = \frac{1}{k!}\del_h^{k}
b^{\vec{n}}_0(h,c)$. Without loss of generality we may then assume that all
entries except the last row are zero. In our example, this procedure
results in
  \be
  N^{(2)} = \left[\begin{array}{cc}
  0 & 0 \\\noalign{\medskip}
  10h^2-16h^3-2h^2c-hc & 20h-48h^2-4hc-c
  \end{array}\right]\,,
  \ee
where $b^{\{1,1\}}_k = \frac{1}{k!}\del_h^{k}(3h)$ and
$b^{\{2\}}_k = \frac{1}{k!}\del_h^{k}(-2h(2h+1))$ upto an overall
normalization. The last step is trying to find simultaneous solutions
for the last row, i.e.\ common zeros of polynomials $\in\mathbb{C}[h,c]$. In our
example, $N^{(2)}_{2,1}=0$ yields $c = 2h(5-8h)/(2h+1)$. Then, the last
condition becomes $N^{(2)}_{2,2} = -2h(16h^2+16h-5)/(2h+1) = 0$ which can
be satisfied for $h\in\{0,\frac{-5}{4},\frac{1}{4}\}$. From this we finally
obtain the explicit logarithmic null vectors at level 2:
  $$
  \begin{array}{c|c}
  (h,c) & \vac{\chi^{(2)}_{h,c}}
    \\\noalign{\smallskip}\hline\hline\noalign{\smallskip}
  (0,0) &
    (3L_{-1}^2-2L_{-2})\Vac{0;a(\theta)}
    \\\noalign{\smallskip}
  (\frac{1}{4},1) &
    (3L_{-1}^2-3L_{-2})\Vac{\frac{1}{4};a(\theta)}
    -4L_{-2}\Vac{\frac{1}{4};\del_{\theta}a(\theta)}
    \\\noalign{\smallskip}
  (\frac{-5}{4},25) &
    (3L_{-1}^2+3L_{-2})\Vac{\frac{-5}{4};a(\theta)}
    -4L_{-2}\Vac{\frac{-5}{4};\del_{\theta}a(\theta)}
  \end{array}
  $$
Note, that according to our formalism, $h=0,c=0$ does not turn out to be
a logarithmic null vector at level 2. Here and in the following the highest
order derivative $\del_{\theta}^k a(\theta)$ indicates the maximal rank
of a logarithmic null vector to be $k$ (and hence the maximal rank of the
corresponding Jordan cell representation to be $r=k+1$). It is implicitly
understood that $a(\theta)$ is then chosen such that the highest order
derivative yields a non-vanishing constant.
 
Here, all null vectors are normalized such that all coefficients are integers.
Clearly, they are not unique since with $\Vac{\chi(\theta)} =
\sum_k \Vac{\chi^{}_k;\del_{\theta}^k a(\theta)}$ every vector
  \be
  \Vac{\chi'(\theta)} = \sum_k
  \Vac{\chi^{}_k;\sum_{l\geq 0}\lambda_{k,l}\del_{\theta}^{k+l}a(\theta)}
  \ee
is also a null vector.
\eq

\noindent It is well known that up to an overall normalization we have
for the coefficients $b_0^{\vec{n}}$ for the part of the null vector
built on the state $\vac{h;1}$ in the Jordan cell
  \be
  b^{\{1,1\}}_0 = 3h\,,\ \ \ \
  b^{\{2\}}_0   = -2h(2h + 1)\,,
  \ee
such that according to the last section we should put 
  \be
  b^{\{1,1\}}_1 =  3\,,\ \ \ \
  b^{\{2\}}_1   = -8h - 2\,,
  \ee
which are the derivatives of the $b_0^{\vec{n}}$ coefficients with respect
to $h$.
The matrix elements $\avac{h}L_2\left.\del_{\theta}^k\vac{\chi^{(2)}_{h,c}}
\right|_{\theta=0}$, $k=0,1$, do give us
further constraints, namely
  \be
  c = -2h\frac{8h-5}{2h+1}\,,\ \ \ \
  0 = -2h\frac{(4h+5)(4h-1)}{2h+1}\,.
  \ee
{}From these we learn that only for $h\in\{0,\frac{-5}{4},\frac{1}{4}\}$
we may have a logarithmic null vector (with $c=0,25,1$ respectively).
Therefore, the level 2 null vector for $h=\frac{-1}{4}$ of the $c=-7$
LCFT is just an ordinary one.
 
Next, we look at the level 4 null vector with the general ansatz
  \begin{eqnarray*}
  %\!\!\!\!\!\!\!\!
  \lefteqn{\vac{\chi^{(4)}_{h,c}}  = }\\
  & & \left(b^{\{1,1,1,1\}}_0L_{-1}^4 + b^{\{2,1,1\}}_0L_{-2}L_{-1}^2
      + b^{\{3,1\}}_0L_{-3}L_{-1} + b^{\{2,2\}}_0L_{-2}^2
      + b^{\{4\}}_0L_{-4}\right)
  \vac{h;a(\theta)}\\
  &+&
  \left(b^{\{1,1,1,1\}}_1L_{-1}^4 + b^{\{2,1,1\}}_1L_{-2}L_{-1}^2
      + b^{\{3,1\}}_1L_{-3}L_{-1} + b^{\{2,2\}}_1L_{-2}^2
      + b^{\{4\}}_1L_{-4}\right)
  \vac{h;\del_{\theta}a(\theta)}\nonumber\,.
  \end{eqnarray*}
Considering the possible matrix elements determines the coefficients up to
overall normalization as
  \bea
  b^{\{1,1,1,1\}}_0 &=&
  h^4(1232h^3-2466h^2-62h^2c+1198h-296hc+13hc^2+5c^3+92c^2\nonumber\\
                    & &+\,128c-144)\,,
  \nonumber\\
  b^{\{2,1,1\}}_0   &=&
  -4h^4(1120h^4-2108h^3+140h^3c+428h^2-66h^2c+338h-323hc\nonumber\\
                    & &+\,90hc^2+60c^2-78+99c)\,,\nonumber\\
  b^{\{3,1\}}_0     &=&
  24h^4(96h^5-332h^4+44h^4c+382h^3-8h^3c+4h^3c^2-53h^2c+12h^2c^2
  \nonumber\\
                    & &-\,235h^2+11hc^2+14hc+65h-6+3c+3c^2)\,,\nonumber\\
  b^{\{2,2\}}_0     &=&
  24h^4(32h^3-36h^2+4h^2c+8hc+22h+3c-3)(3h^2+hc-7h+2+c)
  \,,\nonumber\\
  b^{\{4\}}_0       &=&
  -4h^4(550h+3c^3-224h^2c+66hc^2+748h^3-48+2508h^4+11hc^3\nonumber\\
                    & &+\,41h^2c^2-40h^3c-3008h^5+12h^2c^3+120h^3c^2-184h^4c
                       +102hc+27c^2\nonumber\\
                    & &-\,1698h^2+18c+4h^3c^3+768h^6+448h^5c+76h^4c^2)\,.
  \eea
Even for ordinary null vectors at level 4 we have $p(4)=5$ conditions,
but due to the freedom of overall normalization only 4 conditions have
been used so far. The last, $\avac{h}L_{4}\left.\vac{\chi^{(4)}_{h,c}}
\right|_{\theta=0} = 0$, fixes the central charge as a function of the
conformal weight to
  \be c\in\left\{
  -2\frac{h(8h-5)}{2h+1},
  -\frac{2}{5}\frac{8h^2+33-41h}{3+2h},
  -\frac{3h^2-7h+2}{h+1},
  1-8h
  \right\}\,.
  \ee
If we again put $b^{\vec{n}}_1(h,c)=\del_h b^{\vec{n}}_0(h,c)$ such that
the null vector conditions take on the form of total derivatives with
respect to $h$ we get the
additional constraint $\avac{h}L_{4}\left.\del_{\theta}^{}
\vac{\chi^{(4)}_{h,c}}\right|_{\theta=0} = 0$. That result in the
terribly lengthy polynomial 
  \bea
  \label{eq:loglevel4}
  0 &=&
        -4h^3(-14308h^3c^2+6600h-528c+30hc^3+1239840h^5-113592h^2+5290hc
        \nonumber\\
    & & +\,144c^2+462h^2c^3+4368h^3c^3+275hc^4+360h^2c^4+3296h^4c^3+74240h^6c
        \nonumber\\
    & & +\,25632h^5c^2+67584h^7+595224h^3-25812h^2c-12712h^3c+11574h^2c^2
        \nonumber\\
    & & -\,2475hc^2-1287136h^4+60c^4-249408h^5c+324c^3-12192h^4c^2-504320h^6
        \nonumber\\
    & & +\,187040h^4c+140h^3c^4)
        \,,
  \eea
in which we may insert the four solutions for $c$ to obtain sets of discrete
conformal weights (and central charges in turn). We skip these 
straightforward but tedious
explicit calculations for all the possible cases, which one may find in
the third reference of \cite{Flohr:1996a}. We note that a good check of
whether one has done the calculations right is, as a rule of thumb,
whether this last condition,
which after insertion of $c=c(h)$ is a polynomial solely in $h$, 
factorizes.

\bq
Omitting trivial (non logarithmic) solutions, all logarithmic singular
vectors with respect to the Virasoro algebra at level $n=4$ are:
%{\vskip-0.75ex
  $$
  \begin{array}{c|c}
  (h,c) & \vac{\chi^{(4)}_{h,c}}
    \\\noalign{\smallskip}\hline\hline\noalign{\smallskip}
  (-\frac{1}{4},-7) &
    (315L_{-1}^4+315L_{-2}^2-210L_{-3}L_{-1}-210L_{-4}-1050L_{-2}L_{-1}^2)
    \Vac{\frac{-1}{4};a(\theta)}
    \\\noalign{\smallskip} &
    +\,(-878L_{-3}L_{-1}+2577L_{-1}^4-11830L_{-2}L_{-1}^2+3657L_{-2}^2
    -1718L_{-4})\Vac{\frac{-1}{4};\del_{\theta}a(\theta)}
    \\\noalign{\smallskip}
  (0,-2) &
    (L_{-1}^4 - 2L_{-2}L_{-1}^2 - 2L_{-3}L_{-1})\Vac{0;a(\theta)}
    +2L_{-4}\Vac{0;\del_{\theta}a(\theta)}
    \\\noalign{\smallskip}
  (\frac{3}{8},-2) &
    (1260L_{-1}^4+2835L_{-2}^2+1260L_{-3}L_{-1}-1890L_{-4}-6300L_{-2}L_{-1}^2)
    \Vac{\frac{3}{8};a(\theta)}
    \\\noalign{\smallskip} &
    +\,(3832L_{-3}L_{-1}+2152L_{-1}^4-14120L_{-2}L_{-1}^2+9882L_{-2}^2
    -7008L_{-4})\Vac{\frac{3}{8};\del_{\theta}a(\theta)}
    \\\noalign{\smallskip}
% \end{array}
% $$
% $$
% \begin{array}{c|l}
  (0,1) &
    (-3L_{-1}^4 + 12L_{-2}L_{-1}^2 - 6L_{-3}L_{-1})\Vac{0;a(\theta)}
    +(-16L_{-2}^2+12L_{-4})\Vac{0;\del_{\theta}a(\theta)}
    \\\noalign{\smallskip}
  (1,1) &
    (-60L_{-1}^4+240L_{-2}L_{-1}^2+120L_{-3}L_{-1}-240L_{-4})\Vac{1;a(\theta)}
    \\\noalign{\smallskip} &
    +\,(-89L_{-1}^4+476L_{-2}L_{-1}^2+118L_{-3}L_{-1}-716L_{-4})
    \Vac{1;\del_{\theta}a(\theta)}
    \\\noalign{\smallskip}
  (\frac{9}{4},1) &
    (45L_{-1}^4+405L_{-2}^2+630L_{-3}L_{-1}-810L_{-4}-450L_{-2}L_{-1}^2)
    \Vac{\frac{9}{4};a(\theta)}
    \\\noalign{\smallskip} &
    +\,(1996L_{-3}L_{-1}+110L_{-1}^4-1220L_{-2}L_{-1}^2+1206L_{-2}^2
    -2772L_{-4})\Vac{\frac{9}{4};\del_{\theta}a(\theta)}
    \\\noalign{\smallskip}
  (-\frac{21}{4},25) &
    (-990L_{-1}^4-8910L_{-2}^2-33660L_{-3}L_{-1}-65340L_{-4}
    -9900L_{-2}L_{-1}^2)\Vac{\frac{-21}{4};a(\theta)}
    \\\noalign{\smallskip} &
    \,+(45946L_{-3}L_{-1}+901L_{-1}^4+11650L_{-2}L_{-1}^2+12861L_{-2}^2
    +102234L_{-4})\Vac{\frac{-21}{4};\del_{\theta}a(\theta)}
    \\\noalign{\smallskip}
  (-3,25) &
    (63504L_{-1}^4+254016L_{-2}L_{-1}^2+635040L_{-3}L_{-1}+762048L_{-4})
    \Vac{-3;a(\theta)}
    \\\noalign{\smallskip} &
    +\,(59283L_{-1}^4+110124L_{-2}L_{-1}^2+148302L_{-3}L_{-1}+76356L_{-4})
    \Vac{-3;\del_{\theta}a(\theta)}
    \\\noalign{\smallskip} &
    +\,(-15104L_{-1}^4-186920L_{-2}L_{-1}^2-63504L_{-2}^2-450920L_{-3}L_{-1}
    -575628L_{-4})\Vac{-3;\del_{\theta}^2a(\theta)}
    \\\noalign{\smallskip}
  (-\frac{27}{8},28) &
    (77220L_{-1}^4+173745L_{-2}^2+849420L_{-3}L_{-1}+1042470L_{-4}
    +386100L_{-2}L_{-1}^2)\Vac{\fr{-27}{8};a(\theta)}
    \\\noalign{\smallskip} &
    +\,(269896L_{-3}L_{-1}+71336L_{-1}^4+150760L_{-2}L_{-1}^2-148374L_{-2}^2
    +113616L_{-4})\Vac{\fr{-27}{8};\del_{\theta}a(\theta)}
    \\\noalign{\smallskip}
  (-2,28) &
    (13860L_{-2}L_{-1}^2+27720L_{-3}L_{-1}+27720L_{-4}+6930L_{-1}^4)
    \Vac{-2;a(\theta))}
    \\\noalign{\smallskip} &
    +\,(1577L_{-1}^4-9716L_{-2}L_{-1}^2-3564L_{-2}^2-18640L_{-3}L_{-1}
    -21412L_{-4})\Vac{-2;\del_{\theta}a(\theta)}
    \\\noalign{\smallskip}
  (-\frac{11}{4},33) &
    (208845L_{-1}^4+696150L_{-2}L_{-1}^2+208845L_{-2}^2+1253070L_{-3}L_{-1}
    +1253070L_{-4})\Vac{\fr{-11}{4};a(\theta)}
    \\\noalign{\smallskip} &
    +\,(58354L_{-1}^4-244540L_{-2}L_{-1}^2-304086L_{-2}^2-525036L_{-3}L_{-1}
    -684156L_{-4})\Vac{\fr{-11}{4};\del_{\theta}a(\theta)}
  \end{array}
  $$ 
%}
It is worth mentioning that level $n=4$ is the smallest level where one
finds a logarithmic null vector of higher rank, namely a rank $r=3$
singular vector with $h=-3$ and $c=25$.
\eq

\noindent Here, we are only interested in the null vector for $h=
\frac{-1}{4}$. And indeed, the first two solutions for $c$ admit (among
others) $h=\frac{-1}{4}$ to satisfy (\ref{eq:loglevel4}) with the final
result for the null vector
  \bea
  \lefteqn{\Vac{\chi^{(4)}_{h=-1/4,c=-7}} =}\\
  & &\left(\fr{315}{128}L_{-1}^4-\fr{525}{64}L_{-2}L_{-1}^2
  +\fr{315}{128}L_{-2}^2-\fr{105}{64}L_{-3}L_{-1}
  -\fr{105}{64}L_{-4}\right)
  \Vac{\fr{-1}{4};(\alpha_1\theta^1+\alpha_0\theta^0)}\nonumber\\
  &+&\left(-\fr{2463}{128}L_{-1}^4+\fr{2485}{64}L_{-2}L_{-1}^2
  +\fr{1241}{64}L_{-3}L_{-1}-\fr{1383}{128}L_{-2}^2
  +\fr{821}{64}L_{-4}\right)
  \Vac{\fr{-1}{4};(\alpha_1\theta^0)}\,.\nonumber
  \eea
This shows explicitly the existence of a non-trivial logarithmic null vector
in the rank 2 Jordan cell indecomposable representation with highest weight
$h=\frac{-1}{4}$ of the $c_{3,1}=-7$ rational LCFT. Here, $\alpha_0,\alpha_1$
are arbitrary constants such that we may rotate the null vector arbitrarily
within the Jordan cell. However, as long as $\alpha_1\neq 0$, there is
necessarily always a non-zero component of the logarithmic null vector
which lies in the irreducible sub-representation. Although there is the
ordinary null vector built solely on $\vac{h;0}$, there is therefore no
null vector solely built on $\vac{h;1}$, once more demonstrating the fact
that these representations are indecomposable.

  \subsection{Kac determinant and classification of LCFTs}

As one might extrapolate from the ordinary CFT case, 
it is quite a time consuming
task to construct logarithmic null vectors explicitly. However, if we are
only interested in the pairs $(h,c)$ of conformal weights and central
charges for which a CFT is logarithmic and owns a logarithmic null vector,
we don't need to work so hard.

As already explained, logarithmic null vectors are subject to the condition
that there exist fields in the theory with identical conformal weights.
As can be seen from (\ref{eq:h}), there are always fields of identical
conformal weights if $c=c_{p,q}=1-6\frac{(p-q)^2}{pq}$ is from the minimal
series with $p>q>1$ coprime integers. However, such fields are to be
identified in these cases due to the existence of BRST charges
\cite{Fel89,FFK89}. Equivalently,
this means that there are no such pairs of fields within the truncated
conformal grid
  \be\label{eq:confgrid}
  H(p,q) \equiv \{h_{r,s}(c_{p,q}):0<r<|q|,0<s<|p|\}\,.
  \ee
It is worth noting that explicit calculations for higher level null
vectors along the lines set out above will also produce ``solutions'' 
for the well known null vectors 
in minimal models, but these ``solutions'' never have a non-trivial Jordan
cell structure. For example, at level 3 one finds a solution with
$c=c_{2,5}=-\frac{22}{5}$ and $h=h_{2,1}=h_{3,1}=-\frac{1}{5})$ which, however,
is just the ordinary one. This was to be expected because each Verma module
of a minimal model has precisely two null vectors (this is why all weights
$h$ appear twice in the conformal grid, $h_{r,s}=h_{q-r,p-s}$).
We conclude that logarithmic null vectors can only occur if fields of
equal conformal weight still exist after all possible identifications due
to BRST charges (or due to the embedding structure of the Verma modules
\cite{FeFu83})
have been taken into account. For later convenience, we further define
the boundary of the conformal grid as
  \bea\label{eq:confbound}
  \del H(p,q)   &\equiv& \{h_{r,p}(c_{p,q}):0<r\leq |q|\} \cup
                         \{h_{q,s}(c_{p,q}):0<s\leq |p|\} \,,\\
  \del^2 H(p,q) &\equiv& \{h_{q,p}(c_{p,q})\}\,.\nonumber
  \eea
These three sets are in one-to-one correspondence with the possible three 
embedding structures of the associated 
Verma modules which are of type $III_{\pm}$, $III_{\pm}^{\circ}$,
and $III_{\pm}^{\circ\circ}$ respectively \cite{FeFu83}.

It has been argued that LCFTs are a very general kind of
conformal theories, containing rational CFTs as the special subclass of
theories without logarithmic fields. In the case of minimal models one can
show that logarithmic versions of a CFT with $c=c_{p,q}$ can be
obtained by augmenting the conformal grid. This can formally be achieved
by considering the theory with $c=c_{\alpha p,\alpha q}$. However, it
is a fairly difficult undertaking to calculate explicitly logarithmic
null vectors for augmented minimal models, the reason being
simply that the levels of such null vectors are rather large.
Let us look at minimal $c_{2n-1,2}$ models, $n>1$. Fields within the
conformal grid are ordinary primary fields which do not posses logarithmic
partners. Therefore, pairs of primary fields with logarithmic partners have
to be found outside the conformal grid and, as shown in
\cite{Flohr:1996a} and \cite{Gaberdiel:1996a},
must lie on the boundary $\del H(p,q)$ (note that the corner point
is not an element). Notice that for $c_{p,1}$
models this condition is easily met because the conformal grid $H(p,1)=
\emptyset$. Fields outside the boundary region which have the property
that their conformal weights are $h'=h+k$ with $h\in H(p,q)$, $k\in\mathbb{Z}_+$
do not lead to Jordan cells (they are just descendants of the primary fields).
For example,
the $c_{5,2}=-\frac{22}{5}$ model admits representations with
$h=h_{1,8}=h_{3,2}=\frac{14}{5}$ which do not form a logarithmic pair and
are just descendants of the $h=-\frac{1}{5}$ representation.
Therefore, even for the $c_{2n-1,2}$ models with their
relatively small conformal grid, the lowest level of a logarithmic null
vector easily can get quite large. In fact, the smallest minimal model, the
trivial $c_{3,2}=0$ model, can be augmented to a LCFT with formally
$c=c_{9,6}$ which has
a Jordan cell representation for $h=h_{2,2}=h_{2,4}=\frac{1}{8}$.
The logarithmic null vector already has level 8 and reads explicitly

\bq
  \begin{eqnarray*}
  \lefteqn{\Vac{\chi^{(8)}_{h=1/8,c=0}} =}\\
  & & \left(10800L_{-1}^8-208800L_{-2}L_{-1}^6+928200L_{-2}^2L_{-1}^4
      -1060200L_{-2}^3L_{-1}^2+151875L_{-2}^4+252000L_{-3}L_{-1}^5\right.\\
  & & -\,631200L_{-3}L_{-2}L_{-1}^3+207000L_{-3}L_{-2}^2L_{-1}
      -1033200L_{-3}^2L_{-1}^2+360000L_{-3}^2L_{-2}-1249200L_{-4}L_{-1}^4\\
  & & +\,4165200L_{-4}L_{-2}L_{-1}^2-1133100L_{-4}L_{-2}^2
      +176400L_{-4}L_{-3}L_{-1}+593100L_{-4}^2+624000L_{-5}L_{-1}^3\\
  & & -\,720000L_{-5}L_{-2}L_{-1}-429300L_{-5}L_{-3}+1206000L_{-6}L_{-1}^2
      -455400L_{-6}L_{-2}-206100L_{-7}L_{-1}\\
  & & \left.-\,779400L_{-8}\right)\Vac{\fr{1}{8},a(\theta)}\\
  & \!\!\!\!+\!\!\!\! &
      \left(76800L_{-3}L_{-2}L_{-1}^3+755200L_{-3}L_{-2}^2L_{-1}
      -2596800L_{-3}^2L_{-1}^2+106400L_{-3}^2L_{-2}+179712L_{-4}L_{-1}^4
      \right.\\
  & & +\,123648L_{-4}L_{-2}L_{-1}^2+3621120L_{-4}L_{-3}L_{-1}-857856L_{-4}^2
      +739200L_{-5}L_{-1}^3-5832000L_{-5}L_{-2}L_{-1}\\
  & & \left.+\,992800L_{-5}L_{-3}+3444000L_{-6}L_{-1}^2-154800L_{-6}L_{-2}
      -2210400L_{-7}L_{-1}+488000L_{-8}\right)\Vac{\fr{1}{8},\del a(\theta)}\,,
  \end{eqnarray*}
\eq

\noindent up to an arbitrary state proportional to the ordinary
level 4 null vector.
This shows that minimal models can indeed be augmented to logarithmic
conformal theories. Level 8 is actually the
smallest possible level for logarithmic null vectors of augmented minimal
models. On the other hand, descendants of logarithmic fields are also
logarithmic, giving rise to the more complicated staggered module
structure \cite{Rohsiepe:1996}. 
Thus, whenever for $c=c_{p,q}$ the conformal weight
$h=h_{r,s}$ with either $r\equiv 0$ mod $p$, $s\not\equiv 0$ mod $q$, or
$r\not\equiv 0$ mod $p$, $s\equiv 0$, the corresponding representation is
part of a Jordan cell (or a staggered module structure).

\bq
The question of whether a CFT is logarithmic really makes sense only in the
framework of (quasi-)rationality. Therefore, we can assume that $c$ and all
conformal
weights are rational numbers. It can then be shown that the only possible
LCFTs with $c\leq 1$ are the ``minimal'' LCFTs with $c=c_{p,q}$. Using the
correspondence between the Verma modules $V_{h,c}\leftrightarrow
V_{-1-h,26-c}$ one can further show that LCFTs with $c\geq 25$ might exist
with (formally) $c=c_{-p,q}$. Again, due to an analogous (dual) BRST
structure of these models, pairs of primary fields with logarithmic partners
can only be found outside the conformal grid
$H(-p,q) = \{h_{r,s}(c_{-p,q}):0<r<q,0<s<p\}$, a fact that
can also be observed in direct calculations. For
example, at level 4 we found a candidate solution with
$c_{-3,2}=26$ and $h_{4,1}=h_{1,3}=-4$. But again, the explicit calculation
of the null vector did not show any logarithmic part.
\eq

The existence of null vectors can be seen from the Kac determinant of the
Shapovalov form $M^{(n)} = \avac{h}L_{\vec{n}'}L_{-\vec{n}}\vac{h}$,
which factorizes into contributions for each level $n$.
The Kac determinant has the well known form
  \be\label{eq:kac}
  {\rm det}\,M^{(n)} =
  \prod_{k=1}^n\prod_{rs=k}\left(h-h_{r,s}(c)\right)^{p(n-rs)}\,.
  \ee
A consequence of the general conditions derived earlier is that a 
necessary condition for the existence
of logarithmic null vectors in rank $r$ Jordan cell representations of
LCFTs is that $\frac{\del^k}{\del h^k}\left({\rm det}\,M^{(n)}\right) = 0$
for $k=0,\ldots,r-1$. It follows immediately from (\ref{eq:kac}) that
non-trivial common zeros of the Kac determinant and its derivatives at level
$n$ only can come from the factors whose powers $p(n-rs)=1$, i.e. $rs=n$ and
$rs=n-1$. For example
  \bea\label{eq:dkac}
  \frac{\del}{\del h}\left({\rm det}\,M^{(n)}\right)& = &
  \sum_{n-1\leq rs\leq n}\frac{1}{\left(h-h_{r,s}(c)\right)}{\rm det}\,M^{(n)}
  \nonumber\\ &+&
  \sum_{1\leq rs\leq n-2}\frac{p(n-rs)}{\left(h-h_{r,s}(c)\right)}
  {\rm det}\,M^{(n)}\,,
  \eea
whose first part indeed yields a non-trivial constraint, whereas the second
part is zero whenever ${\rm det}\,M^{(n)}$ is. Clearly (\ref{eq:dkac})
vanishes
at $h=h_{r,s}(c)$ up-to one term which is zero precisely if there is one other
$h_{t,u}(c)=h$. This is the condition stated earlier. Solving it for the
central charge $c$ we obtain
\be\label{eq:csol}
  c = \left\{\begin{array}{l}
  {\displaystyle-\frac{(2t-3u+3s-2r)(3t-2u+2s-3r)}{(u-s)(t-r)}}\\
  {\displaystyle-\frac{(2t-3u-3s+2r)(3t-2u-2s+3r)}{(u+s)(t+r)}}
  \end{array}\right.\,.
  \ee
With an ansatz $c(x)\equiv 1-6\frac{1}{x(x+1)}$ we find
  \be
  x \in \left\{
  \frac{u-s}{t-r+s-u}, \frac{r-t}{t-r+s-u},
  \frac{s+u}{t+r-u-s}, \frac{t+r}{u+s-t-r}
  \right\}\,,
  \ee
i.e.\ $x\in\mathbb{Q}$.
This proves our first claim that logarithmic null vectors only appear in
the framework of (quasi-)rational CFTs. The further claims follow then from
the well known embedding structure of Verma modules for central charges with
rational $x$ (which by the way ensures $c\leq 1$ or $c\geq 25$, where at
the limiting points $x_{c\rightarrow 1}\rightarrow\infty$ and
$x_{c\rightarrow 25}\rightarrow-\frac{1}{2}$).
 
Obviously, null vectors in rank $r$ Jordan cells with conformal weight $h$
require the existence of $r$ different solutions $(r_i,s_i)$ such that
$h_{r_i,s_i}(c)=h$. Up to level 5 there is only one case with $r>2$, namely
the rank 3 logarithmic null vector of the $c=c_{-1,1}=25$ theory with
$h=h_{2,2}=h_{1,3}=h_{3,1}=-3$.

What remains is to find the numbers $r,s,t,u$ (or more generally $r_i,s_i$).
The allowed solutions must satisfy the conditions stated above:
A quadruplet $(r,s,t,u)$ parametrizes a logarithmic null vector, if
with $c=c(r,s,t,u)$ one of the solutions (\ref{eq:csol}) for the central
charge, both $h_{r,s}(c),h_{t,u}(c)\in\del H(c)$
where $H(c)\equiv H(x,x+1)$ is the conformal grid of the Virasoro CFT with
central charge $c=c(x)$. This gives the conformal weights of the ``primary''
logarithmic pairs, the other possibilities are of the form $h
\in\del H(c)$ mod $\mathbb{Z}_+$ and belong to ``descendant'' logarithmic pairs.
We use quotation marks because the logarithmic partner of a primary field
is not primary in the usual sense.
 
As an example, we consider the by now well known models with
$c=c_{p,1}$, $p>1$. Precisely all fields
in the extended conformal grid (obtained by formally considering
$c=c_{3p,3}$) except $h_{1,p}$ and $h_{1,2p}$ as well as their ``duals''
$h_{2,2p}$ and $h_{2,p}$ form triplets $(h_{1,r}=h_{1,2p-r},h_{1,2p+r})$
which constitute a rank 2 Jordan cell with an additional Jordan cell like
module staggered into it (for details see \cite{Rohsiepe:1996}). The excluded
fields form irreducible representations without any null vectors and are all
$\in\del^2 H(p,1)$ mod $\mathbb{Z}_+$. Similar results hold for the
$c=c_{-p,1}$, $p>1$,
models. However, all these LCFTs are only of rank 2.
The only cases of higher rank LCFTs seem to be particular $c=1$ and $c=25$
theories. Notice that such theories are necessarily {\em non-unitary\/}, i.e.\
the Shapovalov form is necessarily not positive definite. However, since
we are able to explicitly construct these theories, e.g.\ the explicit
null vectors in the Appendix, there is no doubt that these theories exist.
The reason is that the $c_{\pm p,1}$, $p>1$, theories still have additional
symmetries such that a truncation of the conformal grid to finite size still
can be constructed, while the $c=1$ and $c=25$ theories presumably are only
quasi-rational, their conformal grid being infinite in at least one direction.

\subsection{The $(h,c)$ plane}
 
It might be illuminating, and the author is fond of plots anyway,
to plot the sets $\del^k H(p,q)$, $k=0,1,2$, for a variety of CFTs.
The product $pq$ is roughly a measure for the size of the CFT since the
size of the conformal grid and thus the field content is determined by it.
Thus, it seems reasonable to plot all sets with $pq\leq n$ where we have
chosen $n=400$.
 
To make the structure of the $(h,c)$ plane better visible,
we transformed the variables via
  \be\label{eq:coord}
  x\mapsto {\rm sign}(x)\log(|x|+1)\ {\rm for}\ x=h,c\,,
  \ee
which amounts in a double logarithmic scaling of the axes both, in positive
as well as in negative direction. The conformal weights are plotted in
horizontal direction, the central charges along the vertical direction.
The following plots show only the part
of the $(h,c)$ plane which belongs to $c\leq 1$ CFTs, i.e.\ minimal models
and $c_{p,1}$ LCFTs, $p>0$. The other ``half'' with $c\geq 25$ shares
analogous features. Due to the map (\ref{eq:coord}) the vertical range of
roughly $[-5.5,1.0]$ corresponds to $-240\leq c\leq 1$, whereas the horizontal
range $[-5.5,5.0]$ does roughly correspond to $-240\leq h\leq 148$.
To guide the eye for better orientation, we give here for the labels
$\pm\{0,1,2,3,4,5\}$ the corresponding values of $h,c$, which are in the same
oder $\pm\{0,1.718,6.389,19.086,53.598,147.413\}$.

If one would put both plots above each other,
one might infer from them that the set of logarithmic representations
precisely lies on the ``forbidden'' curves of the point set of ordinary
highest weight representations. This illustrates the fact that logarithmic
representations appear, if the conformal weights of two highest weight
representations become identical.

\bq
As discussed in \cite{Flohr:1996a}, this
situation can for example arise in the limit of series of minimal models
$c_{p_1,q_1}, c_{p_2,q_2}, c_{p_3,q_3}, \ldots$ with
$\lim_{i\rightarrow\infty}p_iq_i=\infty$. Usually, the field content of
these theories increases with $i$, but it might happen that in the limit
$p_i$ and $q_i$ become almost coprime. More precisely, a sequence such as
for example $\{c_{\alpha p,(\alpha+1)q}\}_{\alpha\in\mathbb{Z}_+}$ converges
to a limiting theory with central charge
$\lim_{\alpha\rightarrow\infty}c_{\alpha p,(\alpha+1)q} = c_{p,q}$. Therefore,
we expect a rather small field content at the limit point since the conformal
weights of the $c_{\alpha p,(\alpha+1)q}$ theories also approach the ones of
the $c_{p,q}$ model (modulo $\mathbb{Z}$). 
A more detailed analysis (second reference
in \cite{Flohr:1996a}) reveals that indeed conformal weights approach each other
giving rise for Jordan cells. Hence, the theory at the limit point, while
having central charge $c_{p,q}$ actually is a LCFT.
The plots presented here clearly visualize this topology of the space of CFTs
in the $(h,c)$ plane of their spectra.
\eq

\WEPSFIGURE{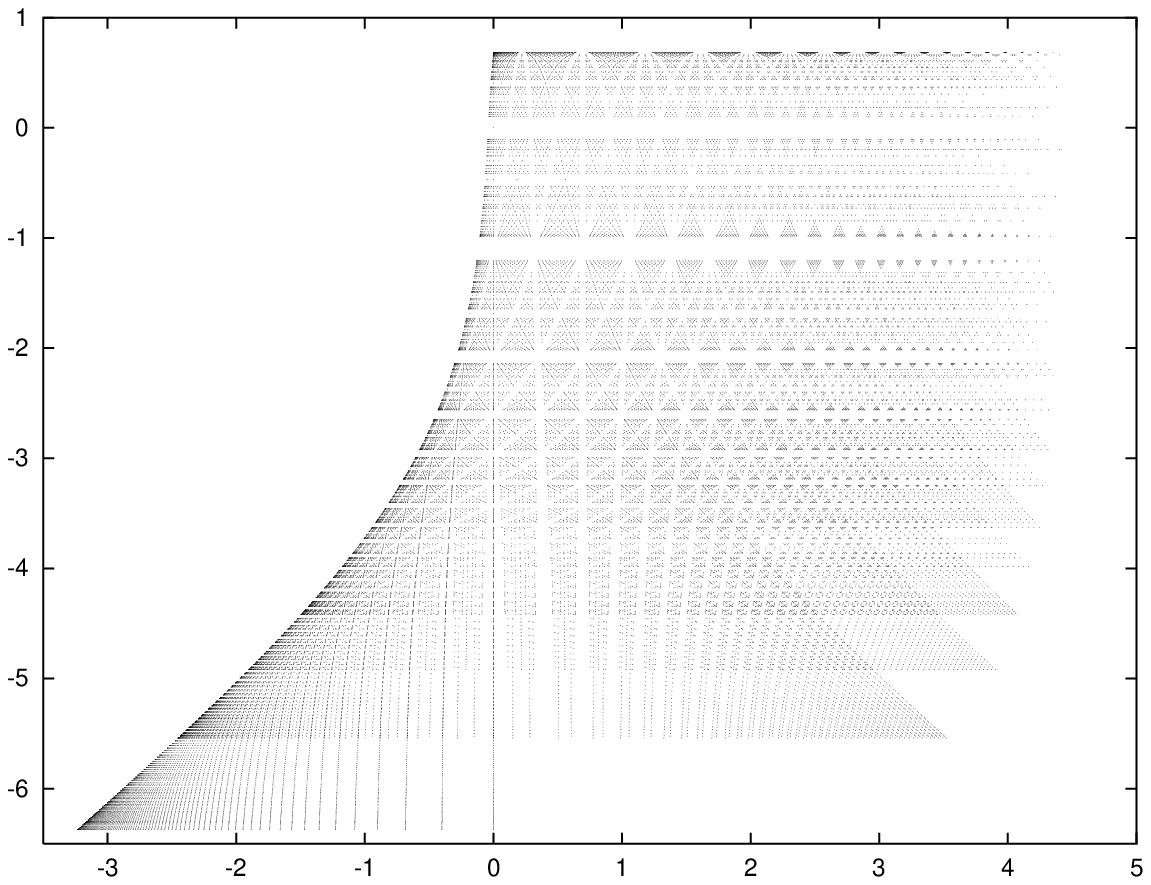,width=9cm}{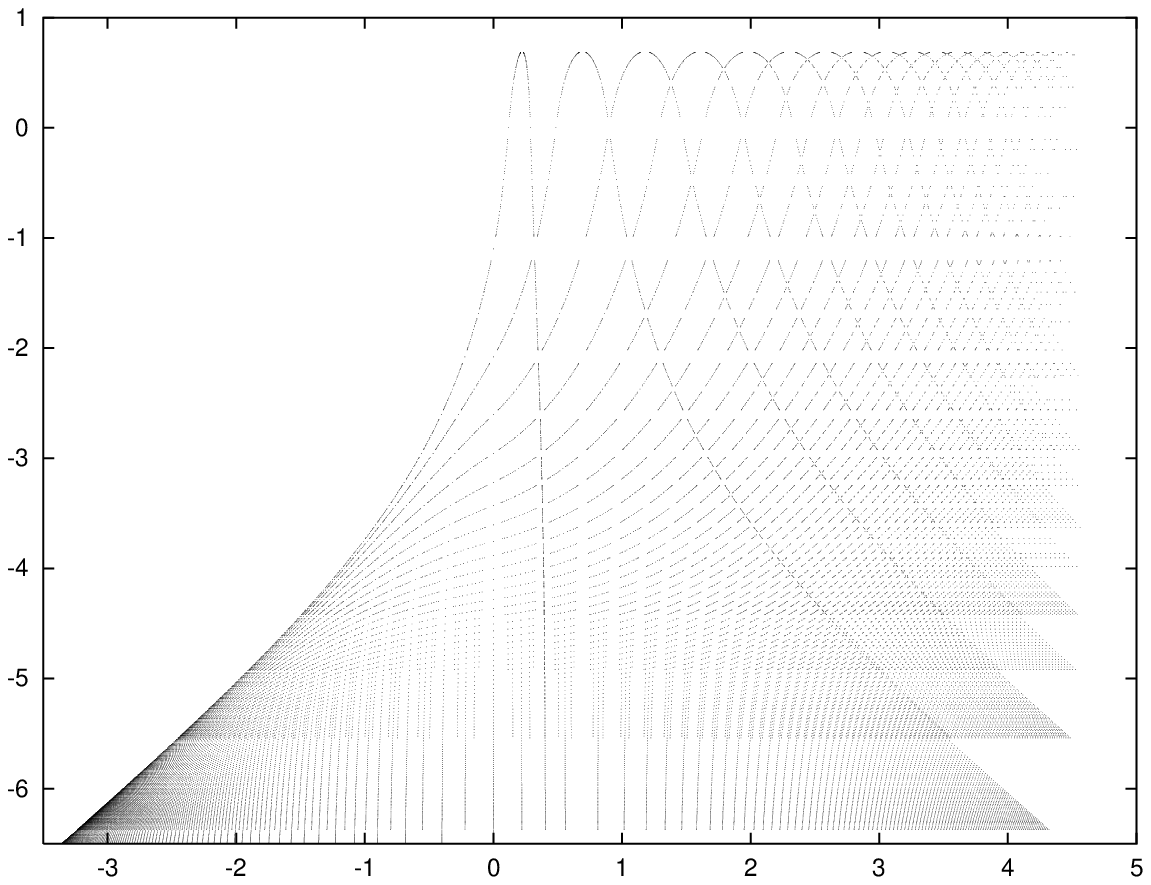,width=9cm}{
{\em Left:} Spectra $(H(p,q),c_{p,q})$ 
for $pq\leq 400$, which constitutes the set of all irreducible highest
weight representations of minimal models. We used a
logarithmic scaling $x\mapsto {\rm sign}(x)\log(|x|+1)$ for $x=h,c$, to
make the pattern structure of the spectra of minimal models better 
visible.
{\em Right:} Spectra $(\del H(p,q),c_{p,q})$ 
for $pq\leq 400$, which constitutes the set of all Jordan cell
representations, i.e.\ all conformal weights where fields with logarithmic
partners exist. The logarithmic scaling is the same as in the left figure
(cf.\ eq.\ \ref{eq:coord}).}
%\DOUBLEFIGURE{H400.eps,width=8cm}{dH400.eps,width=8cm}{
%Spectra $(H(p,q),c_{p,q})$ 
%for $pq\leq 400$, which constitutes the set of all irreducible highest
%weight representations of minimal models. We used a
%logarithmic scaling $x\mapsto {\rm sign}(x)\log(|x|+1)$ for $x=h,c$, to
%make the pattern structure of the spectra of minimal models better 
%visible.}{Spectra $(\del H(p,q),c_{p,q})$ 
%for $pq\leq 400$, which constitutes the set of all Jordan cell
%representations, i.e.\ all conformal weights where fields with logarithmic
%partners exist. The logarithmic scaling is the same as in the other figure
%(cf.\ eq.\ \ref{eq:coord}).}

To summarize, these results strongly suggest that augmented minimal models
form {\em rational\/} logarithmic conformal field theories in the same
sense as the $c_{p,1}$ models do. The only difference between the former and
the latter is that for the $c_{p,1}$ models $H(p,1) = \emptyset$. We know since
BPZ \cite{BPZ} that under fusion $H(p,q)\times H(p,q) \rightarrow H(p,q)$,
and since \cite{Gaberdiel:1996a,Flohr:1996a} 
that under fusion $H'(p,q)\times H'(p,q)
\rightarrow H'(p,q)$ with $H'(p,q)=\del H(p,q)\cup\del^2 H(p,q)$, if we deal
with the full indecomposable representations. Therefore, the only difficulty
can come from mixed fusion products of type $H(p,q)\times H'(p,q)$ which
traditionally (without logarithmic operators) would be zero due to decoupling.
However, the general formalism of OPEs in both, ordinary and logarithmic
CFTs, as presented in the basic CFT survey lectures,
yields non-zero fusion products. This can happen since we
pay the price that representations from $H(p,q)$ appear with non-trivial
multiplicities (because of the fact that the corresponding OPEs yield
fields on the right hand side with $h\in H(p,q)$ mod $\mathbb{Z}$, which have a
non-trivial dependency on the formal $\theta$ variables. In fact,
recent studies of non-trivial $c=0$ theories, which are important for
the description of disorder phenomena in condensed matter physics, 
show that the representations from $H(p,q)=H(2,3)$ do indeed appear in
high multiplicities. It has been observed that an augmented $c=0$ model
(which then is necessarily non-unitary) admits four fields of conformal
weight $h=2$ belonging to a non-trivial enlarged set of $h=0$ 
representations.

\section{Correlation functions}

  As we learned in the CFT lectures, 
  null vectors are the perhaps most important tool in CFT 
  to explicitly calculate correlation functions. In certain
  CFTs, namely the so-called minimal models, a subset of highest-weight
  modules possess infinitely many null vectors which, in principle, allow
  to compute arbitrary correlation functions involving fields only out
  of this subset. It is well known that global conformal covariance
  can only fix the two- and three-point functions up to constants.
  The existence of null vectors makes it possible to
  find differential equations for higher-point correlators, incorporating
  local conformal covariance as well.
  Now, we are going to pursue the question, how this can be
  translated to the logarithmic case.

  For the sake of simplicity, we will concentrate on the case where the
  indecomposable representations are spanned by rank two Jordan cells with
  respect to the Virasoro algebra. The abbreviation LCFT will refer to this
  case.
  To each such highest-weight Jordan cell $\{|h;1\ket,|h;0\ket\}$ belong
  two fields, an ordinary primary field $\Phi_h(z)$, and its logarithmic
  partner $\Psi_h(z)$. We recall that one then
  has $L_0|h;1\ket = h|h;1\ket + |h;0\ket$, $L_0|h;0\ket = h|h;0\ket$.
  Furthermore, the main scope will lie on the evaluation of four-point
  functions, since -- as we have already seen -- the partial differential
  equations induced by the existence of null vectors then reduce to
  ordinary differential equations in the one independent coordinate,
  the harmonic ratio. This is so, because in ordinary CFT, the four-point 
  function is fixed by global conformal covariance up to an arbitrary
  function $F(x,\bar x)$ of the harmonic ratio of the four points, 
  $x=\frac{z_{12}z_{34}}{z_{14}z_{32}}$ with the very common abbreviation
  $z_{ij}=z_i-z_j$. Although LCFTs do not share the property of ordinary
  CFTs that all correlation functions factorize entirely into chiral and
  anti-chiral halfs, it is still possible to consider these halfs
  separately, and we will do so.

\subsection{Consequences of global conformal covariance}

  Before we discuss global conformal covariance, one further remark is
  necessary. We assumed so far silently that operator product expansions
  of primary fields only produce primaries and their descendents on the
  right hand side. We already mentioned that this is not necessarily the
  case, since there are primary fields, the so-called pre-logarithmic fields,
  whose OPE with each other contains a logarithmic field. However, what
  we will continue to assume throughout the remaining part of these notes
  is that primary fields within Jordan cells do indeed only produce
  primaries in OPEs among each other. We will call such primary fields
  {\em proper primary fields}. We note that this is a widely made
  assumption throughout the LCFT literature, and that it is trivially
  true for the Jordan cell containing the identity.

  \bq
  The special case where the Jordan cell is formed out of fields with
  integer conformal weight deserves a comment. A primary field with
  integer conformal weight is typically a chiral local field. In other
  words, $\Psi_{(h;0)}(z)({\rm d}z)^h=\Psi_{(h;0)}(z')({\rm d}z')^h$
  transforms as a $h$-differential. That means
  in particular, that correlation functions involving this primary field
  at, say, coordinate $z$ have trivial monodromy at $z$. In particular,
  $z$ is never a branch point causing a possible multi-valuedness of
  the chiral correlation functions (more correctly of the conformal
  block) at $z$. Now let us consider an $n$-point function with several
  copies of this primary inserted at points $z_i$. Then, all the points
  $z_i$ have trivial monodromy. By contracting these successively via 
  operator product expansions, we never should produce a non-trivial
  monodromy or a branch cut in this process. Therefore, 
  in this particular setting, we can safely conclude
  that the OPE of this primary field with itself will only produce
  other primary chiral local fields and their descendents on the right
  hand side. Thus, primary fields with integer conformal weights are proper
  primaries.

  On the other hand, if $h\not\in\mathbb{Z}$, this is not necessarily
  the case. On the contrary, primary fields with non-integer weight
  do cause non-trivial monodromies around their point of insertion in
  a correlation function. Here, it might very well be possible, that
  several of such primary fields add up under contraction to a logarithmic
  field. For example, the $c=-2$ LCFT possesses a primary field $\mu$ of
  conformal weight $h=-1/8$. Since this field certainly has non-trivial
  monodromy, it should not surprise us that it turns out that its OPE
  with itself contains a logarithmic field. Actually, this field
  exactly behaves as a $\mathbb{Z}_2$ branchpoint. Therefore, insertion of
  $2g+2$ of these fields in a correlator on the complex plane is
  equivalent with considering the original correlator on a genus $g$
  hyper-elliptic curve (since the latter can always be represented as
  a double covering of the complex plane with $g+1$ branch cuts). 
  Contracting several of these fields via OPE eliminates branch cuts or
  leads to degenerate moduli due to infinitely thin handles. These
  in turn manifest themselves as logarithmic divergencies in the
  correlation functions. 
  Indeed, considering a four-point function
  of four $h=-1/8$ fields shows that the monodromy around one point $z$
  essentially is given by $z^{1/4}$. Thus, naively, contracting all four
  fields together to obtain a one-point function would yield a trivial
  monodromy. However, this is not the only possibility, and a single
  branch cut may remain leading to a logarithmic divergency.
  \eq

  Under this assumption that primaries in Jordan cells are proper,
  it was known for some time that in LCFT 
  already the two-point functions behave differently, and the most
  surprising fact is that the propagator of two (proper) primary fields 
  vanishes,
  $\bra\Phi_h(z)\Phi_{h'}(w)\ket = 0$. In particular, the norm of the
  vacuum, i.e.\ the expectation value of the identity, is zero. On the other
  hand, it can be shown (third reference of \cite{Flohr:1996a}) 
  that all LCFTs possess a logarithmic field
  $\Psi_0(z)$ of conformal
  weight $h=0$, such that with $|\tilde 0\ket = \Psi_0(0)|0\ket$ the
  scalar product $\bra 0|\tilde 0\ket = 1$. More generally, we have
  \bea\label{eq:2ptL}
     \bra\Phi_h(z)\Psi_{h'}(w)\ket &=& \delta_{hh'}\frac{A}{(z-w)^{h+h'}}\,,\\
     \bra\Psi_h(z)\Psi_{h'}(w)\ket &=& \delta_{hh'}\frac{
      B - 2A\log(z-w)}{(z-w)^{h+h'}}\,,\nonumber
  \eea
  with $A,B$ free constants. In an analogous way, the three-point functions
  can be obtained up to constants from the Ward-identities generated by
  the action of $L_{\pm 1}$ and $L_0$. Note that the action of the
  Virasoro modes is non-diagonal in the case of an LCFT,
  \be\label{eq:virL}
   L_n \bra\phi_1(z_1)\ldots\phi_n(z_n)\ket =  
    \sum_iz^n\left[z\partial_i + (n+1)(h_i+\hat{\delta}_{h_i})\right]
          \bra\phi_1(z_1)\ldots\phi_n(z_n)\ket
  \ee
  where $\phi_i(z_i)$ is either $\Phi_{h_i}(z_i)$ or $\Psi_{h_i}(z_i)$ and
  the off-diagonal action is described by some kind of step-operator
  $\hat{\delta}_{h_i}\Psi_{h_j}(z) = \delta_{ij}\Phi_{h_j}(z)$ and
  $\hat{\delta}_{h_i}\Phi_{h_j}(z) = 0$.
  Therefore, the action of the Virasoro modes yields additional terms
  with the number of logarithmic fields reduced by one. This action reflects
  the transformation behavior of a logarithmic field under conformal
  transformations,
  \be
    \phi_h(z) = \left(\frac{\partial f(z)}{\partial z}\right)^h
    \left(1 + \log(\partial_z f(z))\delta_h\right)\phi_h(f(z))\,.
  \ee

  An immediate consequence of the form of the two-point functions and
  the cluster property of a well-defined quantum field theory is that
  $\bra\Phi_{h_1}(z_1)\ldots\Phi_{h_n}(z_n)\ket = 0$, if all fields
  are primaries. Actually, this is only true if a correlator is considered,
  where all fields belong to Jordan cells. LCFTs do contain other primary
  fields, which themselves are not part of Jordan cells, and whose
  correlators are non-trivial. These are the twist-fields, which sometimes
  are also called pre-logarithmic fields (see first ref.\ in 
  \cite{Kogan:1998a}). Twist fields
  introduce non-trivial boundary conditions, since they behave exactly like
  branch cuts. Fusion of a twist with the corresponding anti-twist annihilates
  the branch cut but may leave a puncture, where for example screening integral
  contours may get pinched (for details see \cite{Flohr:1998b}).
  As a consequence, operator product expansions of two conjugate
  twist fields will produce contributions from Jordan cells of primary
  fields and their logarithmic partners. However, since the twist fields
  behave as ordinary primaries with respect to the Virasoro algebra, the
  computation of correlation functions of twist fields only can be
  performed as in the common CFT case. The solutions, however, may exhibit
  logarithmic divergences as well. Here, we are interested to compute
  correlators with logarithmic fields, instead.
 
  Another consequence is that
  \bea
    \bra\Psi_{h_1}(z_1)\Phi_{h_2}(z_2)\ldots\Phi_{h_n}(z_n)\ket
    &=&\bra\Phi_{h_1}(z_1)\Psi_{h_2}(z_2)\Phi_{h_3}(z_3)\ldots\Phi_{h_n}(z_n)
       \ket\nonumber\\
    =\ldots &=& \bra\Phi_{h_1}(z_1)\ldots\Phi_{h_{n-1}}(z_{n-1})\Psi_{h_n}(z_n)
    \ket\nonumber\,.
  \eea
  Thus, if only one logarithmic field is present, it does not matter,
  where it is inserted. Note that the action of the Virasoro algebra does
  not produce additional terms, since correlators without logarithmic fields
  vanish. Therefore, a correlator with precisely one logarithmic field
  can be evaluated as if the theory would be an ordinary CFT.

  The conformal Ward identities (\ref{eq:ward}) are now modified
  via the modified action of the Virasoro modes as given in (\ref{eq:virL}).
  This affects the general form to which global conformal covariance fixes
  correlation functions, e.g.\ the two-point function as given in
  (\ref{eq:2ptL}). It is a very good exercise left to the reader to
  compute the generic form of three-point functions for the simple case
  of a rank two LCFT. The general form of one-, two- and three-point
  functions for arbitrary rank LCFTs has been worked out in detail in
  the last ref.\ of \cite{Flohr:1996a}, 
  where also generic operator product expansion for
  logarithmic fields of arbitrary rank LCFTs have been computed.
  For the three-point functions one finds
  \bea\label{eq:r3ptL}
     \bra\Phi_{h_i}(z_i)\Phi_{h_j}(z_j)\Psi_{h_k}(z_k)\ket &=& 
       C_{ijk;1}(z_{ij})^{h_k-h_i-h_j}(z_{ik})^{h_j-h_i-h_k}
       (z_{jk})^{h_i-h_j-h_k}\,,\\
     \bra\Phi_{h_i}(z_i)\Psi_{h_j}(z_j)\Psi_{h_k}(z_k)\ket &=& 
       [C_{ijk;2}-2C_{ijk;1}\log z_{jk}]\nonumber\\
     &\times& (z_{ij})^{h_k-h_i-h_j}
       (z_{ik})^{h_j-h_i-h_k}(z_{jk})^{h_i-h_j-h_k}\,,\nonumber\\
     \bra\Psi_{h_i}(z_i)\Psi_{h_j}(z_j)\Psi_{h_k}(z_k)\ket &=& 
       \left[C_{ijk;3} - C_{ijk;2}(\log z_{ij} + \log z_{ik} 
       + \log z_{jk})^{\phantom{2}}\right.\nonumber\\
     & &{}+C_{ijk;1}(2\log z_{ij}\log z_{ik} 
       + 2\log z_{ij}\log z_{jk}
       + 2\log z_{ik}\log z_{jk}\nonumber\\ 
     & &\left.{}-\log^2 z_{ij} - \log^2 z_{ik}
         - \log^2 z_{jk})\right]\nonumber\\
     &\times& (z_{ij})^{h_k-h_i-h_j}
       (z_{ik})^{h_j-h_i-h_k}(z_{jk})^{h_i-h_j-h_k}\,,\nonumber
  \eea
  where the other two correlation functions with two logarithmic
  fields are given by accordingly made cyclic permutations of the
  middle equation. Note that the structure constants do only depend
  on the total number of logarithmic fields involved, not on the
  positions where these were inserted. These only betray themselves
  through the generic form of the coordinate dependent parts.
  General formul\ae\ for arbitrary rank LCFT, i.e.\ where there
  are more than one logarithmic partner field per primary, can be
  found in in the last ref.\ of \cite{Flohr:1996a}.

\bq
  To simplify matters even further, we have again assumed that
  the logarithmic partner field be quasi-primary, i.e.\ that
  $L_1|h;1\ket = 0$. This is not necessarily the case. However,
  for our discussion, it is sufficient -- even if
  the logarithmic partner is not quasi-primary -- that the state
  $L_1|h;1\ket$ be orthogonal to all states of fields actually 
  occurring within the considered correlation functions. To be
  more specific, if $L_1|h;1\ket\equiv|\xi\ket\neq 0$, but on
  the other hand $\bra h';n|\xi\ket = 0$, $n=0,1$, for all fields
  $\Phi_{h'}$ and $\Psi_{h'}$ occurring in the above considered
  correlation functions, then the non-quasi-primary reminder of the
  action of $L_1$ on $|h;1\ket$ does not affect the behavior of
  the correlation functions in question under global conformal
  transformations. To our best knowledge, this holds true in all
  explicitly known LCFTs where non-quasi-primary logarithmic partner
  fields exist. However, a more detailed discussion of this issue is
  rather technical, and beyond the scope of these notes.
\eq

  We have argued above that the correlation function with only
  one logarithmic field is completely independent of where this
  logarithmic field is inserted. Furthermore, we have seen in the
  above examples of two- and three-point functions that the
  structure constants also do not depend on where logarithmic fields
  are inserted. It can be shown on general grounds that this is 
  indeed always true. But it does {\em not\/} apply to the arbitrary functions 
  of harmonic ratios for higher $n$-point functions (one may think of
  these functions of harmonic ratios as ``structure functions'').
  Hence, it is more difficult to find the general form
  of four-point functions, and the resulting expressions
  are a bit cumbersome, since the number of possible contractions
  of fields leading to logarithmic terms heavily grows with the number
  of logarithmic fields one can insert in a correlation function.
  But let us write them down anyway. With the common solution
  %\begin{equation}\label{eq:muij}
  $\mu_{ij} = H/3-h_i-h_j$, $H=\sum_ih_i$,
  %\end{equation}
  we obtain in condensed notation:
  \bea
    \bra\Phi_i\Phi_j\Phi_k\Psi_l\ket &=&
      \prod_{r<s}z_{rs}^{\mu_{rs}}F^{(0)}(x)\,,\\
    \bra\Phi_i\Phi_j\Psi_k\Psi_l\ket &=&
      \prod_{r<s}z_{rs}^{\mu_{rs}}\left[F^{(1)}_{kl}(x) - 2F^{(0)}(x)
      \log(z_{kl})\right]\,,\\
    \bra\Phi_i\Psi_j\Psi_k\Psi_l\ket &=&
      \prod_{r<s}z_{rs}^{\mu_{rs}}\left[ F^{(2)}_{jkl}(x)\right.
       -\!\!\sum_{{r<s\in\{jkl\}\atop t=\{jkl\}-\{rs\}}}\!\!
       (F^{(1)}_{rt}(x)+F^{(1)}_{st}(x)-F^{(1)}_{rs}(x))\log(z_{rs})\nonumber\\
    & &{}+ F^{(0)}(x)(2\!\!\sum_{{r<s\in\{jkl\}\atop t=\{jkl\}-\{rs\}}}\!\!
       \log z_{rt}\log z_{ts} - \!\!\sum_{r<s\in\{jkl\}}\!\!\log^2 z_{rs})
       \left.\vphantom{\sum}\right]\,,
  \eea
  where
  other choices for the places of insertions of logarithmic fields are simply
  obtained by renaming the indices. Note the occurrence of non-trivial linear
  combinations $F^{(1)}_{rt}(x) + F^{(1)}_{st}(x) - F^{(1)}_{rs}(x)$, which
  is fixed by global conformal invariance. Moreover, we have introduced
  the notation $F^{(\ell)}_{i_1\ldots i_{\ell+1}}(x)$ to denote an
  arbitrary function of the harmonic ration $x$ belonging to a term
  which stems from a correlator of $\ell+1$ logarithmic fields inserted
  at the coordinates $z_{i_1},\ldots z_{i_{\ell+1}}$. Due to the
  off-diagonal action of the Virasoro modes (\ref{eq:virl}), which reduces
  the number of logarithmic fields by one, correlation functions involve
  all such functions $F^{(m)}_{j_1,\ldots,j_{m+1}}(x)$ with
  $m\leq \ell$ and $\{j_1,\ldots,j_{m+1}\}\subset\{i_1\ldots i_{\ell+1}\}$.
  Finally, the four-point function of four logarithmic fields has
  the lengthy form
  \bea
    \bra\Psi_1\Psi_2\Psi_3\Psi_4\ket &=&
      \prod_{i<j}z_{ij}^{\mu_{ij}}\left[
        F^{(3)}_{1234}(x)\right.\nonumber\\
      &-&\!\!\sum_{{r<s\atop \{t,u\}=\{jklm\}-\{rs\}}}\!\!\left({\ts\frac13}(
        2F^{(2)}_{rtu}(x)+2F^{(2)}_{stu}(x)-F^{(2)}_{rst}(x)-F^{(2)}_{rsu}(x))
        \log z_{rs}\right.\nonumber\\
      & &{}-\left.{\ts\frac12}(2F^{(1)}_{rs}(x)
          -F^{(1)}_{rt}(x)-F^{(1)}_{ru}(x)-F^{(1)}_{st}(x)-F^{(1)}_{su}(x))
        \log^2 z_{rs}\right)\nonumber\\
      &-&\!\!\sum_{{\{rst\}\subset\{jklm\}\atop u\in\{jklm\}-\{rst\}}}\!\!
        {\ts\frac12}(
        F^{(1)}_{rs}(x)+F^{(1)}_{st}(x)-F^{(1)}_{rt}(x)-2F^{(1)}_{su}(x))
        \log z_{rs}\log z_{st}\nonumber\\
      &-&\!\!\sum_{{r\neq s\neq t\neq u\atop r<u}}\!\!F^{(0)}(x)(
        2\log z_{rs}\log z_{st}\log z_{tu} - \log^2z_{ru}\log z_{st})\nonumber\\      &+&\!\!\sum_{{r\neq s\neq t\atop r<t}}\!\!\left.2F^{(0)}(x)
        \log z_{rs}\log z_{st}\log z_{tr}\vphantom{\sum}\right]\,.
  \eea
  Therefore, the full solution for the four-point function of an
  LCFT involves twelve (!) different functions $F^{(r)}_{i_1\ldots i_{r+1}}(x)$,
  $0\leq r\leq 3$.\footnote{Due to crossing symmetry, these twelve functions
  are not really all independent of each other. However, at this stage it
  is much simpler to denote the functions in this ``over-counting'' way.}
  In a similar way, one can make an $SL(2,\mathbb{C})$ covariant ansatz
  for a generic $n$-point function of Jordan cell fields. These results
  generalize the expressions obtained in \cite{Kausch:1995} for the $h=0$
  Jordan cell of the identity field.

  \EPSFIGURE{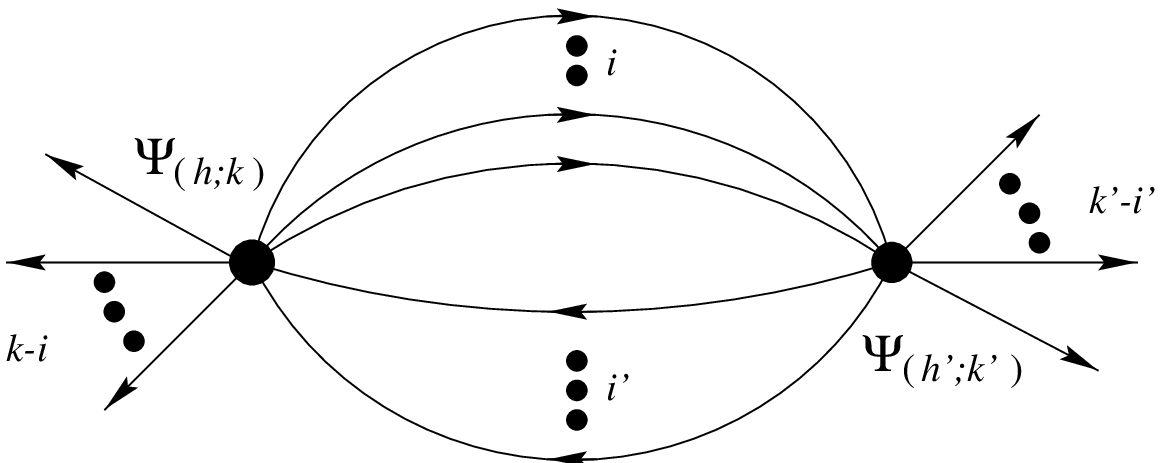,width=6.5cm}{Graphical representation of
    contractions of logarithmic fields leading to a (maximal) power
    $\log(z-z')^{i+i'}$ of logarithmic divergencies.}
  The power $s$ with which a logarithm
  $\log^s z_{ij}$ may occur is determined in the following way:
  Each logarithmic field $\Psi_{(h;\ell)}$ may be thought of as a vertex 
  with $\ell$ outgoing legs if it is the $(\ell+1)$-th field in the Jordan 
  block. That is, a vertex may have $0,\ldots,r-1$ outgoing legs for a rank
  $r$ Jordan block. Moreover, such a vertex may receive the same number of
  incoming legs. Then, the maximal power $s$ is the maximal number
  of lines joining two possible vertices, i.e.\ $\ell_i+\ell_j\leq 2(r-1)$.
  This number may be decreased by the further requirement that in a
  rank $r$ LCFT, precisely $r-1$ legs must remain uncontracted, i.e.\
  must not be joined to any vertex. Thus, an $n$-point function with only
  one logarithmic field is non-zero only if this is the top field in
  the Jordan block $\Psi_{(h;r-1)}$, its legs do not link to any
  other vertex (i.e.\ field), and all other fields do not carry legs,
  because they are all primary. If more than one logarithmic field
  is present, the correlation function will essentially be a sum over
  all possible graphs where all but $r-1$ legs are linked to arbitrary
  vertices. (One may think of the requirement of $r-1$ free legs also
  in the way that $r-1$ legs have to be linked to the point infinity.)
  Unfortunately, this consideration does not give the relative
  numerical factors of the different terms associated to different
  graphs. But we can immediately infer from this consideration that
  the maximal power of any logarithmic term obviously is limited
  by $2(r-1)$, which in the main scope of these notes, $r=2$, is simply
  two. In fact, although the above formula for the four-point function
  involves terms with upto three logarithms, there is no single
  logarithm with a power larger two. Also, since one of the four
  legs must remain unjoined, the total number of logarithms per
  monomial cannot exceed three in a four-point function, or $(n-1)$ in
  an $n$-point function, respectively.

  \EPSFIGURE{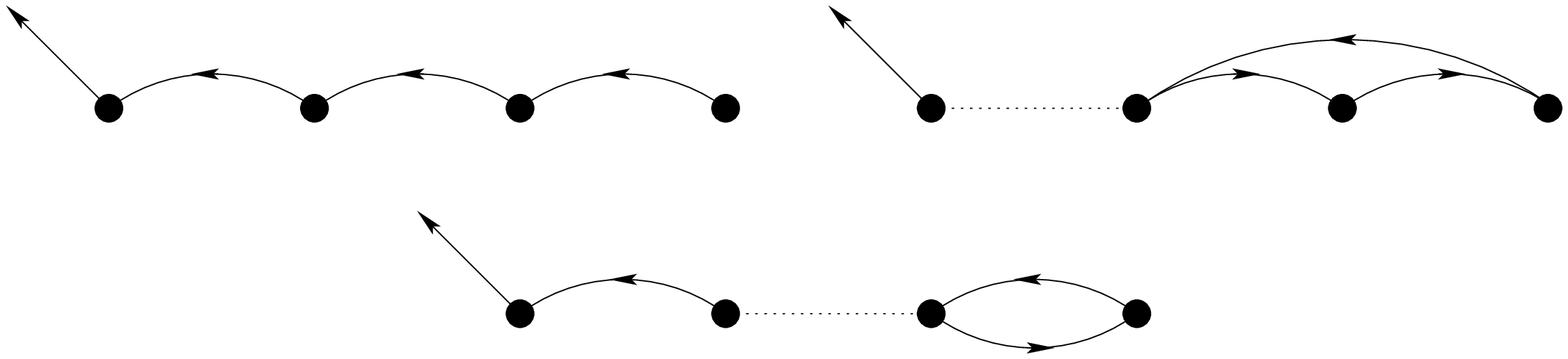,width=6.5cm}{All inequivalent graphical representations
    of contractions in a four-point function with four logarithmic fields.
    All other graphs are equivalent to these by relabeling.}
  It is a useful exercise to draw all the possible graphs for the
  four-point functions of a rank two LCFT along the lines we have
  just discussed. The careful reader will notice that in this simple
  case, neither the arrows nor the remaining free leg are really necessary,
  so that we only need to draw the inequivalent graphs where $n-1$ lines
  join upto $n$ vertices, and no vertex gets more that two lines.
  More generally, the task is to distribute $\sum_{i=1}^nk_i - (r-1)$ lines,
  $0\leq k_i\leq r-1$ being the levels of the (logarithmic) fields within
  the Jordan blocks, onto $n$ vertices such that 
  each vertex $i$ can maximally receive $(r-1)$ legs. The last condition is
  equivalent to saying that the resulting graph must be a $2(r-1)$-vertex
  graph, or that there are at most $(r-1)$ loops.
  However, it is clear that this will become for $r>2$ quite
  a non-trivial combinatorial task. 

  \subsection{Correlation functions, OPEs  and locality}

\bq
  Although we will in practice avoid this issue, we would like to mention
  briefly what it is about. Physical observables should be single valued
  functions of the parameters which can be influenced by the experiment.
  In quantum field theory, correlation functions are the mathematical
  objects which correspond to physical observable entities.

  We almost always have written down, and will continue to do so, only half
  of the theory, since we only denoted how fields depend on $z$, not on 
  $\bar z$. In generic non-logarithmic CFT one has the property that
  correlation functions factorize into holomorphic and anti-holomorphic
  parts such that it is sufficient to look at one half. The complete
  theory can then always easily be reconstructed. It is known that this
  is not any longer true in LCFT. Gurarie pointed out that even quantities
  which in itself do not involve logarithmic fields directly, do not
  factorize \cite{Gurarie:1993}. A thorough study for the $c=-2$ LCFT
  has bee carried out in a series of papers \cite{Gaberdiel:1996a}, where
  a (unique) local $c=-2$ theory has been constructed. 

  The general results on chiral correlation functions we have obtained so far
  are sufficient to suggest a simple recipe for writing down local
  version of them. We briefly recapitulate the results for generic LCFTs
  with Jordan cells built from quasi-primary fields, where primary fields
  are proper primaries. The LCFT is supposed
  to have rank $r$ Jordan cells $\{\Psi_{(h;k)} | 0\leq k\leq r-1\}$. 
  One can show that the rank of the Jordan cell of the identity determines
  the rank of all other Jordan cells, which we therefore assume to be all
  equal to $r$.

  One-point functions
  $\bra\Psi_{(h;k)}(z)\ket\equiv E_{(h;k)}$ must be constant due to
  translational invariance, and are restricted by the condition
  $$
    hE_{(h;k)} + (1-\delta_{k,0})E_{(h;k-1)} = 0\,.
  $$
  Under the usual assumptions, $E_{(h;k)}=0$ for $h\neq 0$ which leaves
  only $E_{(h=0;r-1)}\neq 0$ which we normalize to one.

  Two-point functions can be computed upto structure constants
  $D_{(h_1,h_2;k)}$ by global conformal covariance alone, yielding
  \be\label{eq:L2pt} 
    \bra\Psi_{(h_1;k_1)}(z_1)\Psi_{(h_2;k_2)}(z_2)\ket =
    \delta_{h_1,h_2}\left(\sum_{\ell=0}^{k_1+k_2}D_{(h_1,h_2;k_1+k_2-\ell)}
    \frac{(-2)^{\ell}}{\ell!}\log^{\ell}(z_{12})\right)(z_{12})^{-h_1-h_2}\,.
  \ee
  Here, we have indicated the implicit condition $h_1=h_2$. For a rank $r$
  LCFT, all constants $D_{(h,h;k)}=0$ for $k < r-1$. Note that the 
  structure constants depend only on one label for the level within a
  Jordan cell. In this way,
  the two-point functions define for each possible conformal weight $h$
  matrices $G^{(2)}_{k_1,k_2}$ of size $r\times r$. However, these
  matrices depend only on $2r-r$ yet undetermined constants
  $D_{(h,h;k)}$, $r-1\leq k\leq 2r-2$. Moreover, all entries above the
  anti-diagonal are zero.

  The three-point functions can be fixed along the same lines upto 
  constants $C_{(h_1,h_2,h_3;k)}$.
  A closed formula of the type as given above for the two-point function
  is extremely lengthy. However, the three-point functions can all be
  given in the form:
  \begin{eqnarray}\label{eq:L3pt}
    & & \bra\Psi_{(h_1;k_1)}(z_1)\Psi_{(h_2;k_2)}(z_2)\Psi_{(h_3;k_3)}(z_3)\ket
    = \sum_{k=r-1}^{k_1+k_2+k_3}C_{(h_1,h_2,h_3;k)}
    \sum_{j_1=0}^{k_1}\sum_{j_2=0}^{k_2}\sum_{j_3=0}^{k_3}
    \delta_{j_1+j_2+j_3,k_1+k_2+k_3-k}\phantom{mmmn} \nonumber\\
    & & \phantom{mmmm}\times\
    \frac{1}{j_1!j_2!j_3!}(\partial_{h_1})^{j_1}(\partial_{h_2})^{j_2}
    (\partial_{h_3})^{j_3}\left(z_{12}^{h_3-h_1-h_2}z_{13}^{h_2-h_1-h_3}
    z_{23}^{h_1-h_2-h_3}\right)\,.
  \end{eqnarray}
  The corresponding formula for the two-point function can be rewritten
  in the same manner involving derivatives with respect to the conformal
  weight,
  \begin{eqnarray}
    & & \bra\Psi_{(h_1;k_1)}(z_1)\Psi_{(h_2;k_2)}(z_2)\ket
    = \sum_{k=r-1}^{k_1+k_2}\delta_{h_1,h_2}D_{(h_1,h_2;k_1+k_2-k)}
    \frac{1}{k!}(\partial_{h_2})^{k}(z_{12})^{-2h_2}\,,
  \end{eqnarray}
  which evaluates to exactly the form given in (\ref{eq:L2pt}).
  Note that again the yet free structure constants depend only on the
  total level within the Jordan cells, i.e.\ on the sum of the individual
  levels. This agrees with what one might expect from the total
  symmetry of the three-point structure constants under permutations.
  Differentiation with respect to the conformal weights reproduces
  precisely the logarithmic contributions to satisfy the inhomogeneous
  Ward identities. 

  With the complete set of two- and three-point functions at hand, we can
  now proceed to determine the operator product expansions in their generic
  form. To do this, we first consider the asymptotic limit
  $$
    \lim_{z_1\rightarrow z_2}\bra\Psi_{(h_1;k_1)}(z_1)\Psi_{(h_2;k_2)}(z_2)
    \Psi_{(h_3;k_3)}(z_3)\ket
  $$ 
  and define
  matrices $(G^{(3)}_{k_1})^{}_{k_2,k_3}$ in this limit similar to the
  matrix of two-point functions. This essentially
  amounts to replacing $z_{13}$ by $z_{23}$. Next, we take the matrix of
  two-point functions $(G^{(2)})_{k_1,k_2}$ and invert it to obtain
  $(G^{(2)})^{\ell_1,\ell_2}$. Finally, the matrix product
  \begin{equation}\label{eq:O=CD}
    C_{(h_1,h_2;k_1+k_2)}^{(h_3;k_3)} = (G^{(3)}_{k_1})^{}_{k_2,k}
    (G^{(2)})^{k,k_3}
  \end{equation}
  yields matrices $(C_{(h_1;k_1),h_2}^{\ h_3})_{k_2}^{k_3}$ encoding all the
  OPEs of the field $\Psi_{(h_1;k_1)}(z)$ with fields of arbitrary level
  in their Jordan cells.
  
  This formula can be made a bit more explicit with the help of some
  notation. Let us denote the complete set of two--point functions
  as $\bra\ell,k\ket = \bra\Psi_{(h;\ell)}(z_2)\Psi_{(h;k)}(z_3)\ket$
  and correspondingly the
  three-point functions as $\bra\ell,k_1,k_2\ket =
  \lim_{z_1\rightarrow z_2}\bra\Psi_{(h_1;k_1)}(z_1)\Psi_{(h_2;k_2)}(z_2)
  \Psi_{(h;\ell)}(z_3)\ket$, all essentially
  given by formulae (\ref{eq:L2pt}) and (\ref{eq:L3pt}). Then, the OPEs take
  the structure
  \bea\label{eq:fope}
    & &\Psi_{h_1;k_1}(z_1)\Psi_{h_2;k_2}(z_2)\ =\ \sum_h\sum_{k=0}^{r-1}
    \left(\prod_{i=0}^{r-1}\bra i,r-1-i\ket\right)^{-1} \\
    & &\times\left|\begin{array}{ccccccc}
      \bra 0,0\ket  &\ldots&\bra   0,k-1\ket&\bra   0,k_1,k_2\ket&
          \bra   0,k+1\ket&\ldots&\bra   0,r-1\ket\\
      \vdots        &\ddots&\vdots          &\vdots              &
          \vdots          &\ddots&\vdots          \\
      \bra\ell,0\ket&\ldots&\bra\ell,k-1\ket&\bra\ell,k_1,k_2\ket&
          \bra\ell,k+1\ket&\ldots&\bra\ell,r-1\ket\\
      \vdots        &\ddots&\vdots          &\vdots              &
          \vdots          &\ddots&                \\
      \bra r-1,0\ket&\ldots&\bra r-1,k-1\ket&\bra r-s,k_1,k_2\ket&
          \bra r-1,k+1\ket&\ldots&\bra r-1,r-1\ket
    \end{array}\right|\Psi_{(h;k)}(z_2)\,,\nonumber
  \eea
  which in passing also proves that the matrix of two-point functions can be
  inverted without problems. Of course, the denominator is written here in a
  particularly symmetric
  way, it equals $\bra j,r-1-j\ket^r$ for any $0\leq j\leq r-1$. Note that the
  only non-zero entries above the anti-diagonal stem from the inserted column
  of three-point functions.

  With this result, we obtain in the simplest $r=2$ case the well known OPEs
  \begin{eqnarray}\label{eq:ope2c}
    \Psi_{(h_1;0)}(z)\Psi_{(h_2;0)}(0) &=& \sum_{h}
        \frac{C_{(h_1,h_2,h;1)}}{D_{(h,h;1)}^{}}\Psi_{(h;0)}(0)z^{h-h_1-h_2}
        \,,\nonumber\\
    \label{eq:ope2b}\Psi_{(h_1;0)}(z)\Psi_{(h_2;1)}(0) &=& \sum_{h}
        \left[\frac{C_{(h_1,h_2,h;1)}}{D_{(h,h;1)}^{}}\Psi_{(h;1)}(0)
    %   \right.\nonumber\\
    %& & \phantom{mmn}\left.
           +\frac{D_{(h,h;1)}C_{(h_1,h_2,h;2)}
           -D_{(h,h;2)}C_{(h_1,h_2,h;1)}}{D_{(h,h;1)}^2}\Psi_{(h;0)}(0)\right]
           \!z^{h-h_1-h_2}\,,\nonumber\\
    \Psi_{(h_1;1)}(z)\Psi_{(h_2;1)}(0) &=& \sum_{h}
        \left[\left(\frac{C_{(h_1,h_2,h;2)}}{
           D_{(h,h;1)}^{}}
           - \frac{2C_{(h_1,h_2,h;1)}}{
           D_{(h,h;1)}^{}}\log(z)\right)\Psi_{(h;1)}(0)\right.\nonumber\\
    \lefteqn{\!\!\!\!\!\!\!\!\!\!
           +\left(\frac{D_{(h,h;1)}C_{(h_1,h_2,h;3)}
           -D_{(h,h;2)}C_{(h_1,h_2,h;2)}}{D_{(h,h;1)}^2}\right.
    %     \nonumber}\\
    %& & \phantom{mmn}
           +\frac{2D_{(h,h;2)}C_{(h_1,h_2,h;1)}
           -D_{(h,h;1)}C_{(h_1,h_2,h;2)}}{D_{(h,h;1)}^2}\log(z)
           \nonumber}\\
    &-& \left.\left.\frac{D_{(h,h;1)}C_{(h_1,h_2,h;1)}}{
           D_{(h,h;1)}^2}\log^2(z)\right)\Psi_{(h;0)}(0)
    \right]\!z^{h-h_1-h_2}\,.
  \end{eqnarray}
  Note that, for instance, the OPE of a proper primary with its
  logarithmic partner necessarily receives two contributions. One might
  naively have expected that proper primary fields do not change the
  J-level, although already the OPE of the stress-energy tensor with
  a logarithmic field will have an additional term involving the primary
  field.

  Finally, we want to remark on the question of locality.
  The two- and three-point functions and the OPEs can easily be brought
  into a form for a local LCFT constructed out of left- and right-chiral
  half. The rule for this is simply to replace each $\log(z_{ij})$ by
  $\log|z_{ij}|^2$, and to replace each power $(z_{ij})^{\mu_{ij}}$ by
  $|z_{ij}|^{2\mu_{ij}}$. This yields a LCFT where all fields have the
  same holomorphic and anti-holomorphic scaling dimensions and the same
  level within the respective Jordan cells. 
  Such an ansatz automatically satisfies both, the holomorphic
  as well as the anti-holomorphic Ward identities, if $z$ and $\bar z$
  are formally treated as independent variables. It is important to note,
  however, that the resulting full amplitudes do not factorize into
  holomorphic and anti-holomorphic parts. This is a well known feature of
  LCFTs. For example, the last OPE equation in (\ref{eq:ope2c}) would read
  in its full form
  \begin{eqnarray}\label{eq:ope3}
     & & \Psi_{(h_1;1)}(z,\bar z)\Psi_{(h_2;1)}(0,0) = \sum_{h}
        |z|^{2(h-h_1-h_2)}
        \left[\frac{C_{(2)} - 2C_{(1)}\log|z|^2}{
           D_{(1)}^{}}\Psi_{(h;1)}(0,0)\right.\\
     & & {}+\left(\left.\frac{D_{(1)}C_{(3)}
           -D_{(2)}C_{(2)}}{D_{(1)}^2}
        +\frac{2D_{(2)}C_{(1)}
           -D_{(1)}C_{(2)}}{D_{(1)}^2}\log|z|^2
        -\frac{D_{(1)}C_{(1)}}{
           D_{(1)}^2}\log^2|z|^2\right)\Psi_{(h;0)}(0,0)\right]\nonumber
  \end{eqnarray}
  with an obvious abbreviation for the structure constants. The reader
  is encouraged to convince herself of both, that on one hand this does indeed
  not factorize into holomorphic and anti-holomorphic parts, but that on the
  other hand this does satisfy the full set of conformal Ward identities.
\eq

  \subsection{A note on the Shapovalov form in LCFT}

  It is often very convenient to work with states instead of the fields
  directly, in particular when purely algebraic properties such as null
  states are considered. As usual, we have an isomorphism between the
  space of fields and the space of states furnished by the map
  $|h;k\ket = \Phi_{(h;k)}(0)|0\ket$. Although one does not necessarily
  have a scalar product on the space of states, one can introduce a
  pairing, the Shapovalov form, between states and linear functionals.
  Identifying the out-states with (a subset of) the linear functionals
  equips the space of states with a Hilbert space like structure.
  As in ordinary conformal field theory, we have
  $\bra h;k| = (|h;k\ket)^{\dagger} = \lim_{z\rightarrow 0}\bra 0|
  \Phi_{(h;k)}(1/z)$. Using now that logarithmic fields transform under
  conformal mappings $z\mapsto f(z)$ as
  \begin{eqnarray*}
     \Phi_{(h;k)}(z) &=& \sum_{l=0}^k\frac{1}{l!}\frac{\partial^l}{\partial
     h^l}\left(\frac{\partial f(z)}{\partial z}\right)^h\Phi_{(h;k-l)}(f(z))\\
                     &=& \sum_{l=0}^k\frac{1}{l!}\log^l\left|\frac{\partial
     f(z)}{\partial z}\right|\left(\frac{\partial f(z)}{\partial z}\right)^h
     \Phi_{(h;k-l)}(f(z))\,,
  \end{eqnarray*}
  the out-state can be re-expressed in a form which allows us to apply
  (\ref{eq:L2pt}) from the small print above to evaluate the Shapovalov form. 
  In ordinary conformal
  field theory, we simply get $\bra h| = \lim_{z\rightarrow\infty}\bra 0|
  z^{2h}\Phi_h(z)$ such that $\bra h|h'\ket = \delta_{h,h'}$ upto
  normalization. Interestingly, the transformation behavior of
  logarithmic fields yields a very similar result, canceling all
  logarithmic divergences. Thus, we obtain for the Shapovalov form
  $$
    \bra h;k|h';k'\ket = \delta_{h;h'}D_{(h,h';k+k')}\,,
  $$
  which is a lower triangular matrix. To demonstrate this, we
  consider again the example of a rank two LCFT. Then we clearly have
  $\bra h;0|h;0\ket = 0$, $\bra h;1|h;0\ket=\bra h;0|h;1\ket = D_{(h,h;1)}$
  and with
  $$
    \lim_{z\rightarrow 0}\bra 0|\Phi_{(h;1)}(1/z)
    \Phi_{(h;1)}(0)¦\ket = \lim_{z\rightarrow\infty}\bra 0|
    z^{2h}\left[\Phi_{(h;1)}(z) + 2\log(z)\Phi_{(h;0)}(z)\right]
    \Phi_{(h;1)}(0)|0\ket
  $$
  the desired result $\bra h;1|h;1\ket = D_{(h,h;2)}$. Hence, the
  Shapovalov form is well defined and non-degenerate for the logarithmic
  case much in the same way as it can be defined for ordinary CFTs. Note that
  the definition of the Shapovalov form does not depend on whether the CFT
  is unitary or not.

  For completeness, we mention that the Shapovalov form is not uniquely
  defined in
  LCFTs, because the basis $\{|h;k\ket : k=0,\ldots r-1\}$ of states spanning
  the rank $r$ Jordan cell
  is not unique. The reason is that we always have the freedom to
  redefine the logarithmic partner fields, or their states respectively, as
  $$
    \Phi'_{(h;k)}(z) = \Phi_{(h;k)}(z) + \sum_{i=1}^k\lambda_i\Phi_{(h;k-i)}(z)
  $$
  with arbitrary constants $\lambda_i$. At this state, there are no further
  restrictions from the structure of the LCFT which could fix a basis within
  the Jordan cells. Only the proper primary field, or the proper highest-weight
  state respectively, is uniquely defined upto normalization.

  \subsection{Differential equations from null vectors}
 
  We are now going to use the generalization of null vectors to the 
  logarithmic case at hand, which we did work out before, to
  effectively compute correlation functions
  involving fields from non-trivial Jordan cells.
  As an example, we consider a four-point function with such a primary field
  which is degenerate at level two. To simplify the formul\ae, we fix the
  remaining three points in the standard way, i.e.\ we consider
  $G_4=\bra\phi_1(\infty)\phi_2(1)\Phi_{h_3}(z)\phi_4(0)\ket$. According to
  (\ref{eq:virl}), the level two descendant yields
  \be\label{eq:null}
    \left[\frac{3\,\partial_z^2}{2(2h_3+1)} + \!\sum_{w\neq z}
    \!\left(
    \frac{\partial_w}{w-z}-\frac{h_w+
    \hat{\delta}_{h_w}}{(w-z)^2}\right)\right]G_4=0
    \,,
  \ee
  where again $\phi_i$ may be either a primary $\Phi_{h_i}$ or its logarithmic
  partner $\Psi_{h_i}$.
  If there is only one logarithmic field, $\hat{\delta}_h$ 
  will produce a four-point
  function without logarithmic fields, i.e.\ won't yield an additional term.
  Hence, after rewriting this equation as an ordinary differential equation
  solely in $z$, we can express the conformal blocks in terms of
  hypergeometric functions. Putting without loss of generality the
  logarithmic field at infinity, we can rewrite 
  \be
    G_4  =  z^{p+\mu_{34}}(1-z)^{q+\mu_{23}}F^{(0)}(z)\,,
%   p   &=& {\textstyle
%           \frac16-\frac23h_3-\mu_{34}-\frac16\sqrt{r_4}}\,,\nonumber\\
%   q   &=& {\textstyle
%           \frac16-\frac23h_3-\mu_{23}-\frac16\sqrt{r_2}}\,,\nonumber\\
%   r_i &=& 1-8h_3+16h_3^2+48h_ih_3+24h_i\,,\nonumber
  \ee
  with the notations as in (\ref{eq:4pt-level2}. Then, 
  $F^{(0)}$ is a solution of the hypergeometric system
  ${}_2F_1(a,b;c;z)$ given by (\ref{eq:hypergeom}).
% \bea
%   a &=& {\textstyle
%         \frac12-\frac16\sqrt{r_2}-\frac16\sqrt{r_4}-\frac16\sqrt{r_1}
%         }\,,\nonumber\\
%   b &=& {\textstyle
%         \frac12-\frac16\sqrt{r_2}-\frac16\sqrt{r_4}+\frac16\sqrt{r_1}
%         }\,,\nonumber\\
%   c &=& {\textstyle
%         1 - \frac13\sqrt{r_4}
%         }\,.
% \eea
  Hence, we see that in a rank two LCFT, correlation functions of
  fields from Jordan blocks vanish, if there is no logarithmic partner
  present; and they look exactly as in the ordinary case, if there is
  precisely one logarithmic partner present. This nicely fits with
  our brief discussion on graphs and combinatorics, since there is
  only one leg around, and that one must remain unlinked.
  The next complicated case is the presence of two logarithmic fields.
  The ansatz now reads
  \be
     G_4 = z^{p+\mu_{34}}(1-z)^{q+\mu_{23}}\left(
     F^{(1)}_{ij}(z) - 2\log(w_{ij})F^{(0)}(z)\right)\,.
  \ee
  Surprisingly, if the two logarithmic fields are put at $w_2=1$ and $w_4=0$,
  the additional term in the new ansatz vanishes. However, the $\hat{\delta_h}$
  operators in (\ref{eq:null}) create two terms such that the standard
  hypergeometric equation becomes inhomogeneous,
  \be
    \left[z(1-z)\partial_z^2 + (c-(1+a+b)z)\partial_z - ab\right]
        F^{(1)}_{24}(z) 
    = \frac{{\textstyle\frac23}(2h_3+1)}{z(1-z)}F^{(0)}(z)\,. 
  \ee
  The solution of this inhomogeneous equation cannot be given in
  closed form, it involves integrals of products of hypergeometric functions.
  But for special choices of the conformal weights, simple
  solutions can be obtained. The best known LCFT certainly is the
  CFT with central charge $c=c_{2,1}=-2$. It has the following
  extended Kac table, which formally can be obtained by considering this
  CFT as a ``minimal'' model with $c=c_{6,3}$, i.e.\ where we artificially
  enlarge the Kac table by considering not coprime numbers in the
  minimal series $c_{p,q}=1-6(p-q)^2/(pq)$.
  $$
    \begin{array}{|c||c|c|c|c|c|}
      \hline
      (r,s) &\pp 1\pp&\pp 2\pp&\pp 3\pp&\pp 4\pp&\pp 5\pp\\ 
      \hline\hline
        1   & 0 & -\frac18            & 0 & \phantom{-}\frac38  & 1 \\ 
      \hline
        2   & 1 & \phantom{-}\frac38  & 0 & -\frac18            & 0 \\ 
      \hline
    \end{array}
  $$
  The field of conformal weight
  $h=h_{2,1}=1$ in the Kac table possesses a logarithmic partner, which
  is the (1,5) field in the Kac table.
  Choosing all weights in the four-point function to be equal to $h$, we find
  with ${}_2F_1(-4,-1;-2;z) = A(2z-1) + Bz^3(z-2)\equiv Af_1+Bf_2$ the
  solutions\footnote{We have deliberately chosen this example where the
  hypergeometric functions reduce to simple polynomials.}
  \bea\label{eq:2sol}
    F^{(0)}(z) &=& [z(1-z)]^{-4/3}(Af_1+Bf_2)\,,\\
    F^{(1)}_{24}(z) &=& [z(1-z)]^{-4/3}\left[Cf_1 + Df_2\right.\\
    &+& {\textstyle(\frac23(B-2A)f_2-\frac23Af_1)}\log(z)
    - {\textstyle(\frac23(B-2A)f_2-\frac23Af_1)}\log(1-z)\nonumber\\
    &+& {\textstyle\frac19(6z^2-6z-7)Af_1}
    + \left.{\textstyle(-\frac23z^3+\frac{5}{9}f_1)B}\right]\,.\nonumber
  \eea
  Note that $F^{(0)}$ does not depend on which field is the logarithmic
  one (hence the omitted lower index), since only the contraction
  of {\em two\/} logarithmic fields causes logarithmic divergences.
  A nice example for this is the twist field $\mu(z)$ in the $c=-2$ LCFT,
  which has $h=-1/8$. Although its OPE with itself yields a logarithmic term,
  $\mu(z)\mu(w)\sim \tilde{\mathbb{I}}(w) + \log(z-w)\mathbb{I}$, no
  logarithm shows up in its two-point function. At least four twist fields
  are necessary to get a logarithm in a correlation function, which is
  equivalent to two logarithmic fields, since
  $\tilde{\mathbb{I}}(z)\tilde{\mathbb{I}}(w)\sim
  -2\log(z-w)\tilde{\mathbb{I}}(w) -\log^2(z-w)\mathbb{I}(z)$.
  
  \bq
  In fact, it is well known that $\bra\mu(\infty)\mu(1)\mu(z)\mu(0)\ket$
  is proportional to $[z(1-z)]^{\frac14}{}_2F_1(\frac12,\frac12;1;z)$, since
  the twist field $\mu$ is degenerate of level two. The hypergeometric
  system ${}_2F_1(\frac12,\frac12;1;z)$ has two solutions. For $|z|<1$,
  only one of them can be expanded as a power series in $z$, the other
  has a logarithmic divergency. For the curious, the solutions read
  \bea
    {}_2F_1(\frac12,\frac12;1;z)\phantom{-1} 
    &=& \sum_n \frac{(\frac12)_n(\frac12)_n}{(1)_n(1)_n}z^n\,,\\
    {}_2F_1(\frac12,\frac12;1;z-1) 
    &=& \log(z){}_2F_1(\frac12,\frac12;1;z) +
    \left.\frac{\partial}{\partial_{\epsilon}}
    {}_3F_2(\frac12+\epsilon,\frac12+\epsilon,
    1;1+\epsilon,1+\epsilon;z)\right|_{\epsilon=0}\,,\nonumber
  \eea
  where the last term enjoys a regular power series expansion for
  $|z|<1$ which, however, is too complicated to write down in a simple
  closed formula. As usual, the Pochhammer symbol is defined as
  $(a)_n=\Gamma(a+n)/\Gamma(a)$.
  \eq

  To summarize, we have so far considered correlation functions with
  logarithmic fields, 
  but where the null field condition was exploited for a primary field.
  We found that the off-diagonal action of the differential operators,
  which stem from the Virasoro modes of the null state descendant,
  leads to {\em inhomogeneous\/} differential equations. These can
  be solved in a hierarchical scheme, since the inhomogeneities
  for a given correlation function are determined by the solutions
  for correlation functions with fewer logarithmic fields.

  We also learned that a null vector descendant
  on the full Jordan cell (not on its irreducible sub-representation) is
  more complicated. For example, the logarithmic partner of the $h=1$ field
  in the $c=-2$ LCFT turns out to be the $h=h_{1,5}$ field in the Kac table.
  Indeed, as shown in the third ref.\ of \cite{Flohr:1996a}, 
  there exists a null vector of the form
  \bea\label{eq:l5}
    \lefteqn{|\chi^{(5)}_{h=1,c=-2}\ket =}\\
    & & [{\textstyle\frac{16}{3}}L_{-1}L_{-2}^2 + \textstyle{\frac{52}{3}}
        L_{-2}L_{-3} - 12L_{-1}L_{-4}
        + \textstyle{\frac{148}{3}}L_{-5}]|h;0\ket\nonumber\\
    &\!+\!& [L_{-1}^5 - 10L_{-1}^3L_{-2} + 36L_{-1}^2L_{-3} - L_{-1}L_{-4}
        + 16L_{-1}L_{-2}^2 - 40L_{-2}L_{-3} + 160L_{-5}]
        |h;1\ket\nonumber
  \eea
  The first descendant is precisely the same as for a primary field degenerate
  of level five. However, a remarkable fact in LCFT is that the null
  descendant factorizes,
  \bea
    |\chi^{(5)}_{h=1,c=-2}\ket &=& (\ldots)|h;0\ket + 
    (L_{-1}^3 - 8L_{-1}L_{-2} + 20L_{-3})(L_{-1}^2 - 2L_{-2})
      |h;1\ket\nonumber \\
    &=& (\ldots)|h;0\ket + 
    (L_{-1}^3 - 8L_{-1}L_{-2} + 20L_{-3})|\chi^{(2)}_{h=1,c=-2}\ket
      \nonumber\\
    &"\!=\!"& (\ldots)|h;0\ket + |\chi^{(3)}_{h=3,c=-2}\ket\,,
  \eea
  namely into the level two null descendant times a level three descendant
  which turns out to be the null descendant of a primary field of conformal
  weight $h_{3,1}=3$. Hence, the level two descendant of the logarithmic
  field is a null descendant only up to a primary field of weight
  $h_{3,1}=h_{1,5}+2$.
  
  \bq
  It is worth noting that this is a general LCFT feature:
  Namely, the typical LCFT case is that the
  logarithmic partner constituting a Jordan cell representation is
  degenerate of level $n+k$ with $n$ the level where the primary has its
  null state, and $k>0$. On the other hand, the conformal properties of
  the logarithmic field are identical to the ones of its primary partner
  up to the non-diagonal contributions. Hence the two fields could not be
  distinguished if these additional contributions were ignored. It follows
  that in a correlator without any further logarithmic fields (where
  the off-diagonal part of the null field does not contribute), the
  logarithmic field must behave exactly as its primary partner, i.e.\
  must possess the same null field. The only way this can happen
  consistently is that the diagonal part of the null vector
  factorizes.
  
  Another important point is that the additional descendant on the primary
  partner is not unique. We learned that due to the Jordan cell structure,
  a descendant on the logarithmic partner state necessarily involves
  a descendant part built on the primary highest-weight state.
  However, although this contribution cannot be zero, it is not
  unique. If again the
  logarithmic partner constituting a Jordan cell representation is
  degenerate of level $n+k$,
  then the descendant of the primary field is
  determined only up to an arbitrary contribution $\sum_{|\vec{m}|=k}
  \alpha^{\vec{m}}L_{-\vec{m}}|\chi^{(n)}_{h,c}\ket$, where
  $|\chi^{(n)}_{h,c}\ket$ denotes the ordinary level $n$ null descendant of the 
  highest-weight state.
  \eq

  That the $(1,5)$ entry of the Kac table does indeed refer to the
  logarithmic partner of the $h=1$ primary $(2,1)$ field can be seen from
  the solutions of the homogeneous differential equation resulting from
  (\ref{eq:l5}) when there are no off-diagonal contributions. Of course,
  the resulting ordinary differential equation of degree five has, among
  others, the same solutions as the hypergeometric equation above for the
  $(2,1)$ field. These are the correct solutions, if there is no other
  logarithmic field. The other three solutions turn out to have logarithmic
  divergences. Therefore, they cannot be valid solutions for this case,
  but must constitute solutions for a correlator with two logarithmic fields.
  However, in this case one has to take into account that the full null
  state has an additional contribution from the primary partner of the
  $(1,5)$ field. The full inhomogeneous equation reads (with a 
  particularly simple choice for the primary part of the descendant)
  \bea
    0 &=&\left[z^3(1-z)^3\partial^5 + 8z(z-1)(z^2-z+1)\partial^3\right.
     - 4(2z-1)(5z^2-5z+2)\partial^2 \nonumber\\
    & & {}+ 24(2z-1)^2\partial
     -  \left.48(2z-1)^{\vphantom{2}}\right]F^{(1)}_{34}(z)\nonumber \\
    &+&\left[-{\ts\frac{16}{3}}z(z-1)(2z-1)^2\partial^3\right.
     + {\ts\frac{44}{3}}(2z-1)(5z^2-5z+2)\partial^2\nonumber\\
    & &{\ts{}-\frac{8}{z(z-1)}}(57z^4-114z^3+90z^2-333z+5)\partial\nonumber\\
    & & {}+ {\ts\frac{16}{z(z-1)}}\left.(2z-1)(18z^2-18z+5)\right]F^{(0)}(z)
  \eea
  in the case of one further logarithmic field put at zero. Similar equations
  can be written down for all three choices $F^{(1)}_{3j}(z)$ as well as
  for higher numbers of logarithmic fields. In general, there is one part
  of the differential equation for $F^{(r)}_{I}$ with $I=\{3,i_1,\ldots,i_r\}$,
  and the inhomogeneity is given by $F^{(r-1)}_{I-\{3\}}$. It is clear from
  this that the full set of solutions can be obtained in a hierarchical
  scheme, where one fist solves the homogeneous equations and increases
  the number of logarithmic fields one by one.

  In the example above, $F^{(0)}$ is given as in (\ref{eq:2sol}). Then the
  inhomogeneity reads $80(3z^2-3z+1)A+16z(z^2-9z+3)B$. With this,
  the solution is finally obtained to be given as
  \bea
    F^{(1)}_{34} &=&  C_1f_1 + C_2f_2
     + C_3[3f_1\log({\ts\frac{z}{z-1}}) - 6]
     + C_4[3f_2\log(z-1)-12z^3]\\
    &+& C_5[3(f_1+f_2)\log(z) + 12z(z^2-3z+1)]\nonumber\\
    &+& \left[{\ts\frac29}(3f_1-2f_2)\log(z)\right.
                + {\ts\frac29}(7f_1+2f_2)\log(z\!-\!1)
     + {\ts\frac{1}{27}}\left.(12z^3-18z^2+32z-1)\right]\!A\nonumber\\
    &+& \left[{\ts\frac29}(f_2-f_1)\log(z)\right.
                - {\ts\frac29}(4f_1+f_2)\log(z\!-\!1)
     + {\ts\frac{1}{27}}\left.(36z^2-6z^3-17f_1)\right]\!B\,.\nonumber
  \eea
  As is obvious from the above expression, correlation functions involving
  more than one logarithmic field become quite complicated. Although
  the two logarithmic fields were chosen to be located at $z,0$, the above
  solution also contains terms in $\log(z-1)$. This is a consequence of
  the associativity of the OPE and duality of the four-point function.

  In principle, the full set of four-point functions can be evaluated in
  this way. Care must be taken with the solutions of the homogeneous
  equation. As indicated above, not all of them might be valid solutions.
  If the correlator does contain only one logarithmic field, then
  there cannot be any logarithmic divergences in the solution. However,
  it is instructive to find the reason, why already the homogeneous
  equation admits logarithmic solutions. Firstly, one should remember that
  a similar situation arises in minimal models. All primary fields come
  in pairs in the Kac table, which are usually identified with each other,
  $(r,s) \equiv (q-r,p-s)$ if the central charge is $c=c_{p,q}$. So, in
  principle, one and the same correlator can be evaluated by exploiting
  two different null state conditions, which in general will be of
  different degrees, $rs\neq rs+qp -(qs+pr)$. Therefore, the physical
  solutions are given by the intersection of the two sets of solutions.

  \bq
  In the logarithmic case, the typical BPZ argument that only the common
  set of fusion rules can be non-vanishing \cite{BPZ}, has to be modified.
  The $(2,1)$ field has the formal BPZ fusion rules
  $[(2,1)]\times[(2,1)] = [(1,1)] + [(3,1)]$, but the last term must vanish
  due to dimensional reasons, since $h_{3,1} =3 > 2h_{2,1} = 2\cdot 1$.
  On the other hand, one has in a formal way
  $[(2,1)]\times[(1,5)] = [(1,1)]$, meaning that the OPE of the logarithmic
  field with its own primary partner won't yield a logarithmic dependency.
  Note that a logarithmic field can be considered as the normal ordered
  product of its primary partner with the logarithmic partner of the
  identity, i.e.\ $\Psi_h(z) = \mbox{:$\Phi_h\tilde{\mathbb{I}}$:}(z)$.
  As long as an OPE of such a field with a primary field is considered,
  one can evaluate it in the usual way, and then take the normal ordered
  product of the right hand side with $\tilde{\mathbb{I}}$, since the
  latter field behaves almost as the identity field with respect to fusion
  with primary fields.
  But as soon as the OPE of two logarithmic fields is taken, one gets
  a new term: $[(1,5)]\times[(1,5)] = [(1,1)] + [(1,3)] + [(1,5)]$,
  where all terms are omitted which must vanish due to dimensional reasons.
  Now, the (1,3) field $\tilde{\mathbb{I}}$ itself appears in the OPE, which is
  correct because the OPE of two such normal ordered products will
  involve the well-known OPE $\tilde{\mathbb{I}}(z)\tilde{\mathbb{I}}(w)
  \sim -2\log(z-w)\tilde{\mathbb{I}}(w)$.
  This demonstrates that the logarithmic solutions of the conformal blocks of
  the four-point function can only be valid when sufficiently many
  logarithmic fields are involved.
  \eq
  
  Let us come back to the above mentioned observation that the null state
  of a logarithmic field of level $n+k$ factorizes into the level $n$
  null descendant of its primary partner times the level $k$ null state of
  a primary field of conformal weight $h+n$. Indeed, it is a nice exercise
  to show that in our $c=-2$ example the Virasoro modes of the level two null
  descendant, acting on the logarithmic $\Psi_{h=1}$ field, produce
  a field which transforms as a primary field of conformal weight $h=3$.
  The reason is that the derivative, acting on a logarithmic field, eats
  up the fermionic zero modes. Indeed, in
  \bea
    [L_{-n},\Psi_h(z)] &=& z^n((n+1)h+z\partial)\Psi_h(z)\phantom{mmml} \\
    &=& (n+1)h\Psi_h(z) + \mbox{:$(\partial\Phi_h)\tilde{\mathbb{I}}$:}(z)
    + \mbox{:$\Phi_h(\partial\tilde{\mathbb{I}})$:}(z)\,. \nonumber
  \eea
  where the $\hat{\delta}_h$ part is omitted, the derivative first acts as
  derivative on the primary part of the logarithmic field, and then acts
  on the field $\tilde{\mathbb{I}}$. In the $c=-2$ LCFT this basic logarithmic
  field can be constructed out of two anti-commuting scalar fields,
  \be\label{eq:thetafield}
    \theta^{\alpha}(z) = \sum_{n\neq 0}\theta_n^{\alpha}z^{-n}
                       + \theta_0^{\alpha}\log(z) + \xi^{\alpha}\,,
    \ \ \ \ \alpha=\pm\,, 
  \ee
  whose zero modes are responsible for all the logarithms.
  Then $\tilde{\mathbb{I}}(z) = -\frac12\epsilon_{\alpha\beta}
  \mbox{:$\theta^{\alpha}\theta^{\beta}$:}(z)$. Therefore, the derivative
  will eat up zero modes, e.g.\ $\tilde{\mathbb{I}}(0)|0\ket =
  \xi^+\xi^-|0\ket$ and $\partial\tilde{\mathbb{I}}(0)|0\ket =
  (\theta_{-1}^+\xi^- + \theta_{-1}^-\xi^+)|0\ket$. By considering states,
  one can show that the level two null descendant applied to the state
  $\Psi_{h=1}(0)|0\ket$ yields a state proportional to a
  highest-weight state of weight $h=3$.

  \bq
  Recall, that we mentioned earlier in some of the small print that
  logarithmic partner fields are not necessarily quasi-primary.
  The $h=1$ logarithmic field in the $c=-2$ LCFT is an example for
  just this phenomenon. In fact, the logarithmic $h=1$ field is given
  as $\Psi_{h=1}(z)= \mbox{:$\tilde{\mathbb{I}}\partial\theta^{\alpha}$:}(z)$,
  i.e.\ it really is a doublet. The primary $\Phi_{h=1}(z) =
  \partial\theta^{\alpha}(z)$ is, of course, also a doublet.
  Moreover, $\Psi_{h=1}$ is indeed not quasi-primary, since
  $L_1\Psi_{h=1}(0)|0\ket = \xi^{\alpha}|0\ket$. Note the appearance of
  one of the crucial zero-modes, i.e.\ the state which spoils $\Psi_{h=1}$
  being quasi-primary is a fermionic state. One can show that
  both, the primary $\Phi_{h=1}$ as well as the field of weight $h=3$
  generated by the action of the level two null descendant on 
  $\Psi_{h=1}$, are descendants of the state $|\xi^{\alpha}\ket$.
  \eq

  In order to understand why already the solution of the homogeneous
  fifth-order differential equation does yield logarithmic divergencies,
  we should keep in mind that the regular solutions of the second-order
  differential equations (where the null field was assumed to be
  a descendant of the primary) make only sense if one of the other three
  fields is a logarithmic field. Similarly, the five solutions
  of the fifth-order equation seem to make only sense in the presence of
  one further logarithmic field. However, we just argued that
  the correct solution for
  a four-point function with two logarithmic fields must be obtained
  from an inhomogeneous differential equation, and we also have seen
  that logarithms may arise just from these inhomogeneities, when
  we looked at the level two null field. So, where do these logarithms
  come from in the homogeneous case? 

  It seems that the only possibility left is to assume that two of
  the other three fields must possess an OPE which produces a
  logarithmic field, similarly to twist fields. Hence, we would need
  two seemingly primary $h=1$ fields $\mu_{h=1}$ which, however,
  have an OPE of the form $\mu_{h=1}(z)\mu_{h=1}(w)\sim (z-w)^{-2}[
  \tilde{\mathbb{I}}(w) + \log(z-w)\mathbb{I}]$ or with another pair
  of primary and logarithmic field on the right hand side. In fact,
  such fields may indeed exist, namely $\mu_{h=1}(z)=
  \mbox{:$\frac12\epsilon_{\alpha\beta}\theta^{\alpha}
  \partial\theta^{\beta}$:}(z)$. Expanding this field in modes
  via (\ref{eq:thetafield}) shows that it is not itself logarithmic,
  since there is no mode $\xi^+\xi^-$. However, its OPE with itself
  produces precisely such a term, as well as a term with a 
  logarithmic divergency.

  As a consequence, the level five null state condition captures all
  possibilities of a four-point function with four $h=1$ fields:
  Besides the ``ordinary'' primary field doublet 
  $\Phi^{\alpha}_{h=1}=\partial\theta^{\alpha}$ and its logarithmic
  partner doublet $\Psi^{\alpha}_{h=1}=\mbox{:$\tilde{\mathbb{I}}\partial
  \theta^{\alpha}$:}$, which are fermionic with respect to the number
  of $\xi^{\alpha}$-zero modes, there is also the bosonic primary field
  $\mu_{h=1}=\mbox{:$\frac12\epsilon_{\alpha\beta}\theta^{\alpha}
  \partial\theta^{\beta}$:}(z)$, which is pre-logarithmic. 
  The null vector condition cannot see, of which sort the other three fields
  are, as long as none of them is logarithmic. Only the latter sort
  does produce the betraying inhomogeneity. But the bosonic
  pre-logarithmic primary, contracted with itself, will yield 
  a logarithmic field in the internal channel, and this in turn is
  responsible for a logarithmic divergency, when contracted with the
  proper logarithmic $h=1$ field. Of course, since correlation functions
  are only non-zero if the $\xi$-fermion number is even, the only
  possible choice for this case is one logarithmic field, one proper
  primary field, and two pre-logarithmic fields.

  We leave it as an exercise to the reader to work out a similar
  structure for the $h=0$ fields, and compute all possible
  non-vanishing four-point functions of four fields of weight $h=0$.

\section{Ghost systems}

A very important family of CFTs are the so-called ghost systems.
Mathematically, they are the CFT description of the complex analysis
of $j$-differentials. Thus, one starts with considering a pair of
anti-commuting fields $b(z)$ and $c(z)$ with conformal weights
$j$ and $1-j$ respectively. Indeed,
$b^{(j)} = b(z)({\rm d}z)^j$ and $c^{(1-j)} = c(z)({\rm d}z)^{1-j}$ are
invariant under conformal transformations provided $b(z)$ transforms as
$b(z')=b(z)({\rm d}z'/{\rm d}z)^{-j}$ and analogously for $c(z)$.

Although we will see in a moment that the resulting CFT is not unitary,
it possesses a natural scalar product defined via
\be\label{eq:bc-sp}
  \bra b^{(j)},c^{(1-j)}\ket = \oint b(z)c(z){\rm d}z
  = \oint b(z)({\rm d}z)^jc(z)({\rm d}z)^{1-j}\,.
\ee
If $2j\in\mathbb{Z}$, these fields make sense as chiral fields, meaning that
they behave benign under the monodromy $z\mapsto{\rm e}^{2\pi{\rm i}}z$,
acquiring nothing more than a sign (for $j$ half-integer). Under these
circumstances, they possess a mode expansion
\be\label{eq:bc-modes}
  b(z) = \sum_{n\in\mathbb{Z}}b_nz^{-n-j}\,,\ \ \ {\rm i.e.}\ \ \
  b_n = \oint{\rm d}zb(z)z^{n+1}\,,
\ee
and analogously for $c(z)$. Since the fields are anti-commuting their
modes satisfy the relations
\be
  \{b_m,c_n\} = \delta_{m+n,0}\,,
\ee
with all other anti-commutators vanishing. 

We mention the ghost systems here because they can be viewed as the
non-logarithmic sectors of larger, logarithmic, CFTs. This shall serve
as an explicit example, how CFTs can be enlarged, or augmented, to form
logarithmic CFTs. We will demonstrate this in particular for the
$c=-2$ ghost system.

Let us first extract some general information such as the equations of motion.
The action of the $bc$ system is given by
\be
  S = {\textstyle\frac{1}{2\pi}}\int{\rm d}^2z\,b(z)\bar{\partial}c(z)\,,
\ee
which is conformally invariant by construction due to $j+(1-j)=1$.
The operator equations of motion may be obtained in the usual 
path integral way without any complications, and are
\be
  \left\{\begin{array}{rcl}
    \bar{\partial}c(z) = \bar{\partial}b(z) &=& 0\,,\\
    \bar{\partial}b(z)c(z') &=& 2\pi\delta^2(z-z',\bar z-\bar z')\,,\\
    \bar{\partial}b(z)b(z') = \bar{\partial}c(z)c(z') &=& 0\,. 
  \end{array}\right.
\ee
Since we have not yet fixed $j$ and therefore do not know whether we have
a well-defined mode expansion, we define normal ordering by requiring that
normal ordered objects behave classically. Recalling that
$\bar{\partial}z^{-1}=\partial \bar z^{-1} = 2\pi\delta^2(z,\bar z)$, we
find that the normal ordered product $\nop{bc}$ must read
\be
  \nop{b(z)c(z')} = b(z)c(z') - (z-z')^{-1}\,.
\ee
Again, we may turn this around to identify the singular part of the
corresponding OPE. Combinatorially, normal ordering for the ghost system
is much the same as for the free scalar field, i.e.\ goes with Wick's
theorem, except that interchanging two fields may result in sign flips.
Therefore, when contracting two fields, one should first anti-commute them
until they are next to each other, where each anti-commutation flips the
sign. We thus obtain the following OPEs, where $x\sim y$ means that $x$ is
equal to $y$ upto regular terms:
\be\label{eq:bc-opes}\begin{array}{rclcrcl}
  b(z)c(w)&\sim&{\displaystyle\frac{1}{z-w}}\,,  & & 
  c(z)b(w)&\sim&{\displaystyle\frac{1}{z-w}}\,,\\
  b(z)b(w)& =  &{\cal O}(z-w)\,,  & & c(z)c(w)& =  &{\cal O}(z-w)\,.
\end{array}\ee
Note that there are two sign flips in the second OPE, one from anti-commuting,
and one due to $z\leftrightarrow w$. The last both OPEs are actually not
only holomorphic, but they have a zero due to anti-symmetry (Pauli principle:
expectation values with two identical fermions at the same place must vanish).

The stress energy tensor is obtained via Noether's theorem with respect to
world sheet transformations $\delta z = \varepsilon(z)$, under which 
$\delta b = (\varepsilon\partial + j(\partial\varepsilon))b$ and
$\delta c = (\varepsilon\partial + (1-j)(\partial\varepsilon))c$, such that
\be\label{eq:Tghost}
  T(z) = (1-j)\nop{(\partial b)c} - j\nop{b(\partial c)}\,,\ \ \ \ 
  \bar T(\bar z) = 0\,.
\ee 
The interested reader should work out the OPE of $T(z)$ with the fields
$b(w)$ and $c(w)$ to verify that they have the expected form 
(\ref{eq:opeTprim}). Also, the OPE of $T(z)$ with $T(w)$ is not hard to
work out, it has the standard form (\ref{eq:opeTT}) and it reveals the 
conformal anomaly to be
\be\label{eq:bc-c}
  c = c_{bc}=-2(6j^2 - 6j + 1) < 0\ \ {\rm for}\ \ j\in\mathbb{R}-
  [{\textstyle\frac12(1-\frac{1}{\sqrt{3}}),\frac12(1+\frac{1}{\sqrt{3}})}]\,,
\ee
which is clearly negative for all (half)-integer $j$ except $j=\frac12$.
Obviously, this CFT is purely holomorphic (or actually meromorphic). Of course,
there exists a completely analogous anti-holomorphic CFT with action
$S=\frac{1}{2\pi}\int{\rm d}^2z\bar b\partial\bar c$. But as it stands, this
is a theory which is completely left-chiral, the right-chiral part being
the trivial CFT with $c=0$.

The $bc$ system admits a {\em ghost number\/} symmetry $\delta b=-{\rm i}
\varepsilon b$, $\delta c = {\rm i}\varepsilon c$. It stems from a global
$U(1)$ symmetry of the action under the transformation
$b(z) \mapsto \exp(-{\rm i}\alpha(z))b(z)$, $c(z) \mapsto 
\exp({\rm i}\alpha(z))c(z)$ for arbitrary holomorphic $\alpha(z)$.
The corresponding
Noether current is simply $j(z) = -\nop{bc}(z)$. Thus we may expect to
have a quantum number with respect to the corresponding conserved Noether 
charge, the ghost number. Again, it is defined for the left-chiral sector,
and an analogous definition holds for the right-chiral sector, both being
separately conserved. If one computes the OPE of $T$ with $j$, one finds
that
\be
  T(z)j(w) \sim \frac{1-2j}{(z-w)^3} + \frac{1}{(z-w)^2}j(w) + \frac{1}{(z-w)}
  \partial_wj(w)\,,
\ee
meaning that $j(w)$ is not a primary conformal field. Under conformal
mappings, $j(w)$ thus transforms as
\be
  \delta j(w) = \left(-\varepsilon(w)\partial_w - (\partial_w\varepsilon(w))
                + {\textstyle\frac12}(2j-1)\partial_w^2\right)j(w)\,.
\ee

One particular case is $j=1-j$, i.e.\ $j=\frac12$. The central charge
(\ref{eq:bc-c}) is then $c=1$. It is customary, to use the notion
$b=\psi$, $c=\bar{\psi}$ in this case. It is then easy to see that this CFT
can be split into two identical copies by writing
$\psi=\frac{1}{\sqrt{2}}(\psi_1+{\rm i}\psi_2)$ and
$\bar{\psi}=\frac{1}{\sqrt{2}}(\psi_1-{\rm i}\psi_2)$,
such that
\bea
  S &=& {\textstyle\frac{1}{4\pi}}\int{\rm d}^2z\left(\psi_1\bar{\partial}\psi_1
            + \psi_2\bar{\partial}\psi_2\right)\,,\\
  T &=& -{\textstyle\frac12}\left(\psi_1\partial\psi_1 
            + \psi_2\partial\psi_2\right)\,.
\eea
Each of the $\psi_i$ theories has central charge $c=\frac12$, and can be
recognized as the CFT of a free fermion. This theory corresponds to the
case $m=3$ in (\ref{eq:cseries}) and is the first non-trivial example of
a so-called {\em minimal model}, which are CFTs with only finitely many
Virasoro conformal families (primaries with all their descendants).
It will not concern us further, but it should at least be noted that
it possesses only three primary fields of conformal weights $h_{1,1}=h_{2,3}=0$,
$h_{1,2}=h_{2,2}=\frac{1}{16}$, and $h_{2,1}=h_{1,3}=\frac12$ 
according to (\ref{eq:hseries}), which perfectly coincides
with the two order parameters of the two-dimensional Ising model (plus
the identity), the spin $\sigma$ and the energy $\epsilon$, and their
critical exponents.
Another important value is
$j=2$, for which we get $c_{bc}=-26$, and which is important in bosonic
string theory.

%\subsection{$\beta\gamma$ systems}
%
%\bq
%We can redo everything from the last section with a pair of fields 
%$\beta(z)$ and $\gamma(z)$, which behave exactly as in the $bc$ system
%except that they are {\em commuting}. The only differences are that the
%OPEs now read $\beta(z)\gamma(w)\sim-(z-w)^{-1}$ and $\gamma(z)\beta(w)\sim
%(z-w)^{-1}$. Note the differing sign. The stress energy tensor looks exactly
%as in the $bc$ system, but the different sign under commutation yields now
%the central charge $c_{\beta\gamma}=2(6j^2-6j+1) = -c_{bc}$. The system with
%$j=\frac32$ has $c_{\beta\gamma} = 11$ is important for the superstring.
%\eq

\subsection{Mode expansions}

We will assume for now that $j\in\mathbb{Z}$. Then we have well-defined
mode expansions (\ref{eq:bc-modes}), i.e.\
\be
  b(z)=\sum_{n\in\mathbb{Z}} b_n z^{-n-j}\,,\ \ \ \
  c(z)=\sum_{n\in\mathbb{Z}} c_n z^{-n-(1-j)}\,.
\ee
The anti-commutators can be obtained from the OPE, and turn out to be
$\{b_m,c_n\}=\delta_{m+n,0}$ with all other anti-commutators vanishing.
It seems sensible to impose highest-weight conditions, and to consider
states which are annihilated by all modes $b_{n^{}}$ and $c_{n'}$ with 
$n,n'>0$.
But what about the zero modes? It turns out that we have now pairs 
$|+\ket$, $|-\ket$ of highest-weight states with the properties
\be
  \left\{\begin{array}{rclcrclcr}
    b_0|-\ket &=&0\,,      & &b_0|+\ket &=&|-\ket\,,    & & \\
    c_0|-\ket &=&|+\ket\,, & &c_0|+\ket &=&0\,,         & & \\
    b_n|-\ket &=&b_n|+\ket &=&c_n|-\ket &=&c_n|+\ket\,, & & n>0\,.
  \end{array}\right.
\ee
We may construct Verma modules on these highest-weight states by acting
with the modes $b_{-n^{}}$ and $c_{-n'}$ with $n>0$. We now have to
fix notation by convention, saying that $b_0$ be an annihilator, and that
$c_0$ be a creator. This singles out $|-\ket$ as the ghost vacuum 
$|0\ket^{(-)}$.
Note, however, that for consistency we must require that ${}^{(+)}\bra 0|=
{}^{(-)}\bra 0|c_0$
be the correct out-vacuum such that ${}^{(+)}\bra 0|0\ket^{(-)} = 1$. In this
way we guarantee that the conditions defining the in-vacuum $|0\ket^{(-)}$
are dual to those defining the out-vacuum ${}^{(+)}\bra 0|$. However, this
is a further example for the situation that the ``metric on field space'',
the two-point structure constants $\bra \alpha|\beta\ket=D_{\alpha\beta}$,
is not diagonal.

Let us now introduce a grading or particle number operator, the ghost number
operator $N_g$ for the resulting Fock space. We define its action on the
vacua as $N_g|0\ket^{(\mp)} = \mp\frac12|0\ket^{(\mp)}$, and further define
that it counts the modes as $N_g(b_n)=-b_n$ and $N_g(c_n)=+c_n$. This definition
is cooked up in such a way that the scalar product (\ref{eq:bc-sp}) is
non-vanishing only if the total ghost number is zero. For instance,
${}^{(-)}\bra 0|0\ket^{(-)}=0$ since the total ghost number is $N_g=-1$.
Indeed, $|0\ket^{(-)}=b_0|0\ket^{(+)}$, and since $b_0^{\dag}=b_0^{}$, we
see that ${}^{(-)}\bra 0|b_0=0$.

Next, we consider the mode expansion of $T(z)$. Since the stress energy
tensor is made up from the $bc$ system, its Virasoro modes will have
the form
\be
  L_m \propto \sum_{n\in\mathbb{Z}} (mj-n)\nop{b_nc_{m-n}} + \delta_{m,0}
  {\cal N}_{bc}\,,
\ee
where there might be an additional term due to normal ordering, which can
only be a constant since the anti-commutators are $c$-numbers. Note that
this is mode normal ordering, i.e.\ normal ordering of creation
operators left to annihilation operators, which should not be confused with 
field normal ordering. The constant ${\cal N}_{bc}$ is easily computed 
by checking the consistency condition that
\be
  2L_0|-\ket = [L_1,L_{-1}]|-\ket = (jb_0c_1)((1-j)b_{-1}c_0)|-\ket
  = j(1-j)|-\ket \stackrel{!}{=}0\,.
\ee
Thus, we learn that ${\cal N}_{bc}=\frac12j(1-j)$ and hence
\be
  L_m = \sum_{n\in\mathbb{Z}} (mj-n)\nop{b_nc_{m-n}} + {\textstyle\frac12}
  j(1-j)\delta_{m,0}\,.
\ee
The non-vanishing constant ${\cal N}_{bc}$ hints at the fact that 
mode normal ordering and field normal ordering are not identical in the
ghost system. One can show that the difference amounts to
\be
  (\nop{b(z)c(z')})_{{\rm field\ ordering}} -
  (\nop{b(z)c(z')})_{{\rm mode\ ordering}} = \frac{1}{(z-z')}\left(\left(
  \frac{z}{z'}\right)^{1-j} - 1\right)\,.
\ee

\bq
The reader should convince herself that the corresponding ordering
constant ${\cal N}_{\phi}$ in the free bosonic CFT is zero, i.e.\ that the 
Virasoro modes are given simply by
\be
  L_m = {\textstyle\frac12}\sum_{n\in\mathbb{Z}}\nop{a_{m-n}a_n}
\ee
without an additional term ${\cal N}_{\phi}\delta_{n,0}$.
This can be done in complete analogy to the ghost system, i.e.\ by
checking that $L_0|0\ket = \frac12[L_1,L_{-1}]|0\ket = 0$.
The fact that there is no ordering constant is coincident with the fact
that mode normal ordering and field normal ordering are equivalent for
the free bosonic theory.
\eq

Let us return to the ghost number current $j=-\nop{bc}$ with its charge
\be
  N_g = {\textstyle\frac{1}{2\pi{\rm i}}}\int_0^{2\pi}{\rm d}w 
  j_{{\rm cyl}}(w) = 
  \sum_{n>0}(c_{-n}b_n - b_{-n}c_n) + c_0b_0 - \frac12\,,
\ee
which indeed satisfies $[N_g,b_n]= -b_m$ and $[N_g,c_n]=+c_n$. It therefore
counts the number of $c$ excitations minus the number of $b$ excitations
of a given state. The constant is necessary to reproduce our definition of
the action of $N_g$ on the ground states $N_g|\mp\ket=\mp\frac12|\mp\ket$.

Note that we have defined the ghost number for the physically relevant
cylinder (the string world-sheet). Since the ghost current is not a
primary field, the translation to the complex plane has to be performed
carefully. Recalling that $z={\rm e}^w$ mediates the map between
cylinder and complex plane, we find
\be
  (\partial_z w)j_{{\rm cyl}}(w) = j(z) + (j-{\textstyle\frac12})
  (\partial_z^2 w)/(\partial_z w) = j(z) + (j-{\textstyle\frac12})z^{-1}\,.
\ee
This is quite similar to the effect that the zero mode of the Virasoro
algebra, $L_0$, receives a shift by $-c/24$ when we map the theory from the
cylinder to the complex plane. Thus, the ghost number also receives a shift,
namely $N_{g,{\rm plane}} = \oint {\rm d}zj(z) = N_g+Q_j$ with $Q_j=
j-{\textstyle\frac12}$. The above definitions led to unusual vacuum states,
which are {\em not\/} the $SL(2,\mathbb{C})$-invariant vacua introduced
earlier. This disadvantage is the price paid for treating the ghost
system in a way where ordering prescriptions are more or less independent
of the spin $j$ of the system.

\subsection{Ghost number and zero modes}

The above approach is sometimes not useful, especially if a particular
ghost system is considered. Then, it is more natural to use the
$SL(2,\mathbb{C})$-invariant vacuum.
Let us now be specific and put $j=2$. For this value, the $bc$ system thus
consists out of a spin-two field and a vector field, and has central
charge $c=-26$. The string theorists tell us, that this ghost
system is particularly important for the bosonic string.

The mode expansions read in this specific case simply
\be
  b(z) = \sum_n b_n z^{-n-2}\,,\ \ \ \ c(z) = \sum_n c_n z^{-n+1}\,.
\ee
We now wish to reproduce the canonical field normal ordering by a mode
normal ordering prescription. The natural way to do this for a chiral local
field $\Phi_h(z)$, $2h\in\mathbb{Z}$, with mode expansion $\Phi_h(z)=
\sum_n\phi_n z^{-n-h}$ is to call all modes with $n>-h$ annihilators, and all
other modes creators, i.e.\ by imposing highest weight conditions
$\phi_n|0\ket = 0$ for $n>-h$. In our example, we thus would like to impose
\be
  b_n|0\ket = 0\ \ \forall\ n\geq -1\,,\ \ \ \ c_n|0\ket = 0\ \ 
  \forall\ n\geq 2\,.
\ee
In this way, the vacuum $|0\ket$ is indeed the $SL(2,\mathbb{C})$-invariant
vacuum. The corresponding conditions for the out-vacuum then read
\be
  \bra 0|b_{-n} = 0\ \ \forall\ n\geq -1\,,\ \ \ \ \bra 0|c_{-n} = 0\ \
  \forall\ n\geq 2\,.
\ee
But now, we have to keep in mind that the modes $b_{-n}$ are conjugate to
the modes $c_n$, since we have the canonical commutation relations
$\{b_n,c_m\} = \delta_{n+m,0}$. Both highest-weight conditions together
tell us that the three modes $b_{-1},b_0,b_{0}$ are annihilators in both
directions, i.e.\ they annihilate to the right as well as to the left.
On the other hand, the three modes $c_{-1},c_0,c_{1}$ are creators in both
directions, i.e.\ they neither annihilate to the right nor to the left.

As a consequence, we find that $\bra 0|0\ket = \bra 0|\{b_0,c_0\}|0\ket = 0$.
Even more strangely, also $\bra 0|c_i|0\ket = 0$ for $i\in\{-1,0,1\}$.
In fact, the first non-vanishing expression is $\bra 0|c_{-1}c_0c_{1}|0\ket$,
i.e.\ we need at least three $c$-modes. One sees this by inserting a one
in the form $1=\{b_i,c_{-i}\}$ for $i\in\{-1,0,1\}$. For example,
$\bra 0|c_0c_1|0\ket = \bra 0|\{b_1,c_{-1}\}c_0c_1|0\ket = 0$. Of course,
this does not any longer work for the correlator $\bra 0|c_{-1}c_0c_{1}|0\ket$,
since we are forced to insert the one as $1=\{b_n,c_{-n}\}$ with $n>1$, which
does not annihilate anymore. The three $c$-modes are necessary to eat
up the three zero modes of the field $b(z)$. One might hide them in a
redefinition of the out-vacuum as $\bra\tilde 0| = \bra 0|c_{-1}c_0c_1$ such
that $\bra\tilde 0|0\ket = 1$.

We therefore find that the ghost system correlators can only be non-zero,
if the total ghost number, i.e.\ the number of $c$-fields minus the
number of $b$-fields is exactly three, $N_g=\#c-\#b=3$. The reader should note
that this differs from our discussion in the preceeding section, since we
made a different choice of vacuum. The vacuum used now is the physical
vacuum.

We did go into some length here to show some features of ghost systems,
which should definitely remind us in typical LCFT features. Indeed, the
zero-modes appearing in the ghost systems are very reminiscent of
the zero modes in logarithmic CFTs.

\subsection{Correlation functions}

The above discussion can immediately applied to calculate correlation
functions of the $bc$ ghost system. We already know that, for instance,
$\bra c(z)c(w)\ket = 0$. The first non-trivial correlator is
\bea
  \bra c(z_1)c(z_2)c(z_3)\ket &=& \bra 0|\sum_n\sum_m
    c_{-n}z_1^{+n+1}c_{n-m}z_2^{-(n-m)+1}c_mz_3^{-m+1}|0\ket\nonumber\\
  &=& \sum_{n\leq -1}\sum_{m\leq 1}\bra 0|c_{-n}z_1^{+n+1}c_{n-m}z_2^{-(n-m)+1}
    c_{m}z_3^{-m+1}|0\ket\,,
\eea
where we inserted the mode expansion and used the highest-weight condition
of the vacuum states. There are only two summations here, since the total
level (with respect to the $L_0$ grading) must be zero, which fixes the
mode of the third field, if the modes of the other two fields are given.
Since all the modes $c_k$ anti-commute with each other,
it is easy to see that the only non-vanishing choices are $m,n\in\{-1,0,1\}$.
This leads to the six terms
\bea
  &\bra 0|&\left(c_{-1}c_1c_0 z_1^2z_3 +
                 c_{-1}c_0c_1 z_1^2z_2 +
                 c_0c_{-1}c_1 z_1z_2^2\right.\nonumber \\ & & \left.{}+
               c_0c_1c_{-1} z_1z_3^2 +
               c_1c_{-1}c_0 z_2^2z_3 +
               c_1c_0c_{-1} z_2z_3^2\right) |0\ket \nonumber\\
  & =& \bra 0|c_{-1}c_0c_0\left(-z_1^2z_3 +  
                            z_1^2z_2 
                           -z_1z_2^2 +
                           z_1z_3^2 +
                           z_2^2z_3 
                           -z_2z_3^2\right)|0\ket\,,
  \nonumber
\eea
where the signs come from anti-commuting the modes. Collecting terms 
results in the simple expression
\be
  \bra c(z_1)c(z_2)c(z_3)\ket = (z_1-z_2)(z_1-z_3)(z_2-z_3)\,,
\ee
which indeed satisfies the Pauli principle. In the same manner, all
correlation functions can be obtained. Firstly, it is clear that an
arbitrary correlation function must have first order zeroes for each
pair of coordinates, where two $c$-fields coincide. The same is true for
each pair of coordinates, where two $b$-fields approach each other.
Only when a $c$-field approaches a $b$-field, the singular OPE 
(\ref{eq:bc-opes}) will lead to a first order pole. The only non-trivial
feature is that the number of $c$-fields must exceed the number of 
$b$-fields by precisely three. Thus, in all generality we find
\be
  \bra0|\prod_{i=1}^pc(z_i)\prod_{j=1}^qb(w_j)|0\ket = \prod_{i<i'}(z_i-z_{i'})
  \prod_{j<j'}(w_j-w_{j'})\prod_{i,j}(z_i-w_j)^{-1}\delta_{p,q+3}\,.
\ee

\subsection{The logarithmic $c=-2$ theory}

The $c=-2$ theory has been extensively studied (see e.g.\ \cite{Cappelli:1998,
Flohr:1996a,Gurarie:1993,Gurarie:1997,Walgebra,Kausch:1995}).
Here we want to give a very brief self-contained account
which includes all of the developments relevant to
our discussion of ghost systems.
 
The $c=-2$ theory can be represented as a pair of
ghost fields, or anti-commuting
fields $\theta$, $\bar \theta$ with the standard action \cite{Gurarie:1993}
\begin{equation}
  S=\int \partial \theta \bar \partial \bar \theta\,.
\end{equation}
This action has an $SU(2)$ (actually even an $SL(2,\mathbb{C})$)
symmetry which becomes evident if we
introduce the `spin-up' and `spin-down' fields $\theta^+ \equiv \theta$
and $\theta^- \equiv \bar \theta$ in terms of which the action is
\begin{equation}\label{action}
  S \propto \int \epsilon_{\alpha \beta} \partial
  \theta^{\alpha} \bar \partial \theta^{\beta}\,,
\end {equation}
where $\epsilon$ is the antisymmetric tensor.
Acting on $\theta$ by $SU(2)$ matrices  does not change the action.
The $SU(2)$ algebra is generated by the
$SU(2)$ triplet of generators
\be\label{walgebra}
  W^{\alpha \beta} \propto \partial \theta^\alpha \partial^2 \theta^\beta +
                   \partial \theta^\beta  \partial^2 \theta^\alpha
\ee
of dimension 3,
which form a ${\cal W}$-algebra rather than a Kac-Moody algebra
\cite{Walgebra}.\footnote{As has been noted in a number of publications,
and also somewhere above,
the dimension 1 fields $\theta \partial \theta$ have logarithms in their
correlations functions and therefore do not form a Kac-Moody algebra.}

The fields $\theta$ are complex. Nevertheless writing down the full action
\begin{equation}\label{complex}
  S \propto {\rm i} \int  \epsilon_{\alpha \beta} \partial \theta^\alpha
  \bar \partial \theta^\beta - {\rm i} \int
  \epsilon_{\alpha \beta} \partial {\theta^\alpha}^{\dagger} \bar \partial
  {\theta^\beta}^{\dagger}
\end {equation}
shows that $\theta^\dagger$ decouple from $\theta$ and we
can consider them independently. If, on the other hand, we include
them, the central charge for the theory
(\ref{complex}) is $c=-4$. We emphasize that $\bar \theta$ is
not a complex conjugate of $\theta$, but is just another field.
Alternatively, we could take $\theta$, $\bar \theta$ to
be real fields with an $SL(2,\mathbb{R})$ symmetry. The potentially
misleading notation $\theta,\bar\theta$ is, however, conventional and
commonly used.

To quantize the theory (\ref{action}) we have to compute the fermionic
functional integral
\be
  \label{Ferm}
  \int {\cal D} \theta {\cal D} \bar \theta \exp (-S)\,,
\ee
We note that computed formally this fermionic path integral
is equal to zero due to the ``zero modes'' or constant parts of
the fields $\theta$ which do not enter the action (\ref{action}).
We will meet these zero modes below in more detail and see how these
are connected to the above mentioned zero-modes of the spin one-zero
ghost system.
To make the path integral non-zero we have to insert the fields $\theta$ 
into the
correlation functions (compare with ref. \cite{Friedan}), as in
\be
  \int {\cal D} \theta {\cal D} \bar \theta \, \bar \theta (z)
  \theta(z) \exp (-S) = 1\,.
\ee
Therefore, the vacuum $|0\rangle$
of this theory is somewhat unusual. Its norm
is equal to zero,
\be
  \left\langle 0 | 0 \right \rangle = 0\,,
  \label{strange1a}
\ee
while the explicit insertion of the fields $\theta$ produces nonzero
results
\be
  \label{strange1b}
  \VEV{\bar \theta(z) \theta(w)}=1\,.
\ee
Furthermore, if we want to compute correlation functions of the
fields $\partial \theta$ we also need to insert the zero modes
explicitly,
\bea \label{strange2}
  \VEV{\partial \theta(z) \partial \bar \theta(w) } &=& 0\,, 
  \ \ \ \ {\rm but} \\
  \VEV{\partial \theta(z) \partial \bar \theta(w) \bar \theta(0)  \theta(0)} 
  &=& - {1 \over (z-w)^2}\,.
\eea
Here, the second correlation function is computed by analogy to the free
bosonic field. All this strongly reminds us in some of the typical
features of LCFTs. And indeed, from the point of view of conformal field
theory, the strange behavior of (\ref{strange1a}), (\ref{strange1b}), and 
(\ref{strange2})
can be explained in terms of the logarithmic operators which
naturally appear at $c=-2$. As was discussed in \cite{Gurarie:1993}, the
theory with central charge $c=-2$ must necessarily possess an operator 
$\tilde{\mathbb{I}}$ of scaling dimension
zero, in addition to the unit operator $\mathbb I$, such that
\be \label{jacobi}
  [L_0, \tilde{\mathbb I}] = \mathbb I
\ee
(where $L_0$ is the Hamiltonian).
Moreover, we know from the basic (L)CFT introductionary lectures that
it follows by general arguments such as conformal invariance and
the operator product expansion, that
the property (\ref{jacobi}) necessarily implies the correlation functions
\bea
\label{Kogan}
\VEV{\mathbb I \mathbb I}&=&0\,, \nonumber\\
\VEV{\mathbb I(z) \tilde{\mathbb I}(w)}&=& 1\,, \\
\VEV{\tilde {\mathbb I}(z) \tilde {\mathbb I}(w)} &=& - 2 \log(z-w)\,.\nonumber
\eea
These relations force us to conclude that the operator $\tilde{\mathbb I}$
must be identified with the normal ordered product of $\theta$ and
$\bar \theta$,\footnote{
The author is grateful to A.B.\ Zamolodchikov for pointing
out that $\tilde{\mathbb I}$, as well as any other
local field, can be expressed in terms of the
fundamental fields $\theta$ and $\bar \theta$ of the theory.}
\be
  \tilde{\mathbb I} \equiv  - \mbox{:$\theta \bar \theta$:} = - {1 \over 2}
  \epsilon_{\alpha \beta} \theta^\alpha \theta^\beta
\ee
The stress energy tensor of the theory (\ref{action}) is given by
\be\label{eq:Tthth}
  T=\mbox{:$\partial \theta \partial \bar \theta$:}\,,
\ee
and it is easy to see that its expansion with $\tilde{\mathbb I}$ 
is indeed given by
\be
  \label{expansion}
  T(z) \tilde{\mathbb I}(w)= {1 \over (z-w)^2} + 
  {\partial \tilde{\mathbb I} \over z-w} + \dots
\ee
indicating that $\tilde{\mathbb I}$ is indeed not a primary field.
 
Now, the mode expansion of the fields $\theta$ has to be written in the
form
\be\label{modes}
  \theta(z) = \sum_{n \not = 0} \theta_n z^{-n} + \theta_0 \log(z) + \xi\,,
\ee
where $\xi$ are the crucial zero modes (they disappear in the
expansion for $\partial \theta$). Here $n \in \mathbb Z$ in the
untwisted sector (ie. with periodic boundary conditions)
and $n \in \mathbb Z + \frac{1}{2}$ in the twisted sector
(anti-periodic boundary conditions).

To be consistent with the earlier results (\ref{strange2}) and 
(\ref{expansion})
we have to impose the following anti-commutation relations (the interested
reader might wish to compare these with a slightly different realization
of the $c=-2$ LCFT in terms of so-called symplectic fermions of scaling
dimension one \cite{Kausch:1995})
\bea \label{anticom}
  \left\{ \theta_n, \bar \theta_m \right\} &=&  
  {1 \over n} \delta_{n+m,0} \ \ \ \ \forall\ n \neq 0\,, \\
  \left\{ \theta_0, \bar \theta_0 \right\} &=& 0\,,\nonumber\\ 
  \left\{ \theta_m, \theta_n \right\} = \left\{ \bar \theta_n, \bar
  \theta_m \right\} &=& 0\,, \nonumber\\ 
  \left\{ \xi, \bar \xi \right\} &=& 0\,, \nonumber\\ 
  \left\{ \xi, \bar \theta_0 \right\} &=& 1\,, \nonumber\\
  \left\{ \theta_0, \bar \xi \right\} &=& -1\,.\nonumber
\eea
The last two relations are absolutely crucial in keeping
(\ref{expansion}) intact. The mode expansion $\theta_n$ should
not be confused with the notations $\theta^{\alpha}$ and $\theta^{\beta}$
introduced earlier. To avoid confusion we will primarily use
the $\theta, \bar \theta$ notation.

It is now very important to note that the modes
$\xi$ become the creation operators for logarithmic states. Indeed,
\be
  \theta_n | 0\rangle =0\ \ \ \ \forall\ n \geq 0\,,
\ee
and
\be
  \tilde{\mathbb I} |0 \rangle = \bar \xi \xi | 0\rangle\,.
\ee
The mode expansion (\ref{modes}) together with (\ref{anticom}) and
\be
  \VEV{0|0}=0\,,\ \ \ \ \VEV{\bar \xi \xi }=1
\ee
can be
used to compute any correlation function in the theory.
 
For instance, we can reproduce the typical LCFT correlation functions of
a Jordan block pair of fields such as (\ref{Kogan})
\be
  \VEV {\mathbb I(z) \tilde{\mathbb I}(w) } = \VEV {\bar \theta(w) \theta(w)} =
  \VEV{\bar \xi \xi} = 1\,.
\ee
while on the other hand
\bea
\label{Kogan1}
  \VEV{\tilde{\mathbb I} (z) \tilde{\mathbb I} (w)} &=& 
  \VEV{ \mbox{:$\bar \theta(z)\theta(z)$:}\, \mbox{:$\bar\theta(w)
  \theta(w)$:} } \nonumber\\
  &=&
  \VEV{ \bar \xi \theta(z) \bar \theta (w) \xi } + \VEV { \bar \theta(z) \xi
  \bar \xi \theta(w) } = - 2 \log(z-w)\,.
\eea
The last line of (\ref{Kogan1}) can be computed either directly
in terms of modes or by comparison with (\ref{strange2}).

\bq
As has been discussed at length in the literature, the fields $W$ introduced
in (\ref{walgebra}) form a ${\cal W}$-algebra and in fact all the states
of the $c=-2$ theory can be classified according
to various representations of that algebra. A clear review can be found in
\cite{Kausch:1995}. Six representations are listed in that paper. They
can easily be represented in terms of the fields of our theory.
We have the unit operator $\mathbb I$, the logarithmic operator 
$\tilde{\mathbb I} = - \mbox{: $\theta
\bar \theta$}:$, the $SU(2)$ doublet of dimension 1 fields $\partial \theta$ and
$\partial \bar \theta$, the twist field $\mu$ of dimension $-1/8$,
a doublet of twist fields
$\sigma_{\alpha}\equiv {\left( \theta_{\alpha} \right)}_{- {1 \over 2}} \mu$
of dimension\footnote{$\theta_{-{1\over 2}}$
is the mode expansion (\ref{modes}) for $\theta$ where $n\in 
\mathbb Z+{1 \over 2}$
to reproduce the twisted sector. The zero modes are naturally absent
in that sector.} $3/8$,
and finally a structure of fields $\theta$, $\partial \theta$ and
$\theta \partial \theta$ connected with each other by the action
of the Virasoro generators $L_n$.
\eq

With all the preliminaries completed we can proceed to construct
the correlation functions of the fields $\theta$. 
For example, the correlation
function
\be
\label{Haldane}
\VEV {\partial \theta(z_1) \partial \bar \theta(w_1) \dots
\partial \theta(z_n) \partial \bar \theta(w_n) \tilde{\mathbb I} } =
\sum_{\sigma} {\rm sign}{\sigma} \prod_{i=1}^n {1 \over (z_i -
w_{\sigma(i)})^2}\,,
\ee
where $\sigma(i)$ is the permutation of the numbers $1$, $2$, $\dots$, $n$,
reproduces the Haldane-Rezayi wave function which was proposed for
the fractional quantum Hall effect at filling $\nu=5/2$.
Note the explicit insertion of the logarithmic operator 
$\tilde{\mathbb I}=\mbox{:$\bar
\theta \theta$:}$ to make (\ref{Haldane}) non-zero. 
For convenience,
we express the correlation functions in this section
in `$z$-$w$' notation in which the $\theta$'s are at the
points $z_i$ and the $\bar \theta$'s are at the $w_i$'s,
which makes some of the formul\ae\ more transparent.

The correlation functions in the twisted sector can be found
by splitting the logarithmic operator into two twist fields $\mu$
according to the general formula (see for example \cite{Gurarie:1993})
\be
  \mu(z) \mu(w) \approx \mathbb I \log (z-w) + \tilde{\mathbb I}\,,
\ee
and is equal to
\bea
\label{twisted}
\lefteqn{\VEV {
\partial \theta(z_1) \partial \bar \theta(w_1) \dots
\partial \theta(z_n) \partial \bar \theta(w_n)
\mu(\eta_1) \mu(\eta_2) } = }\\  & & 
{\left( \eta_1-\eta_2 \right)}^{1 \over 4} \sum_{\sigma}
{\rm sign\sigma}\prod_{i=1}^n {
(z_i-\eta_1)(w_{\sigma(i)}-\eta_2)+(z_i-\eta_2)(w_{\sigma(i)}-\eta_1) \over
(z_i-w_{\sigma(i)})^2 \sqrt{ (z_i-\eta_1) (z_i-\eta_2)(w_{\sigma(i)}-\eta_1)
(w_{\sigma(i)}-\eta_2)} }\,.\nonumber
\eea
Note that we do not need the logarithmic operator any more. It has
been split into two twist fields. Alternatively, we can say that in
the twisted sector the summation in (\ref{modes}) is over half integer numbers
and the zero modes no longer
enter the expansion for the fields $\theta$.

Correlation functions of the type
(\ref{twisted}) are, for instance, useful for
constructing the bulk excitations in the Haldane-Rezayi description of
the $\nu=5/2$ fractional quantum Hall effect.
However, the twist fields are not the only way of
doing it. We could also split the logarithmic
operator according to the
operator product expansion
\be
  \tilde{\mathbb I}(z) \tilde{\mathbb I}(w) = -2 \log(z-w) 
  \tilde{\mathbb I} + \dots\,,
\ee
which follows from (\ref{Kogan}). Thus, we can easily compute other
correlation functions such as the following one:
\be
\VEV{
\partial \theta(z_1) \partial \bar \theta(w_1) \dots
\partial \theta(z_n) \partial \bar \theta(w_n)
\tilde{\mathbb I} (u_1) \tilde {\mathbb I}(u_2) }\,.
\ee
It can be computed by either solving the
differential equations of conformal field theory, or by the straightforward
mode expansion (\ref{modes}) and (\ref{anticom}). Either method results in
\bea
\label{newpart}
\lefteqn{\VEV{
\partial \theta(z_1) \partial \bar \theta(w_1) \dots
\partial \theta(z_n) \partial \bar \theta(w_n)
\tilde{\mathbb I} (u_1) \tilde{\mathbb I}(u_2) } = }\\
&-& 2 \log(u_1-u_2)
\sum_{\sigma} {\rm sign}{\sigma} \prod_{i=1}^n {1 \over (z_i -
w_{\sigma(i)})^2} \nonumber\\ &-& 
\sum_{\sigma} {\rm sign}{\sigma} \sum_{k=1}^{n} \left\{
\prod_{i \not = k} \left( {1 \over (z_i -
w_{\sigma(i)})^2 }\right)  {(u_1-u_2)^2 \over (u_1-z_{k}) (u_1-w_{\sigma(k)})
(u_2-z_k) (u_2-w_{\sigma(k)})} \right\}\,.\nonumber
\eea
We see that it splits into two terms. One is the product of the
Haldane-Rezayi wave function (\ref{Haldane}) and the logarithm. The other
is a nontrivial expression. In fact, it is easy to get rid of the
trivial part by taking one of the logarithmic operators to infinity.
In doing so we have to remember the transformation law for the
logarithmic fields which follows from (\ref{jacobi}),
\be
  \tilde{\mathbb I}(f(z))=\tilde{\mathbb I} (z) 
  + \log \left( {\partial f \over \partial z } \right)\,.
\ee
According to the standard procedure, taking the position
of the field $\tilde{\mathbb I}(z)$ to infinity
corresponds to taking the position of the
field $\tilde{\mathbb I}(1/z)=\tilde{\mathbb I}(z) - 2 \log(z)$ to the origin.
Therefore the trivial part of (\ref{newpart}) disappears.

We could have computed all these correlation functions also by
using Wick's theorem for anti-commuting fields together with the
fundamental contractions
\bea\label{eq:thth}
  \bra\theta(z)\bar\theta(w)\ket &=& -\log(z-w)\,,\\
  \bra\theta(z)\theta(w)\ket\ =\ 
  \bra\bar\theta(z)\bar\theta(w)\ket &=& 0\,.\nonumber
\eea
So far, we only looked at correlation functions with derivatives of
$\theta$ fields and explicit insertions of the logarithmic $\tilde{\mathbb I}$
operator. It turned out that this CFT, although logarithmic, possesses
many correlation functions which are entirely void of any logarithms.
We only have to confine ourselves to a certain subset of all possible 
fields (in particular the derivative fields $\partial\theta^\alpha$ will do)
together with a minimal insertion of operators such that the resulting
object is non-zero. If we only use derivative fields $\partial\theta^\alpha$,
we have to make sure that the zero modes $\bar\xi\xi$ are somehow
inserted. The minimal way to do this is to put one field $\tilde{\mathbb I}$
at infinity.

It is a highly instructive exercise to redo some of the above outlined
calculations with the slight modification that we take correlators
of fields $\partial\bar\theta$ and $\theta$, i.e.\ we allow that 
the $\theta$ field be inserted without derivative, but not so for
the $\bar\theta$ field. Note that this
means that the $\xi$ zero mode is then present, but not the
$\bar\xi$ zero mode. As the attentive reader might already have guessed,
we furthermore may suggestively identify\footnote{The identification can
be made mathematically rigorous, if in addition the zero-mode $\theta_0$ is
put to zero. Otherwise, the mode expansion of $\theta(z)$ would contain
the term $\log(z)\theta_0$ absent in the mode expansion of $c(z)$.
However, even if this mode is present, it does not affect any of the
correlation functions with $\del\bar\theta$ fields.}
\be
  b(z)\equiv\partial\bar\theta\,,\ \ \ \ c(z)\equiv\theta(z)\,.
\ee 
The conformal dimensions do indeed coincide, if we consider a 
spin one-zero ghost system of central charge $\left.-2(6j^2-6j+1)\right|_{
j=1}=-2$. That alone is, of course, not sufficient to justify this
identification, but it is easy to see that all correlation functions
in the $(j=1,0)$ $bc$ system can be reproduced exactly, provided we
evaluate all the expressions in $\partial\bar\theta$ and $\theta$
between the states $|0\ket$ and $\bra\bar\xi|$. The non-trivial
out-state is necessary to provide the still missing zero mode to
ensure that the correlation function does not vanish if the number
of $\theta$ fields does exceed the number of $\del\bar\theta$ fields
by precisely one. Thus, we find that
\bea
  \bra b(z_1)\ldots b(z_p)c(w_1)\ldots c(w_q)\ket &=&
  \bra\bar\xi|\del\bar\theta(z_1)\ldots\del\bar\theta(z_p)
  \theta(w_1)\ldots\theta(w_q)|0\ket\\
  &=& \!\!\prod_{1\leq i<i'\leq p}\!\!(z_i-z_{i'})
      \!\!\prod_{1\leq j<j'\leq q}\!\!(w_j-w_{j'})
      \!\prod_{{1\leq i\leq p\atop 1\leq j\leq q}}\!
      \frac{1}{(z_i-w_j)}\delta_{p+1,q}\,.\nonumber
\eea
Note further that the definition of the stress energy tensor (\ref{eq:Tghost})
within the $(j=1,0)$ $bc$ ghost system does exactly agree with the definition
of $T$ within the $\theta,\bar\theta$ system (\ref{eq:Tthth}).  This
completes the identification of the $bc$ ghost system with a sub-sector
of the logarithmic $c=-2$ theory given by the $\theta,\bar\theta$ system.
Thus, we can say that the $c=-2$ LCFT is an augmentation of the
ghost system in the above described sense.

\bq
Knizhnik considered a long time ago how to put CFTs on general
Riemann surfaces. He considered ghost systems and described non-trivial
Riemann surfaces as branched coverings of the complex plane (or
Riemann sphere). He showed that the branch points can be simulated
by certain conformal fields, so-called twist fields. In case of
a hyper-elliptic surface, where all branch points have ramification
number two, only $\mathbb Z_2$ twists arise. Strikingly, these
are precisely provided by the field $\mu$ introduced earlier. As we know
now, after the advent of LCFT, twist fields may produce logarithms,
and we already saw that $\bra\mu\mu\mu\mu\ket$ does indeed produce
a logarithmic divergency. Since Knizhnik did, at that time, only
consider the twist fields together with the $bc$ system, he was
badly surprised by the appearance of logarithms. Nowadays, we would
simply say that, after including twist fields to the $bc$ system,
we already have enlarged the CFT to a logarithmic one, since
the logarithmic fields can be obtained from the OPE 
$\mu(z)\mu(w)=(z-w)^{1/4}[\tilde{\mathbb I}(w)+\log(z-w)\mathbb I]$.
Hence, primary fields with this property are now called
pre-logarithmic fields.
\eq

\subsection{Remarks on the Haldane-Rezayi fractional quantum Hall state}

In this section, we briefly discuss one application of LCFT, actually the
only one application we will mention explicitly in these lectures.
Unfortunately, space-time limits do not permit to give any introduction 
to the (fractional) quantum Hall effect and its theoretical description.
The quantum Hall effect is essentially a $2\!+\!1$ dimensional problem.
It can be shown that, in the particular circumstances relevant for the
effect, the Chern-Simons term dominates the standard Maxwell term in the
action accounting for the universality of the effect. We further know that
there is a one-to-one correspondence of Chern-Simons theories in the 
$2\!+\!1$ dimensional bulk (usually a filled cylinder) and unitary
CFTs on the boundary ($S^1\times\mathbb{R}$), a deep result due to Witten.
In the quantum Hall effect, the boundary CFT describes the gapless
edge excitations of the quantum Hall state, which is usually considered
to be some kind of incompressible quantum fluid, the Hall droplet.
The issue which concerns us here is of a different nature. If one
considers the bulk theory without intrinsic time, i.e.\ as a pure quantum 
mechanical problem, then the resulting bulk wave functions show a
striking similarity with CFT correlators of free field type. What is
so far missing is a physical explanation for this resemblance. The issue
is complicated by the fact that there is usually no principle which
selects the correct CFT among an often large variety of possible
``solutions'', i.e.\ of possible candidate CFT which all somehow
reproduce the expected wave function in terms of certain of their
correlation functions. The situation is a bit more promising in the
case of the so-called Haldane-Rezayi state, since for this state 
most CFT candidates can be ruled out right from the start due to
several restrictions such as topological ordering.

We have already seen that the ground state for the exceptional fractional
quantum Hall effect at filling factor $\nu=5/2$, as proposed by 
Haldane and Rezayi, is given as 
\bea
  \Psi_{{\rm HR}} &=& {\rm Pf}\left(\frac{u_iv_j - v_iu_j}{(z_i-z_j)^2}\right)
  \prod_{j<i}(z_i-z_j)^2\,{\rm e}^{-\frac{1}{4\ell_0^2}\sum|z_i|^2}\\
                  &\cong&\sum_{\sigma}\prod_{i=1}^n\frac{1}{
                  (z_i-w_{\sigma(i)})^2}\nonumber
\eea
upto the non-holomorphic exponential factor. This factor ensures that the
probability density of the wave function falls off fast enough for large
arguments. However, after compactifying the plane to the Riemann sphere
(where the homogeneous magnetic field is mapped to the magnetic field of
a monopole in the center of the sphere), this factor is obsolete. In the
above formula, $u_i$ and $v_i$ denote up- and down-spin states of the
$i^{{\rm th}}$ electron, respectively, and $\ell_0$ is the magnetic length.

A very important concept in the theoretical study of the fractional
quantum Hall effect is the so-called {\em topological ordering}, which
refers to the fractional statistics of quasi-particles, and was introduced
by X.G.~Wen. Among other things, this property yields a precise prediction
on the degeneracy of the ground state wave function on a torus geometry.
This in particular allows to test a CFT proposal for such a ground state
bulk wave function. The startling prediction for the $\nu=5/2$ state now
is, after a trivial reduction, that the degeneracy is five-fold. Most
attempts to describe the ground state wave functions in terms of CFT
correlators only yield smaller degeneracies. On the torus, the ground
state reads
\be
  \Psi_{{\rm HR}}^{a,b} = {\rm Pf}\left(\frac{(u_iv_j-v_iu_j)\vartheta_a
    (z_i-z_j)\vartheta_b(z_i-z_j)}{\vartheta_1^2(z_i-z_j)}\right)
    \prod_{j<i}\vartheta_1^2(z_i-z_j)\prod_{k=1}^2\vartheta_1(\sum_i z_i
    -\zeta_k)\,,
\ee
where $\zeta_k$ are two arbitrary complex numbers. Since there is a
linear relationship between $\vartheta_2^2,\vartheta_3^2,\vartheta_4^2$,
we only get five different ground state wave functions with $a,b=2,3,4$,
not taking into account the trivial degeneracy due to the free choice of
the complex center of mass coordinates. The reader unfamiliar with
standard elliptic $\vartheta$-functions should consult any textbook
on elliptic functions. These are the standard double-periodic functions.
As a rule of thumb, a torus correlator is obtained from a plane
correlator by replacing any occurrence of a $(z_i-z_j)$ factor by 
an appropriate double-periodic version of it, essentially given by
$\vartheta(z_i-z_j)$. Furthermore, a torus correlation function always
receives an additional factor for the center of mass coordinate.

The key point is that -- if a CFT description is to be correct -- this
ground state degeneracy must be reproduced by the independent ways how
the identity propagator can be built by the creation of two fields,
which are then taken around a homology cycle in opposite directions to
annihilate themselves when they come together again. These pairs of fields
are then interpreted as quasi-hole-quasiparticle pairs. Thus, the ground state
degeneracy on a torus is equal to the number of {\em distinct\/} bulk
excitations. This number is determined by the number of linear independent
monodromies (or braidings) the quantum Hall state admits.

Supposing that the $c=-2$ CFT is the correct description for the bulk
ground state, we have to count the ways to produce the identity propagator
from OPEs of other fields. If we describe the torus by a branched covering
of the complex plane, we have to insert precisely four branch point
vertex operators $\mu(e_i)$, $i=1,\ldots,4$, into a complex plane
correlator. We already know from the preceeding section that these four
twist fields will ensure that the correlator is non-vanishing. Moreover,
the OPE of two such twist fields contains the logarithmic $\tilde\mathbb{I}$
field, and we need at least one of these logarithmic fields to get a
non-zero correlator. Now, it is merely a matter of counting various
contractions which still yield a torus correlation functions.
The generic one is $\bra\mu\mu\mu\mu\ldots\ket$. Inserting the OPE for
two of these fields, we get two further possibilities,
$\bra\mathbb{I}\mu\mu\ldots\ket$ and $\bra\tilde\mathbb{I}\mu\mu\ldots\ket$.
The last two possibilities come form the excited twists $\sigma^{\pm}$ and
the $h=1$ current field $J=\theta^{\mp}\partial\theta^{\pm}$, which
appears in the OPE of $\sigma^+\sigma^-$. Thus, we also have
$\bra\sigma^+\sigma^-\mu\mu\ldots\ket$ and $\bra J\mu\mu\ldots\ket$.
It needs a bit more work to see that $\bra\sigma\sigma\sigma\sigma\ldots\ket$
and $\bra JJ\ldots\ket$ does not yield different torus correlators.
A naive and handwaving way to see this is the following: we need one
$\tilde\mathbb{I}$ operator, which we may put at infinity, and it does
not matter how we create this (first) one logarithmic operator. 
Having four branch points for a torus, two of them are already accounted
for by this requirement. Thus, we can only take the other two branch points
and see, in which ways we can account for them. The five possibilities
for a bulk excitation are therefore the trivial $\mathbb{I}$ (no excitation
at all) and $\mu\mu$, $\tilde\mathbb{I}$, $\sigma^+\sigma^-$, and $J$.

We will see in the next section that the modular properties of
the $c=-2$ LCFT precisely support this picture. This theory admits
a five dimensional representation of the modular group. Furthermore,
the rank of the fusion matrix $N_{\tilde\mathbb{I}i}^{\ \,j}=
N_{ij}^{\ \,\tilde\mathbb{I}}$ is also five, i.e.\ these fusion matrices
have maximal rank\footnote{As the discussion in the next chapter shows, there
are six representations, but only five of them are linearly independent.}. 
The dimension of the representation of the modular
group coincides with the number of distinct torus partition functions. Thus,
the dimension of the $\mathbb{P}SL(2,\mathbb{Z})$ representation precisely
counts the ways of writing distinct torus propagators. This is 
exactly what the
different ground state excitations are -- formulated in terms of a CFT
description.

\bq
For completeness, we mention the five different ground state wave functions. 
There is an additional trivial two-fold degeneracy, which
stems from the two completely filled Landau levels of the $\nu=5/2$ state.
It is incorporated in the following formul\ae\ by the choice $p,p_{\pm}=0,1$.
We already computed three of these wave functions, namely $\Psi_{\mathbb{I}}$
in (\ref{Haldane}), $\Psi_{\mu\mu}$ in (\ref{twisted}), and 
$\Psi_{\tilde\mathbb{I}}$ in (\ref{newpart}). These results are recast here
in the spinor notation. The interested reader should check, that both
version do indeed coincide.
\begin{eqnarray*}
\Psi_{\mathbb I} &=&
{\rm Pf}\left(\frac{{u_i}{v_j}-{v_i}{u_j}}{({z_i} - {z_j})^2}\right)
{\prod_i}{\left({z_i}-\eta\right)^p}
\,{\prod_{j<i}}{\left({z_i} - {z_j}\right)^2}
\,,\\
\Psi_{\mu\mu} &=&
{\left({\eta_+}-{\eta_-}\right)^{3/8}}
{\rm Pf}\left(\frac{\left({u_i}{v_j}-{v_i}{u_j}\right)
\left(\left({z_i} - {\eta_+}\right)
\left({z_j} - {\eta_-}\right) + i\!\leftrightarrow\! j\right)}
{({z_i} - {z_j})^2}\right)
{\prod_{i,\pm}}{\left({z_i}-{\eta_\alpha}\right)^{p_\pm}}
\,{\prod_{j<i}}{\left({z_i} - {z_j}\right)^2}
\,,\\
\Psi_{\tilde\mathbb I} &=&
{\cal A}\left(\frac{\left({u_1}{v_2}-{v_1}{u_2}\right)
{\left({\eta_+}-{\eta_-}\right)^2}}
{\left({z_1}-{\eta_+}\right) \left({z_1}-{\eta_-}\right)
\left({z_2}-{\eta_+}\right) \left({z_2}-{\eta_-}\right)}
\,\frac{{u_3}{v_4}-{v_3}{u_4}}
{({z_3} - {z_4})^2}\,\ldots\,\right)
{\prod_{i,\pm}}{\left({z_i}-{\eta_\pm}\right)^{{p_\pm}+1}}
\,{\prod_{j<i}}{\left({z_i} - {z_j}\right)^2}
\,,\\
\Psi_{\sigma^+\sigma^-} &=&
{\left({\eta_+}-{\eta_-}\right)^{19/8}}
{\cal A}\left(
\frac{\left({u_1}{v_2}+{v_1}{u_2}\right)\left({z_1}-{z_2}\right)}
{\left({\eta_+}-{z_1}\right)\left({\eta_-}-{z_1}\right)
 \left({\eta_+}-{z_2}\right)\left({\eta_-}-{z_2}\right)}
 \,\frac{\left({u_3}{v_4}-{v_3}{u_4}\right)
 \left(\left({z_3} - {\eta_+}\right)
 \left({z_4} - {\eta_-}\right) + 3\!\leftrightarrow\! 4\right)}
 {({z_3} - {z_4})^2}\,\ldots\,\right)
\\ &\times&
 {\prod_{i,\pm}}{\left({z_i}-{\eta_\pm}\right)^{p_\pm}}
 \,{\prod_{j<i}}{\left({z_i} - {z_j}\right)^2}
\,,\\
\Psi_J &=&
{\cal A}\left(\frac{1}{(\eta-{z_1})^2}\,
\frac{{u_2}{v_3}-{v_2}{u_3}}{({z_2} - {z_3})^2}\,\ldots\,\right)
{\prod_i}{\left({z_i}-\eta\right)^p}
\,{\prod_{j<i}}{\left({z_i} - {z_j}\right)^2}
\,.
\end{eqnarray*}
In the above formul\ae, ${\rm Pf}$ denotes the Pfaffian, and ${\cal A}$
denotes complete anti-symmetrization. Furthermore, the insertion points
of the excitation operators indicated in the labels of the corresponding wave
functions are the coordinates $\eta$ or $\eta_\pm$, respectively. Note that
insertion of a second logarithmic field makes it necessary to explicitly
refer to the coordinates of the first one. More details on this construction
can be found in \cite{Gurarie:1997}. More recent works in the still
ongoing investigation in a CFT description of the Haldane-Rezayi state
are, for example, \cite{Cappelli:1998,Read:1999} and references therein.

The reader might note that the Haldane-Rezayi fractional quantum Hall state
is successfully described by a CFT which is a ghost or spin $(0,1)$ system
with $c=-2$.
This coincides nicely with the observation that the $\nu=5/2$ fractional
quantum Hall state is made out of spin-singlet pairs of electrons, i.e.\
anti-commuting spin $j=0$ states. The full CFT description should also
account for the two fully filled Landau levels. These should be filled
by completely polarized electrons, and indeed, the ghost or spin
$(\frac12,\frac12)$ system with $c=1$ precisely contains two free
Dirac spin fields. Hence, we not only have a ``fit'' of CFT data such
as conformal weights and correlators reproducing the Haldane-Rezayi state
and its excitations, we also have a natural geometrical interpretation
for the particular CFT candidate, namely that it directly describes the
correct spin system in the presence of a magnetic field. The flux quanta
of the magnetic field, which yield the quasi-particle excitations with their
fractional statistics, effectively amount to replacing the plane of the
quantum Hall semi-conductor sample by a ramified double covering of itself
due to the effect of the flux quanta on the paired electron singlet states. 
Each of the flux quanta can then be considered as a branch point.
Thus, the Haldane-Rezayi quantum Hall state beautifully connects
experimentally observable physics, spin systems on Riemannian surfaces
and logarithmic conformal field theory with each other.
\eq

\section{Modular invariance}

So far, we have considered CFT on the simplest possible worldsheet, the
cylinder, which we have mapped by a conformal transformation to the
punctured complex plane. In string theory, the cylinder is the world
sheet of one freely moving non-interacting closed string. Interaction
of several strings yields
world sheets which might be any Riemann surface. It is intuitive to
use the genus of the Riemann surface as an order count, since it
directly corresponds to the loop order of the Feynmann diagram of the
low-energy effective field theory, where the extent of the string
becomes invisible. So, to zero-th order, we have a Riemann sphere with
a number of tubes attached, one for each string which interacts with
the others. To first order, we find a torus, again with a number of
tubes attached, and so on. 

The tubes of the incoming and outgoing strings, if these are considered
to be otherwise non-interacting, can be thought of asymptotically as
infinitely long and infinitely thin spikes. In effect, these tubes
can be replaced by punctures of the Riemann surface, where an appropriate
vertex operator carrying the right momentum and quantum numbers is placed.
What remains is the non-trivial topology of the Riemann surface.

So far, we have described CFT algebraically by a set of highest-weight
states $|h,\bar h\ket = \Phi_{h,\bar h}(0,0)|0\ket$, on which the left-
and right chiral Virasoro algebra acts. In the case of a logarithmic CFT,
we extended this to Jordan cells spanned by several states, 
$|h;i,\bar h;\bar\imath\ket = \Phi_{(h;i),(\bar h;\bar\imath)}(0,0)|0\ket$,
of which only one, $|h;0;\bar h;0\ket$, behaves as a proper highest weight
state.
The question which naturally
arises is which combinations of such ground states actually occur in the 
CFT. If we know this, we have a complete characterization of the
physical states in the theory, namely all the admissible ground states plus
all their descendants created by the generators of the Virasoro algebras,
minus all null states.

Crossing symmetry, or equivalently duality, has already given us
some constraints, but these were constraints for the complex plane only.
Do different Riemann surfaces yield different constraints? And is it
possible to have a theory consistent on any arbitrary Riemann surface?
The answer to both questions is yes, and we will sketch a bit of the
answer in the following. As a general result, one can show for a large
class of CFTs that crossing symmetry of correlators 
on the complex plane and modular invariance of the partition function on 
the torus is sufficient to make the theory consistent on arbitrary
Riemann surfaces. This is one of the motivations why modular invariance
on the torus is often considered to be a fundamental requirement for
CFT.

Interestingly, also condensed matter physicists are very fond of 
modular invariance. To understand this, first note that we usually
consider CFTs in complex variables and, thus, automatically as
Euclidean field theory. Time is then commonly interpreted as
temperature, and partition functions are well defined objects. 
Now, let us conformally map the complex plane (with variable
$z$) with the origin deleted onto a strip of width $L$ (with variable $u$).
This map is given by the exponential $z=\exp(2\pi{\rm i} u/L)$. It is a well
known technique in statistical physics to consider the system on a 
periodic strip, here with width $L$, and to introduce the transfer matrix
$$
  {\cal T} = \exp\left\{-\frac{2\pi}{L}(L_0+\bar L_0-\frac{c}{12})\right\}\,.
$$
Here $L_0+\bar L_0$ serves as Hamiltonian, since this linear
combination generates time translations.\footnote{The reader should take
care that $L_0+\bar L_0$, considered on the $z$-plane, generates
dilatations. Only in the $u$-strip does it generate time translations.}
The additional term involving
the central charge comes from the used conformal map. This map is not
one-to-one, and introduces a conformal anomaly. The reader might convince
herself first that the stress energy tensor on the strip is related to
the one on the plane via 
$$
  T_{{\rm strip}}(u) = -(2\pi/L)^2\left[T_{{\rm plane}}(z)z^2 - \frac{1}{24}c
  \right]\,,
$$
and then that with $\bra T_{{\rm plane}}(z)\ket = 0$ one must have
$\bra T_{{\rm strip}}(u)\ket=\frac{1}{24}c(2\pi/L)^2$. Hence, the above
mentioned shift in the transfer matrix.

\bq
  The OPE of the stress energy tensor with itself tells us how the
  stress energy tensor reacts to conformal transformations. It is
  not an entirely trivial task to explicitly work out the transformation
  of $T(z)$, but the result can be cast in the formula
  $$
    T(z){\rm d}z^2 = T'(z'){\rm d}z'^2 + \frac{c}{12}\{z',z\}{\rm d}z^2\,,
  $$
  where the so-called Schwarzian derivative of the map 
  $z\mapsto z'=f(z)$ is defined as
  $$
    \{z',z\}=\frac{f'''}{f'} - \frac32\left(\frac{f''}{f'}\right)^2
  $$
  The conformal anomaly mentioned above can now be computed easily by
  making use of the just given transformation law of $T$ for 
  $f(z) = -{\rm i}\frac{L}{2\pi}\log(z)$.
\eq

We may now further confine the system to a box of size $L,M$, with
periodic boundary conditions on both sides. Then the partition function
of such a system reads
\be\label{eq:partfunct}
  Z = Z(L,M) = {\rm tr}\,\exp\left\{-2\pi\frac{M}{L}
  (L_0+\bar L_0-\frac{c}{12})\right\}\,.
\ee
A box with periodic boundary conditions has the topology of a torus.
The central observation is now that, since we
deal with a Euclidean theory, space and time are completely symmetric to
each other. It follows that in such a framework a physical sensible
partition function should satisfy $Z(L,M) = Z(M,L)$.

\bq
More generally, one could consider a periodicity, where a time translation
by $M$ is always accompanied by a space translation, generated by 
${\rm i}(L_0-\bar L_0)$.\footnote{On the $z$-plane, ${\rm i}(L_0-\bar L_0)$
generates rotations.} Let us assume that this addition space translation
is by $N$. Then the partition function would read
$$
  Z = Z(L,M,N) = {\rm tr}\,\exp\left\{-2\pi\frac{M}{L}
  (L_0+\bar L_0-\frac{c}{12})+2\pi{\rm i}\frac{N}{L}(L_0-\bar L_0)\right\}\,.
$$
Introducing complex numbers $\omega_1=L$, $\omega_2=N+{\rm i}M$,
$\tau = \omega_2/\omega_2$, one can rewrite this with 
$q=\exp(2\pi{\rm i}\tau)$ and $\bar q=\exp(-2\pi{\rm i}\bar\tau)$ elegantly as
$$
  Z(\tau,\bar\tau) = {\rm tr}\left(q^{L_0-c/24}\bar q^{\bar L_0-c/24}\right)\,.
$$
\eq

\subsection{Moduli space of the torus}

\EPSFIGURE{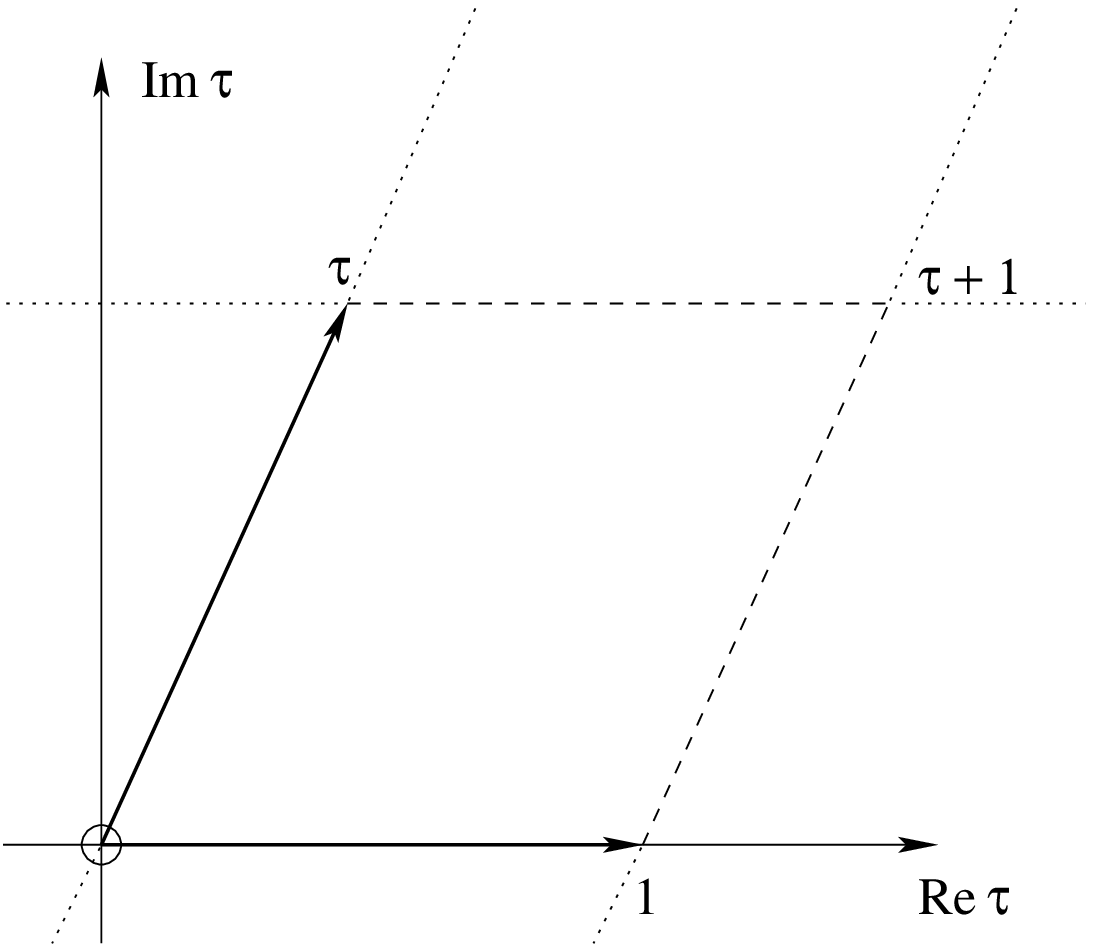,width=6.5cm}{The upper half plane and the modular
parameter $\tau$ defining a lattice, i.e.\ torus.}
As a general rule of thumb, one usually assumes that all states in a 
theory contribute to loop diagrams. This may be seen as a motivation,
why we expect that it is useful to study CFT on the simplest loop
diagram, the torus. Essentially, a torus is a cylinder whose ends have
been sewn together. Mathematically, it is usually described as the complex
plane modulo a lattice. Let the lattice be spanned by two basic
lattice vectors, $\omega_1$ and $\omega_2$. Then two points $z,z'$ in the 
complex plane are identified with each other, if there exist two integers
$n_1,n_2$ such that $z'=z+n_1\omega_1+n_2\omega_2$. Since the overall
size and orientation of the torus shouldn't matter 
(due to global scaling, translational and rotational invariance
of the CFT), we may choose more conveniently one of the base lattice
vectors to lie on the real axis with length one, starting at the origin,
and the other can without loss of generality be taken to lie in the upper 
half plane, $\tau \sim \omega_2/\omega_1$, $\Im{\rm m}\tau > 0$. In effect,
the entire lattice is described by one complex number $\tau\in\mathbb{H}$.

The key observation is now that the lattice, and consequently the torus,
does not change at all if we replace $\tau$ by $\tau+1$, since this
spans the same lattice. Such a transformation is called unimodular.
In the same manner, the lattice does not change if we replace 
$\tau$ by $\-1/\tau$, where we implicitly have to rescale the lattice,
though (the overall since of the torus is irrelevant). Since $\tau\sim
\omega_2/\omega_1$, we see that $-1/\tau$ basically interchanges the
role of $\omega_2$ and $\omega_1$. The group spanned by these
transformations $T:\tau\mapsto\tau+1$, $S:\tau\mapsto-\frac{1}{\tau}$
is called the modular group $PSL(2,\mathbb{Z})$ and is the set of
all $2\times 2$ matrices $M={a\ b\choose c\ d}$ with $a,b,c,d\in\mathbb Z$
and $\det M = ad-bc=+1$. The action of this group on $\tau$ is given by
$M(\tau)=\frac{a\tau+b}{c\tau+d}$ which explains why we restrict the sign
of the determinant and identify matrices $\pm M$ with each other
(this is what the $P$ stands for: $PSL(2,\mathbb Z)=SL(2,\mathbb Z)/
\mathbb Z_2$).

Since the torus does not really change under a $PSL(2,\mathbb Z)$
transformation of its modulus $\tau$, we should expect that a 
physical sensible theory does not change under such a transformation
either, as we have motivated in the preceeding section. Thus we impose
as a condition on our (L)CFT that its partition function be modular
invariant. In the following, we often use the variables
$q={\rm e}^{2\pi{\rm i}\tau}$ and $\bar q={\rm e}^{-2\pi{\rm i}\bar\tau}$
instead of $\tau$ and $\bar\tau$. A series expansion in $q,\bar q$
is then an expansion around the point $\tau=+{\rm i}\infty$, i.e.\ where
the torus is more like a cylinder.

We so far have made elaborate use of the fact that much in conformal
field theory can be considered separately for holomorphic and
anti-holomorphic fields, or left-chiral and right-chiral fields,
respectively. Although one of the not so nice features of
LCFT is that correlation functions do not any longer factorize
into holomorphic and anti-holomorphic parts, we still can consider
most entities in factorized form, as long as we do not impose the
physical constraint that observables should be single-valued.
This is particularly true for the representation theory of the
CFT under consideration. We call a CFT rational, if it has only
finitely many highest-weight representations. Then, as Cardy observed
a long time ago, the partition function of such a ration CFT
can be written as a sesqui-linear form over the characters of these
representations. Thus, denoting the finite set of representations by
${\cal R}$, the partition function takes the form
\be
  Z(\tau,\bar\tau)=\sum_{h,\bar h\in{\cal R}}N_{h\bar h}
  \chi^{}_h(\tau)\chi^*_{\bar h}(\tau)\,,
\ee
where $N_{h\bar h}$ is a certain matrix with non-negative integer entries.
Here, the character of the highest-weight representation
$M_{h,c}$ is defined as usual,
\be
  \chi_h(\tau)={\rm tr}_{M_{h,c}}q^{L_0-c/24}\,,
\ee
and analogously for $\chi^*_{\bar h}(\tau)$.

Since the partition function is modular invariant, the characters from which
it is built must transform covariantly under the modular group.
Therefore, in the present setting of a rational theory, i.e.\
$|{\cal R}|<\infty$, they form a finite-dimensional representation of the
modular group. As a consequence, the transformations $S:\tau\mapsto-1/\tau$
and $T:\tau\mapsto\tau+1$ are represented as matrices acting on the characters,
that is,
\bea
  \chi_h(-\frac{1}{\tau}) &=& \sum_{h'\in{\cal R}} S_{h}^{\ h'}
  \chi_{h'}(\tau)\,,\\
  \chi_h(\tau+1) &=& \sum_{h'\in{\cal R}} T_{h}^{\ h'}\chi_{h'}(\tau)\,.
\eea
One of the most astonishing deep results in CFT is that the $S$-matrix
fulfills a certain algebraic property, which on first glance seems to be
pure magic. Eric Verlinde \cite{Ver88} 
suggested namely, that the $S$-matrix also
yields the so-called fusion rules, which essentially count the
multiplicities of representations appearing on the right hand side
of the fusion product of two representations. The latter is, in analytical
terms, provided by the OPE, and might be thought of as some kind of
tensor product algebraically. To ease notation, let us arbitrarily
enumerate the weights $h\in{\cal R}$ as $h_i$, $i=0,\ldots,|{\cal R}|-1$ with
the convention that $h_0$ refers to the vacuum representation.
Then, the seminal so-called Verlinde formula reads
\be\label{eq:verlinde-formula}
  [h_i]*[h_j] = \sum_{k}N_{ij}^{\ \,k}[h_k]\ \ \ \
  {\rm with}\ \ \ \ N_{ij}^{\ \,k} = \sum_r\frac{S_i^{\ r}S_j^{\ r}
  (S^{-1})_r^{\ k}}{S_0^{\ r}}\,.
\ee
Although, the entries of the $S$-matrix may be very complicated
algebraic numbers (made out of $\exp(2\pi{\rm i}\rho)$ expressions with
$\rho$ rational numbers), the $N_{ij}^{\ k}$ are always non-negative
integers.

In the following, our task will be to generalize this setup to the
logarithmic case. We will take an approach which on one hand tries to stay as
close as possible to the generic case, but on the other hand does 
disentangle the indecomposable representations and
their irreducible highest-weight sub-representation as much as possible.
The lectures of Matthias Gaberdiel will follow a different approach,
which avoids many of the difficulties we will encounter, but where
indecomposable representations are only treated as a whole, losing
any backtrace of their inner structure. In particular, our approach
will always keep the irreducible highest-weight sub-representations
with their corresponding proper primary fields as valid representations
to be separately included into the set ${\cal R}$ of representations.
The price we have to pay for this on first sight quite natural
approach is that in order to account for the states from the full
indecomposable representations, we are forced to generalize the
definition of characters beyond the immediately physical meaningful.

Our explicit character formul\ae, $S$-matrices and fusion rules
will be worked out for the series of pseudo-minimal models
with central charge $c=c_{p,1}=13-6p-6\frac1p$, which all constitute
LCFTs.\footnote{As minimal models, these CFTs do not exist, because
their conformal grids (Kac tables) would be empty.} These models happen
to be the best known LCFTs, with the particular prominent prime 
example of the $c_{2,1}=-2$ theory which we already encountered
several times.

\subsection{The $c_{p,1}$ models}

\bq
In a work of H.G.~Kausch \cite{Walgebra} the possibility to extend
the Virasoro algebra by a multiplet of fields of equal conformal
dimension has been considered. Besides some sporadic solutions he found
a series of algebras extended by a singlet or triplet
of fields of odd dimension which resemble a $SO(3)$ structure. The
operator product expansion is given by
\begin{equation}\label{eq:walg}
  W^{(j)}(z)W^{(k)}(\zeta) =
    \frac{c}{\Delta}\delta^{jk}\frac{1}{(z-\zeta)^{2\Delta}}
  + C_{\Delta\Delta\Delta} i\varepsilon^{jkl}
    \frac{W^{(l)}(\zeta)}{(z-\zeta)^{\Delta}}
  + {\it descendant\ fields}\,,
\end{equation}
where $c = c_{p,1}$ and $\Delta=2p-1$.
Note, that
for the singlet algebra there is no term proportional to the field $W$.
These CFT posses infinitely many degenerate representations with integer
conformal weights
\begin{equation}\label{eq:hwertel}
  h_{2k+1,1} = k^2 p + kp - k\,.
\end{equation}
These representations correspond to a set of relatively local chiral
vertex operators. But there is a peculiarity: The energy operator $L_0$
is no longer diagonal on these degenerate representations, but is given
in a Jordan normal form with non-trivial blocks.

A standard free field construction \cite{BPZ,DoFa84}
shows that the degenerate fields have conformal weights $h_{m,n} =
\frac{\alpha_{m,n}^2}{4} + \frac{c_{p,1}-1}{24}$, where
$\alpha_{m,n} = m\sqrt{p} - n\sqrt{p}^{-1}$. The fundamental region
of the minimal models unfortunately is empty: $\{m,n|1\leq m<1,\,1\leq n<p\} =
\emptyset$. But without loss of generality we can reduce the
labels $(m,n)$ to the region $0<m, 0<n\leq p$,
since $\alpha_{m,n} = -\alpha_{-m,-n}$ and $\alpha_{m,n} = \alpha_{m+1,n+p}$.
Moreover, we
have the following abstract fusion rules which result from the
conditions for the existence of well defined chiral vertex operators
\cite{Kausch:1995}:

For $c = 13-6(p+p^{-1})$ with $p\in\BN$,
there exist well defined chiral vertex operators for triplets of
Virasoro highest weight representations to $(h_{m_1,n_1}$, $h_{m_2,n_2}$,
$h_{m_3,n_3})$ with
$0< m_i$ and $0<n_i\leq p$ iff $|m_1-m_2| <m_3<m_1+m_2$ and
$|n_1-n_2|<n_3\leq\min(p,n_1+n_2-1)$, and moreover
$m_1+m_2+m_3-1\equiv n_1+n_2+n_3-1\equiv 0$ {\rm mod} $2$.

The screening charges have a special meaning. With $\alpha_{\pm} =
\alpha_0 \pm \sqrt{1 + \alpha_0^2}$ and $\alpha_0^2 = (1-p)^2/4p$ the
first of them is given by
$$
  Q =\int_{\Omega_1}\frac{dz}{2\pi{\rm i}}V_{\alpha_+}(z)\,,
$$
where $\Omega_1$ encircles the origin counterclockwise in the standard way.
$Q$ has trivial monodromy on the Fock spaces $\EF_{m,n}$ of the free field
construction on the weights $h_{m,n}$, and therefore is by itself a well
defined local chiral vertex operator
$Q:\EF_{m,n}\rightarrow\EF_{m-2,n}$. This screening charge is
exactly responsible for the multiplet structure of the chiral fields.
We have $Q^m=0$ on $\EF_{m,n}$.
The other screening charge (to the ``power'' $k$) is
$$
  \tilde{Q}^k = \int_{\Omega_k}\frac{dz_1}{2\pi{\rm i}}\ldots\frac{dz_k}
  {2\pi{\rm  i}}V_{\alpha_-}(z_1)\ldots V_{\alpha_-}(z_k)\,,
$$
where the integration path is radially ordered, $|z_1|> \ldots >|z_k|$,
and encircles the origin. It is well defined on $\EF_{m,n}$ iff
$0<k=n<p$. $\tilde{Q}^p$ vanishes identically on $\EF_{m,p}$.
The BRST-identity is $\tilde{Q}^{p-n}\tilde{Q}^n = 0$, such that we have
the following embedding structure of Fock spaces (see
\cite{Fel89,FFK89}) induced by the exact sequence
$$
  \ldots\stackrel{\tilde{Q}^{p-n}}{\longrightarrow}\EF_{m-2,n}
        \stackrel{\tilde{Q}^{n}}{\longrightarrow}\EF_{m-1,p-n}
        \stackrel{\tilde{Q}^{p-n}}{\longrightarrow}\EF_{m,n}
        \stackrel{\tilde{Q}^{n}}{\longrightarrow}\EF_{m+1,p-n}
        \stackrel{\tilde{Q}^{p-n}}{\longrightarrow}\EF_{m+2,n}
        \stackrel{\tilde{Q}^{n}}{\longrightarrow}\ldots
$$
The Virasoro modules are then given by $\EH_{m,n} =
{\rm ker}_{\EF_{m,n}}\tilde{Q}^n$.
The fields $\phi_{2k+1,1}\equiv V_{\alpha_{2k+1,1}}$,
$k\in\BN$, all have integer dimensions $h_{2k+1,1} = k^2p+kp-k$, such
that one is tempted to extend the local chiral algebra by them.
Indeed, it follows from the abstract fusion rules that the local chiral algebra
generated by only the stress energy tensor and the field $\phi_{3,1}$
closes, since no other fields can contribute to the singular part of the
OPE. The multiplet structure is obtained by repeated application of $Q$,
$W^{(j)} = Q^j\phi_{3,1}$. Indeed, this yields three fields with
SO(3)-structure \cite{Walgebra}, and therefore a $\w(2,2p-1,2p-1,2p-1)$-algebra.
With $W = \sum_{j}W^{(j)}$ we get the symmetric singlet
algebra $\w(2,2p-1)$.

With the BRST-structure given above one can construct exactly $2p$
(regular) representations of the fully extended chiral algebra by taking
into account the multiplets generated by the $Q$-operator\footnote{The
operators $Q$ and
$\tilde{Q}^k$ generate four two-dimensional complexes of the $\EF_{m,n}$,
one for $m$ even and odd respectively, and one for $n=p$ and $n\neq p$
respectively \cite{Kausch:1995}.}.
Formally we can write these $\w$-modules as
\begin{eqnarray}              \label{eq:HWp}
  {\EH}^{\w}_{n,+} = \bigoplus_{j=0}^{\infty}\bigoplus_{m=0}^{2j-1}Q^m
  {\EH}_{2j+1,n}\,,\\      \label{eq:HWm}
  {\EH}^{\w}_{n,-} = \bigoplus_{j=1}^{\infty}\bigoplus_{m=0}^{2j-2}Q^m
  {\EH}_{2j,n}\,,
\end{eqnarray}
with $1\leq n\leq p$. The corresponding conformal weights are $h_{1,n}$ and
$h_{2,n}$ respectively. The $\w$-representations for
$h_{1,n}$ are singlets, the ones for $h_{2,n}$ doublets.
There also exist special representations for the weights $h_{0,n},
1\leq n<p$. Their highest weight vectors are singular vectors in
$\EF_{1,p-n}$, which have the {\em same} highest weights. The
corresponding chiral vertex operators are degenerated. For instance,
there are besides the identity $p-1$ additional vertex operators of
conformal weight zero, which map $\EF_{0,n}$ to
$\EF_{1,p-1}$. Consequently, also the descendant fields of the
identity family are degenerated, in particular the Virasoro field itself.
This forces the existence of non-trivial Jordan cells for $L_0$, i.e.\ $L_0$
no longer is diagonalizable. Moreover, the multiplicities of states in
the Virasoro modules must change. We have, in sloppy terms, a $p$-fold
degenerate identity, which will lead to a multiplicity of $p$ in the
characters of the highest weight representations $h_{0,n}$.
\eq

\subsection{Representations and characters}

Let us assume that the Hilbert space $\EH\otimes\bar{\EH}$ is a
direct sum of irreducible highest weight representations (HWR) with
respect to the chiral symmetry algebra $\w$,
\begin{equation}
  \EH\otimes\bar{\EH} = \bigoplus_{\lambda\in\Lambda}
  \EH^{(\lambda)}
  \otimes
  %\bigoplus_{\bar{\lambda}\in\bar{\Lambda}}
  \bar{\EH}^{(\bar{\lambda})}\,.
\end{equation}
Further we assume that $\w$ is maximal such that $\Lambda =
\bar{\Lambda}$ is the set of all $\w$ HWRs, i.e.\
the theory is {\em symmetric}. We decompose $\EH^{(\lambda)}$ into
Virasoro HWRs, the set of them we denote with $N_{\lambda}$,
\begin{equation}\label{eq:WDecomp}
  \EH\otimes\bar{\EH} = \bigoplus_{\lambda\in \Lambda}\left(
  \bigoplus_{\nu\in N_{\lambda}}\EH^{(\lambda)}_{\nu}
  \otimes\bigoplus_{\nu\in N_{\lambda}}
  \bar{\EH}^{(\lambda)}_{\nu}\right)\,.
\end{equation}

\bq
A CFT is said to be {\em rational}, iff $|\Lambda|<\infty$. It is called
{\em quasi-rational}, if $\Lambda$ is countable and only finitely many
terms appear in each fusion product.
The Cartan sub-algebra $\cal C$ is spanned by $L_0$, the central
extension $C$ and the zero modes of the simple primary fields
$\phi_i\in{\cal B}_{{\cal W}}$ which generate the $\w$-algebra.
We denote the highest weight state (HWS) of $\EH^{(\lambda)}_{\nu}$ by
$\vak{h^{(\lambda)}} = \vac{c,h_{\nu},w^{(\lambda)}_1,w^{(\lambda)}_2,\ldots}$
where ${\bf h}\in{\cal C}^*$ is the highest weight vector (HWV).
A regular HWR $M_{\vak{h}}$ of a $\w$-algebra to a HWS
$\vak{h} = \vac{c,h,w_1,w_2,\ldots}$ is then defined to satisfy the
following conditions:
  \begin{eqnarray*}
    C\vak{h} & = & c\vak{h}\,,\nonumber\\
    L_0\vak{h} & = & h\vak{h}\,,\nonumber\\
    \phi_{i,0}\vak{h} & = & w_i\vak{h}\ \ \forall\phi_i\in{\cal B}_{\w}\,,
    \nonumber\\
    L_n\vak{h} & = & 0\ \ \forall n>0\,,\nonumber\\
    \phi_{i,n}\vak{h} & = & 0\ \ \forall\phi_i\in{\cal B}_{\w}\
    {\rm and}\ \forall n>0\,,\nonumber\\
    M_{\vak{h}} & = & U(\w)\vak{h}\,,
  \end{eqnarray*}
where $U(\w)$ denotes the universal enveloping algebra of $\w$. Moreover,
we call a HWR $V_{\vak{h}}$ {\em Verma module}, iff the sequence
  \begin{equation}
    V_{\vak{h}} \longrightarrow M_{\vak{h}} \longrightarrow 0
  \end{equation}
is exact for all HWRs $M_{\vak{h}}$.
The Verma module $V_{\vak{h}}$ has a natural gradation
\begin{equation}
  V_{\vak{h}} = \bigoplus_{n\in\BZ_+}V_{\vak{h}}^n\,,
\end{equation}
where $V_{\vak{h}}^n$ is the $L_0$ eigenspace with eigenvalue $h + n$.
\eq

Let us now assume that there exist HWRs, whose $L_0$ eigenvalues differ
by integers. We must distinguish two cases. If the difference $\Delta h$
of the $L_0$ eigenvalues of two HWRs is always non zero, or the highest
weights differ in at least one component, it still is
possible to diagonalize $L_0$, even if $\Delta h\in\BZ$. Moreover, there
are no logarithmic operators necessary. The reason is that the differential
equations for the conformal Ward identities do not degenerate in this
case. This is different to the case of the modular differential equation
to be satisfied by the characters, which is only sensible modulo integers.
Examples of such rational CFTs with HWRs with $\Delta h\in\BZ$ can
be found in \cite{Flo93}.

Therefore, we now assume the existence of $n+1>1$ HWRs such that
${\bf h}_i-{\bf h}_j=0$ for $1\leq i,j\leq n+1$, i.e.\ we consider a LCFT.
We already learned that we have to modify the definition
of HWRs in the following way: The HWS is replaced by a non-trivial
Jordan cell of $L_0$ of dimension $n+1$, which is spanned by
$\{\vac{{\bf h};0}=\vak{h},\vac{{\bf h};1},\ldots,\vac{{\bf h};n}\}$.
We then will call $M(\vak{h;m})_{0\leq m\leq n}$ a {\em logarithmic\/} HWR
of a $\w$-algebra to the highest weight $L_0$-Jordan cell of rank $n+1$,
$(\vak{h;m} = \vac{c,h,w_1,w_2,\ldots;m})_{0\leq m\leq n}$, if
it satisfies the following conditions:
  \begin{equation}
    \begin{array}{rcl}
      L_0\vac{{\bf h};m} & = & h\vac{{\bf h};m} + \vac{{\bf h};m-1}\,,
      \ \ m>0\,,\\
      L_0\vac{{\bf h};0} & = & h\vac{{\bf h};0}\,,\\
      \phi_{i,0}\vac{{\bf h};m} & = & w_i\vac{{\bf h};m} + \ldots\,,
      \ \ m>0\,,\ \ \forall\phi_i\in{\cal B}_{\w}\,,
    \end{array}
  \end{equation}
and otherwise the conditions of the original definition. The dots in the
last condition represent possible non-diagonal contributions.
In addition, there is in general no orthogonal system of states within the
Jordan cell, i.e.\ $\vev{{\bf h};k}{{\bf h};l}\neq 0$ even for $k\neq l$.
Since the other properties of HWRs remain unchanged, it makes sense to
consider such logarithmic HWRs if the whole Jordan cell structure is taken
into account for the definition of $\w$-families.

Next, we want to discuss the consequences for the characters.
For simplicity, we consider a Jordan cell of form
${h\ 1\choose 0\ h}$, i.e.\ we have two HWSs,
$\vac{h;0}$ and $\vac{h;1}$, on which the action of $L_0$ is given by
$L_0\vac{h;0} = h\vac{h;0}$ and $L_0\vac{h;1} = h\vac{h;1} + \vac{h;0}$.
The off-diagonal element could be any non-zero number, since a Jordan
cell decomposition is just one particular choice. The physical correct
decomposition will be fixed later by modular invariance.

The HWS $\vac{h;0}$ is an ordinary $L_0$-eigenstate, such that the
character of the corresponding HWR should be defined in the usual manner.
The other state, $\vac{h;1}$ is not a $L_0$-eigenstate, application of
$L_0$ generates a new state, which also is not contained in the standard
Verma module. If we apply $L_0$ once again, this state is recovered plus
an additional one, etc. Thus, the operator $L_0$, acting on the Jordan
cell, may be written as
$L_0 = {L_{0;0}\ \ 1\choose 0\ \ L_{0;1}}$, where the second label $j$
refers to the Verma like modules on which the $L_{0;j}$ operators act.

The character of a HWR on a HWS $\vak{h}$ is usually defined as
\be\label{eq:trace}
  \chi_{\vak{h}}(q) = {\rm tr}_{M_{\vak{h}}}q^{L_0-c/24}
  % = \sum_{\psi\in {\cal B}(M_{\vak{h}})}\vev{\psi}{q^{L_0-c/24}\psi}
  \,,
\ee
where $q=\exp(2\pi{\rm i}\tau)$ and the trace
is taken over the module which is created by action of $U({\cal W})$ on
$\vak{h}$. Using our $L_0$ matrix, and treating infinite series in $q$
in a formal way without consideration of their convergence
properties, we obtain
\begin{eqnarray}
  q^{L_0} & = & \sum_{n=0}^{\infty} \frac{(2\pi{\rm i}\tau)^n}{n!}\left(
  \begin{array}{cc} L_{0;0} & 1\\
                          0 & L_{0;1}
  \end{array}\right)^n\nonumber\\
          & = & \sum_{n=0}^{\infty} \frac{(2\pi{\rm i}\tau)^n}{n!}\left(
  \begin{array}{cc} L_{0;0}^n & nL_{0;0}^{n-1}\\
                            0 & L_{0;1}^n
  \end{array}\right)^n\nonumber\\
          & = & \left(
  \begin{array}{cc} q^{L_{0;0}} & 2\pi{\rm i}\tau q^{L_{0;0}}\\
                              0 & q^{L_{0;1}}
  \end{array}\right)\,.
\end{eqnarray}
Since formally $2\pi{\rm i}\tau = \log(q)$, we see that a non-trivial Jordan
cell may generate logarithmic terms in the character expansions. This is
completely analogous to the logarithms in the correlation functions of
certain operators, which stem from the degeneracy of the conformal Ward
identity differential equations: We obtain essentially the same
degeneracies in the modular differential equations for the characters,
which force additional solutions with logarithms. We will continue
to call modular function containing $\log(q)$ terms characters, although,
strictly speaking, such functions do not meet all requirements one usually
imposes on characters. In particular, the formal series expansion of
a character allows to extract the multiplicities of states at a certain
level $n$ in a module from the coefficients of the corresponding $n$-th
term in the expansion. This does not make sense for functions which
are of the form $2\pi{\rm i}\tau\,q^\gamma\sum_n a_na^n$.

\bq
The careful reader may wonder, how the logarithmic terms 
$\log(q)\equiv 2\pi{\rm i}\tau$ can show up in the
characters. Usually, traces (\ref{eq:trace}) over modules are well defined,
since the complete Hilbert space is a direct sum of modules and
$L_0$ can be uniquely restricted to one of the modules.
Now, if $L_0$ has non trivial Jordan form, modules
$M_{\vac{{\bf h};k}}$ and $M_{\vac{{\bf h};l}}$ are not orthogonal.
Therefore, the characters depend on the choice of a basis of generating
states, while the sum $\sum_{k=0}^n \chi_{\vac{{\bf h};k}}(q)$ is invariant
under any base change $\vac{{\bf \tilde h};k} = B^k_{\phantom{k}l}
\vac{{\bf h};l}$.
Only this sum is a trace of a well defined restriction of $q^{L_0-c/24}$
and does never contain any logarithmic parts. But the characters can:
For example change of the basis $\{\vac{h;0},\vac{h;1}\}$ to
$\{\vac{\tilde h;0}=\vac{h;0}+\vac{h;1},
\vac{\tilde h;0}=-\vac{h;0}+\vac{h;1}\}$ yields
$$
  q^{L_0} = \frac{1}{2}\left(
  \begin{array}{cc}
    (1-2\pi{\rm i}\tau)q^{L_{0;0}}+q^{L_{0;1}} &
    (1+2\pi{\rm i}\tau)q^{L_{0;0}}-q^{L_{0;1}}\\
    (1-2\pi{\rm i}\tau)q^{L_{0;0}}-q^{L_{0;1}} &
    (1+2\pi{\rm i}\tau)q^{L_{0;0}}+q^{L_{0;1}}
  \end{array}\right)\,.
$$
The generalization to larger Jordan cells is straightforward.

However, in a mathematical rigorous framework, this is unsatisfactory.
A character should have the interpretation that it counts states at a given
level. This interpretation clearly is lost when $\log(q)$ terms are present.
The lectures of Matthias Gaberdiel (see also \cite{Gaberdiel:1996a}) will
follow a different approach avoiding many of the mathematically disturbing
issues raised in our treatment. We note that, for historical reasons, we
call the modular functions calculated below characters, although they do
not all allow this interpretation. On the other hand, one might consider
torus zero- and one-point functions, in particular so-called torus
partition functions. In ordinary rational CFT, these usually coincide with the
characters of the theory. This is no longer true for LCFTs, and we believe
that the modular functions with $\log(q)$ terms should correctly be
considered as torus partition functions rather than characters. What is
striking, however, is the fact that the torus partition functions do
indeed coincide with characters calculated from first principles along
the lines of \cite{Gaberdiel:1996a}, when a certain limit is taken, as
will be described in more detail in the following. Even more puzzling
is the fact that we can compute well defined characters of the irreducible
sub-representations of the indecomposable representations which, however,
lead via their modular transforms to modular functions without a well
defined interpretation as characters.
\eq

Since the characters of a CFT can be viewed as the zero-point-functions
on a torus with modular parameter $\tau$, they in general turn out to be
certain modular functions whose Fourier expansions around $\tau=+{\rm i}\infty$
are just the $q$-series. 
One of the most powerful tools in CFT is the
modular invariance of the partition function
\begin{equation}\label{eq:PartFct}
  Z(\tau,\bar{\tau}) = (q\bar q)^{-\frac{c}{24}}
  {\rm tr}(q^{L_0}\bar q^{\bar L_0})\,.
\end{equation}
Since the partition function of a rational CFT is a
quadratic form in the characters, modular invariance puts severe
restrictions on the modular behavior of the (generalized) characters.
It will turn out that modular invariance uniquely determines a basis
of HWSs within each Jordan block and therefore all characters, i.e.\
that LCFTs are similarly constraint by modular invariance as generic
CFTs.

We now fix some notations for the following.
We will very often use the so called {\em elliptic functions\/} or
{\em Jacobi-Riemann $\Theta$-functions\/} which are modular forms of
weight $1/2$, defined as
\begin{equation}\label{eq:thetadef}
  \Theta_{\lambda,k}(\tau) = \sum_{n\in\BZ}q^{(2kn + \lambda)^2/4k}\,,\ \
  \lambda \in \BZ/2\,,\ k \in \BN/2\,.
\end{equation}
We call $\lambda$ the {\em index\/} and $k$ the {\em modulus\/} of the
$\Theta$-function. The $\Theta$-functions obey
$\Theta_{\lambda,k} = \Theta_{-\lambda,k}
= \Theta_{\lambda+2k,k}$, and $\Theta_{k,k}$ has, as power series in
$q$, only even coefficients. We also need the {\em Dedekind
$\eta$-function\/} which is defined as $\eta(\tau) =
q^{1/24}\prod_{n\in\BN}(1-q^n)$. The modular properties of these
functions are for $\lambda,k\in\BZ$
\begin{eqnarray}\label{eq:theta}
  \Theta_{\lambda,k}({\ts-\frac{1}{\tau}}) &=& {\ds
  {\ts\sqrt{\frac{-{\rm i}\tau}{2k}}}\,\sum_{\lambda'=0}^{2k-1}
  e^{{\rm i}\pi\frac{\lambda\lambda'}{k}}\Theta_{\lambda',k}(\tau)}\,,\\
  \Theta_{\lambda,k}({\ts\tau + 1}) &=& {\ds
  e^{{\rm i}\pi\frac{\lambda^2}{2k}}\Theta_{\lambda,k}(\tau)}\,,\\
  \eta({\ts-\frac{1}{\tau}}) &=& {\ds\sqrt{-{\rm i}\tau}\,\eta(\tau)}\,,\\
  \eta({\ts\tau + 1})        &=& {\ds e^{\pi{\rm i}/12}\,\eta(\tau)}\,.
\end{eqnarray}
To prove these formul\ae, one has
to make use of the Poisson resummation formula.
The functions $\Lambda_{\lambda,k}(\tau) = \Theta_{\lambda,k}(\tau)/\eta(\tau)$
are then modular forms of weight zero to a particular main-congruence
subgroup $\Gamma(N)\subset{\rm PSL}(2,\BZ)$, e.g.\ $N$ is the least
common multiple of $4k$ and $24$ for $k\in\BZ$.

As we have seen above, the (generalized) characters for 
logarithmic CFTs are functions
in the ring $\BZ[[q]][\log q]$. Therefore we introduce the following
additional functions:
\be
  (\partial\Theta)_{\lambda,k}(\tau) \propto
  \frac{\partial}{\partial\lambda}\Theta_{\lambda,k}(\tau) =
  \frac{\pi{\rm i}\tau}{k}\sum_{n\in\BZ}(2kn+\lambda)q^{(2kn + \lambda)^2/4k}\,,
\ee
where we made explicit that new linear independent solutions of
degenerate differential equations can be obtained by a formal derivation
of the degenerate solution with respect to its parameter. As long as
modular covariance is not concerned, there is no reason why $\tau$ could
not appear as a factor. We introduce the so-called
{\em affine\/} $\Theta$-functions
\begin{equation}
  (\partial\Theta)_{\lambda,k}(\tau) =
  \sum_{n\in\BZ}(2kn+\lambda)q^{(2kn + \lambda)^2/4k}\,,
\end{equation}
which play an important r{\^o}le in the character formul\ae\ for the affine
$\widehat{\euf{su}(2)}$-algebra. They are odd,
i.e.\ $(\partial\Theta)_{-\lambda,k} =
-(\partial\Theta)_{\lambda,k}$. Moreover, per definitionem
$(\partial\Theta)_{0,k} = (\partial\Theta)_{k,k} \equiv 0$.
Their modular behavior is
\begin{equation}\label{eq:dtheta}
  \begin{array}{rcl}
    (\partial\Theta)_{\lambda,k}({\ts-\frac{1}{\tau}}) &=& {\ds
    {\ts(-{\rm i}\tau)\sqrt{\frac{-{\rm i}\tau}{2k}}}\,\sum_{\lambda'=1}^{2k-1}
    e^{{\rm i}\pi\frac{\lambda\lambda'}{k}}
    (\partial\Theta)_{\lambda',k}(\tau)}\,,\\
    (\partial\Theta)_{\lambda,k}({\ts\tau + 1}) &=& {\ds
    e^{{\rm i}\pi\frac{\lambda^2}{2k}}
    (\partial\Theta)_{\lambda,k}(\tau)}\,.\\
  \end{array}
\end{equation}
Since they are no longer modular forms of weight $1/2$ under
$S:\tau\mapsto -1/\tau$, we have to add further functions
\begin{equation}
  (\nabla\Theta)_{\lambda,k}(\tau) =
  \frac{\log q}{2\pi{\rm i}}\sum_{n\in\BZ}(2kn+\lambda)q^{(2kn + \lambda)^2/4k}
\end{equation}
in order to obtain a closed finite dimensional representation of the
modular group. It is clear that $S$ interchanges these two sets of
functions, while $T:\tau\mapsto\tau+1$ transforms $(\nabla\Theta)_{\lambda,k}$
into $(\nabla\Theta)_{\lambda,k}+
(\partial\Theta)_{\lambda,k}$. Therefore, the linear combination
$$
  (\partial\Theta)_{\lambda,k}(\tau)(\nabla\Theta)_{\lambda,k}^*({\tau}) -
  (\nabla\Theta)_{\lambda,k}(\tau)(\partial\Theta)_{\lambda,k}^*({\tau}) =
  (\tau - \bar{\tau})|(\partial\Theta)_{\lambda,k}|^2
$$
is modular covariant of weight 1/2!

Of course, the modular differential equation (see below) could be degenerate of
higher degree, and one had to introduce generalizations
$(\partial^n\Theta)_{\lambda,k}$ and
$(\nabla^n\Theta)_{\lambda,k}$ (the expression $(\tau - \bar{\tau})^n$
is modular covariant of weight $-2n$ for all $n\in\BZ_+$).
One can show \cite{Eho00} that regular rational theories with
$c_{{\rm eff}} \leq 1$ can only have one power $\eta(\tau)\eta(\bar{\tau})$
in the denominator of the partition function. Regular means that the characters
are modular forms.

\bq
Now, the modular behavior of characters of logarithmic CFTs
is almost the one of modular forms, except the possibility to
expand into a power series in $q$. In particular, the asymptotic properties
needed in the proof \cite{Eho00} are only affected in an analytic way by
logarithmic corrections: In fact, although the modular differential equation
makes only sense for particular isolated points in parameter space,
$(c,{\bf h}_1,{\bf h}_2,\ldots)\in\bigoplus{\cal C}^*$,
where the corresponding CFT is rational, it can be regarded as a
differential equation depending on continuously variable parameters --
once it has been written down. The characters of our theories in question
are solutions of certain degenerate modular differential equations,
obtained in a unique way by analytic continuation.
Therefore, we conjecture that the result of \cite{Eho00} should also hold
for logarithmic rational CFTs. Thus, we should only be concerned with $n=1$ in our
case.
\eq

We conclude this introduction of the general setup with a remark on
the classification of rational CFTs. Whence the finite set ${\cal R}$ of
representations and their characters is known, a particular finite
dimensional representation of the modular group is fixed in terms of
multiplicative systems of modular functions (such that the matrices
$S$ and $T$ have constant coefficients). The definitions of multiplicative
systems of modular functions given above were all cooked up from ratios
of modular forms of weight $1/2$. In particular, the denominator was
always chosen to be the Dedekind $\eta$-function. It is possible to relate
the maximal power of $\eta$-functions in the denominator of a character
to the effective number of degrees of freedom, $c_{{\rm eff}}$ of the CFT.
Now, if $c_{{\rm eff}}\leq 1$, this power is at most one such that the
numerator can only be given by a modular form of weight $1/2$. In this
case, the Serre-Stark theorem provides us with a complete set of all
possible such forms. It turns out that they are all of the type 
(\ref{eq:thetadef}) together with one other type, namely
$$
  \tilde\Theta_{\lambda,k}(\tau) = \sum_{n\in\mathbb{Z}} (-)^n
  q^{(2kn+\lambda)^2/4k}\,,\ \
  \lambda \in \BZ/2\,,\ k \in \BN/2\,.
$$
There are no other linearly independent modular forms of weight $1/2$.
This theorem forms the basis of the complete classification of {\em all\/}
rational conformal field theories with $c_{{\rm eff}}\leq 1$. These are
the $c=1$ Gaussian models at compactification radii $2R^2=p/p'\in\mathbb{Q}$,
the minimal models, and the $c=1-6k$, $k\in\mathbb{N}$ series \cite{Flo93}. 
The $c=1$ theories were first classified by P.~Ginsparg. The completeness
of this classification was proven by E.~Kiritsis using the Serre-Stark
theorem. In essence, one formulates some conditions for a potential
partition function of a rational CFT to be physical senseful, e.g.\ that
its Fourier expansion around $q=0$, i.e.\ $\tau\rightarrow+{\rm i}\infty$,
has non-negative integer coefficients, that the ground state has
multiplicity one, etc. These conditions
are then checked for arbitrary finite linear combinations $Z=\sum_iZ[x_i]$,
$x_i=p_i/p'_i\in\mathbb{Q}$, of the basic modular invariant entity
$$
  Z[p/p'](q,\bar q) = \frac{1}{\eta(q)\eta(\bar q)}\sum_{\lambda=0}^{2pp'-1}
  \Theta_{\lambda,pp'}(q)\Theta_{\lambda',pp'}(\bar q)\,.
$$
Here, $\lambda'$ is given in terms of $\lambda$ in the following way: 
For $p,p'$ coprime, there exists always a
representation $\lambda=rp-sp'$ mod $2pp'$ with
integers $r,s$. Then the value $\lambda'$ is given by $\lambda'=rp+sp'$ mod
$2pp'$. The physical conditions restrict the possible linear combinations
to a surprisingly small set of a few series. Besides the known solutions
yielding $c=1$ rational Gaussian models or the minimal models, Kiritsis
found one further possibility for a series of physical partition 
functions, which could be identified with a series of non-unitary rational
CFTs by the present author. Modular invariant partition functions for $c=1$
models and for the minimal models have a beautiful classification pattern
resembling the $A$-$D$-$E$-classification of finite subgroups of $SU(2)$.
For the minimal models, this was shown by Cappelli, Itzykson, and Zuber 
(their work, as well as the classification of $c=1$ models by
Ginsparg, can be found in \cite{ISZ}).

On the other
hand, we are going to show that the logarithmic CFTs of the $c_{p,1}$ series
are at least very close to rationality, and they also have $c_{{\rm eff}}=1$.
In light of the Serre-Stark theorem, it is a surprising and unexpected
result that this other class of CFTs exists. Their existence does not
contradict Serre-Stark. In our approach below, we allow characters which
do not have a homogeneous modular weight such that their transforms 
include unphysical $\log(q)$ terms (which we have to get rid of at the
end by a limiting procedure). The approach in the lectures of Gaberdiel
does not violate Serre-Stark either, since he obtains results which
coincide with characters and partition functions of certain $c=1$ models. 
It should be emphasized in this context that a set of characters and
their partition function does by no means fix an underlying CFT 
uniquely. LCFTs are a particular strong example for this, since their
inner structure is very different from the $c=1$ models with equivalent
partition functions.

\subsection{Characters of the singlet algebras $\w(2,2p-1)$}  

\bq
We are now going to derive the characters of the $c_{p,1}$ models viewed
as $\w(2,2p-1)$ algebras. In particular, we will show that
the singlet models $\w(2,2p-1)$ are not rational since the chiral
symmetry algebra is too small for that.

The additional primary field of the $\w(2,2p-1)$-algebra is just the
symmetric singlet of the $\euf{su}(2)$ triplet of primary fields which
generate the $\w(2,2p-1,2p-1,2p-1)$. One way to obtain the characters is
to explicitly calculate the vacuum character and then get the others by
modular transformations. From the embedding structure of Virasoro Verma
modules for the values $c=c_{p,1}$ of the central charge
\cite{FeFu82,FeFu83,Fel89,FFK89} we learn that the Virasoro character
for the HWR on $\vac{h_{2n+1,1}}$, $n\in\BZ_+$, is given by
\begin{equation}
  \chi^{{\rm Vir}}_{2n+1,1}(\tau) = \frac{q^{(1-c)/24}}{\eta(\tau)}\left(
  q^{h_{2n+1,1}} - q^{h_{-2n-1,1}}\right)\,.
\end{equation}
Therefore \cite{Flo93}, the character of the ${\cal W}$-algebra vacuum
representation is
\be
  \chi^{{\cal W}}_{0}(\tau) = \sum_{n\in\BZ_+}\chi^{{\rm Vir}}_{2n+1,1}(\tau)
                            = \frac{q^{(1-c)/24}}{\eta(\tau)}
             \sum_{n\in\BZ}{\rm sgn}(n)q^{\frac{(2pn+p-1)^2}{4p}}\,,
\ee
where we defined ${\rm sgn}(0) = 0$.
It is convenient to rewrite the signum function as
${\rm sgn}(n+\frac{p-1}{2p})$. This character seems (up to the signum
function) to be quite similar to the classical $\euf{su}(2)$-$\Theta$-function
$\Theta_{p-1,p}(\tau,0,0)$ divided by $\eta$.
Note, that the classical $\euf{su}(2)$-$\Theta$-functions
$\Theta_{\lambda,k}(\tau,z,u)$, coincide for
$z=u=0$ with the elliptic functions defined in (\ref{eq:thetadef}).
They are the building stones for the characters of the
$\widehat{\euf{su}(2)}$ Kac-Moody-algebra. We therefore define
\begin{equation}
  \Xi_{n,m}(\tau) = \sum_{k\in\BZ+\frac{n}{2m}}{\rm sgn}(k)q^{mk^2}\,.
\end{equation}
But the modular transformation behavior is quite different from
(\ref{eq:theta}), while the presence of the signum function does not
change the behavior under $T$,
$\Xi_{n,m}(\tau+1) = \exp({\rm i}\pi\frac{n^2}{2m})\Xi_{n,m}(\tau)$.
In order to get the behavior under $S$, we rewrite the functions
$\Xi_{n,m}$ as linear combinations of
$\Theta_{\lambda,k}$ functions. For this we introduce
\begin{equation}
  \sigma(x,y) = \lim_{\varepsilon\rightarrow 0}\frac{1}{\sqrt{2\pi}}
  \int_{-\infty}^{\infty}\frac{e^{-2\pi{\rm i}yp^2}}{p+{\rm i}
  \varepsilon^2}\left(
  e^{{\rm i}px} - e^{-{\rm i}px}\right)dp\,,
\end{equation}
such that $\sigma(x,0) = {\rm sgn}(x)$. In the following we omit the
obvious limiting procedure. We find
\begin{eqnarray}
  \Xi_{n,m}(\tau) &=& \sum_{k\in\BZ+\frac{n}{2m}}\sigma(k,0)q^{mk^2}\nonumber\\
                  &=& \sum_{k\in\BZ+\frac{n}{2m}}\frac{1}{\sqrt{2\pi}}
                      \int_{-\infty}^{\infty}\frac{dp}{p}\left(
                      e^{2\pi{\rm i}kp} - e^{-2\pi{\rm i}kp}\right)
                      q^{mk^2}\\
                  &=& \frac{1}{\sqrt{2\pi}}\int_{-\infty}^{\infty}
                      \frac{dp}{p}\left(
                      \Theta_{n,m}(\tau,p,0)-\Theta_{n,m}(\tau,-p,0)\right)
                      \nonumber\,.
\end{eqnarray}
Therefore, by linearity of the $S$-transformation, we can write
\begin{equation}
  \Xi_{n,m}(-\frac{1}{\tau}) = \widetilde{\Xi}_{n,m}(\tau) =
  \sqrt{\frac{-{\rm i}\tau}{2m}}
  \sum_{n'\,{\rm mod}\,2m}\sin(-2\pi\frac{nn'}{2m})\Xi_{n',m}(\tau)\,,
\end{equation}
where $\widetilde{\Xi}_{n,m}$ is given by
\be
  \widetilde{\Xi}_{n,m} = \sum_{k\in\BZ+\frac{n}{2m}}
                          \sigma(k,-\frac{1}{2m\tau})q^{mk^2}
                        = \sum_{k\in\BZ+\frac{n}{2m}}
                          {\rm erf}\left(\sqrt{\frac{-m\tau}{4\pi i}}k
                          \right)q^{mk^2}\,.
\ee
Here, ${\rm erf}(x)$ denotes the usual Gauss error function up to
normalization. To derive the last equality, one has to use the scaling
invariance of the integral measure $\frac{dp}{p}$. Although the set of
functions $\Xi_{n,m}$ and $\widetilde{\Xi}_{n,m}$ closes under the
$S$-transformation, they do not form a representation of the full
modular group, since the $\widetilde{\Xi}_{n,m}$ do not close under $T$.
This means that they do not have a good power series expansion in $q$
with integer coefficients and powers which differ by integers only. From
this follows that the modular group forms an infinite dimensional
representation by repeated action of $T$ on
$\widetilde{\Xi}_{n,m}$. Therefore we conclude that the
${\cal W}(2,2p-1)$-algebras do not yield rational CFTs.
  
Similar to the case of the elliptic functions $\Theta_{\lambda,k}$, one may
introduce additional variables which correspond to additional quantum
numbers. For example we could write
\begin{equation}
  \Xi_{n,m}(\tau,z) = \sum_{k\in\BZ+\frac{n}{2m}}\sigma(k,z)q^{mk^2}\,.
\end{equation}
The variable $z$ could belong to the eigenvalue of the additional element
$W_0$ of the Cartan sub-algebra, actually to its square, since only the
latter can be determined. From the transformation behavior of the
$\euf{su}(2)$-$\Theta$-functions \cite{KaPe84} we get
\begin{eqnarray}
  \Xi_{n,m}(\tau+1,z) &=& e^{\frac{\pi in^2}{2m}}\Xi_{n,m}(\tau,z)\,,\\
  \Xi_{n,m}(-\frac{1}{\tau},z\tau^2-\frac{\tau}{2m}) &=&
    \sqrt{\frac{-{\rm i}\tau}{2m}}
    \sum_{n'\,{\rm mod}\,2m}\sin(-2\pi\frac{nn'}{2m})\Xi_{n',m}(\tau,z)\,.
\end{eqnarray}
Indeed, this set of functions forms a finite dimensional representation
of the modular group. But the presence of an additional quantum number
indicates that the chiral symmetry algebra is not yet maximally extended.
Some further remarks on this may be found in \cite{FHW93,FrWe93}.
\eq

\subsection{Characters of the triplet algebras $\w(2,2p-1,2p-1,2p-1)$}

We now view the $c_{p,1}$ models with respect to their maximally extended
chiral symmetry algebra, which we briefly mentioned in some small-print
further up. The typical recipe is ti construct the $\w$-characters
by summing up the Virasoro characters of degenerate representations
whose highest weights differ by integers. In addition, we have to take
care of multiplicities coming from the $\euf{su}(2)$ symmetry.
Using the isomorphism between fields and Fourier modes which span the
Hilbert space of the vacuum representation, one easily can show that the
multiplicity of the Virasoro HWR on $\vac{h_{2k+1,1}}$ is $2k+1$.
In particular, the multiplicity for $h_{3,1} = 2p-1$, the dimension of
the additional primary fields, is 3 as it should be. 
The Virasoro characters are due to Feigin and Fuks \cite{FeFu83}
\begin{equation}
  \chi^{{\rm Vir}}_{2k+1,1} = \frac{1}{\eta(q)}\left(q^{h_{2k+1,1}} -
  q^{h_{2k+1,-1}}\right)\,,
\end{equation}
since there is precisely one singular vector in these representations.
The vacuum representation of the $\w$-algebra is then the Hilbert space
\begin{equation}
  {\EH}^{\w}_{\vac{0}} = \bigoplus_{k\in\BZ_+} (2k+1){\EH}^{{\rm
  Vir}}_{\vac{h_{2k+1,1}}}\,.
\end{equation}
Therefore, the vacuum character is
\begin{eqnarray}\label{eq:chi-vac}
  \chi^{\w}_{0} &=& \sum_{k\in\BZ_+}(2k+1)\chi^{{\rm Vir}}_{2k+1,1}\nonumber\\
                &=& \frac{q^{(1-c)/24}}{\eta(q)}\left(
                    \sum_{k\geq 0}(2k+1)q^{h_{2k+1,1}} -
                    \sum_{k\geq 0}(2k+1)q^{h_{-(2k+1),1}}\right)\nonumber\\
                &=& \frac{q^{(1-c)/24}}{\eta(q)}\left(
                    \sum_{k\geq 0}(2k+1)q^{h_{2k-1,1}} +
                    \sum_{k\geq 1}(-2k+1)q^{h_{-2k+1,1}}\right)\\
                &=& \frac{q^{(1-p)^2/4p}}{\eta(q)}
                    \sum_{k\in\BZ} (2k+1)q^{[(1 - (2k+1)p)^2 -
                    (1-p)^2]/4p}\nonumber\\
                &=& \frac{1}{\eta(q)}
                    \sum_{k\in\BZ}(2k+1)q^{(2pk+(p-1))^2/4p}\nonumber\,.
\end{eqnarray}
This can be expressed in terms of $\Theta$-functions and affine
$\Theta$-functions as
\begin{equation}
  \chi^{\w}_{0} = \frac{1}{p\eta(\tau)}\left((\partial\Theta)_{p-1,p}(\tau)
                  + \Theta_{p-1,p}(\tau)\right)\,.
\end{equation}
But now we are in trouble here, since only the functions
$\Lambda_{\lambda,k} = \Theta_{\lambda,k}/\eta$ are modular forms of
weight zero, while the terms
$(\partial\Lambda)_{\lambda,k} = (\partial\Theta)_{\lambda,k}/\eta$ have
the modular weight 1.
  
Let us consider the modular transformation behavior of
$(\partial\Lambda)_{\lambda,k}$ under $S$ and $T$. From
(\ref{eq:theta}) we get the relations
\begin{eqnarray}
  (\partial\Lambda)_{\lambda,k}(\tau + 1) & = &
  \exp\left(2\pi{\rm i}\left(\frac{\lambda^2}{4k} - \frac{1}{24}\right)\right)
  (\partial\Lambda)_{\lambda,k}\,,\\
  (\partial\Lambda)_{\lambda,k}(-\frac{1}{\tau}) & = &
  (-{\rm i}\tau)\sqrt{\frac{2}{k}}\sum_{1\leq\lambda'\leq k-1}
  \sin\left(\frac{\pi\lambda\lambda'}{k}\right)
  (\partial\Lambda)_{\lambda',k}\,.
\end{eqnarray}
Note the occurrence of a term $\tau$, which cannot be written as a power
series in $q$. We define
$(\nabla\Lambda)_{\lambda,k} \equiv -\tau(\partial\Lambda)_{\lambda,k}$,
which have the modular properties
  \begin{eqnarray}
  (\nabla\Lambda)_{\lambda,k}(\tau + 1) & = &
  \exp\left(2\pi{\rm i}\left(\frac{\lambda^2}{4k} - \frac{1}{24}\right)\right)
  \left((\nabla\Lambda)_{\lambda,k}-(\partial\Lambda)_{\lambda,k}\right)\,,\\
  \label{eq:nlambda}
  (\nabla\Lambda)_{\lambda,k}(-\frac{1}{\tau}) & = &
  -{\rm i}\sqrt{\frac{2}{k}}\sum_{1\leq\lambda'\leq k-1}
  \sin\left(\frac{\pi\lambda\lambda'}{k}\right)
  (\partial\Lambda)_{\lambda',k}\,.
\end{eqnarray}
It is remarkable, that the $T$-transformation is no longer diagonal.
In some cases the $h$-values of the allowed HWRs are
explicitly known. These are
$\w(2,3,3,3)$ at $c = -2$, with the only possible highest weights
$h\in\{-1/8,0,3/8,1\}$, and $\w(2,5,5,5)$ at $c = -7$, which has HWRs
for $h\in\{-1/3,-1/4,0,5/12,1,7/4\}$ only.
With these data one can solve the modular differential
equation to find the characters. 
The result is up to base changes the same.

\bq
The modular differential
equation is a condition which must be satisfied by any finite dimensional
representation of the modular group in terms of forms. One introduces the
so-called modular covariant derivation ${\rm cod}$,
$$
  {\rm cod}_{(s)} = \frac{1}{2\pi{\rm i}}\partial_{\tau} 
  - \frac{1}{12} s G_2(\tau) \,,
$$
which increases the weight of a modular form by two. Here, $G_2$ denotes
the second Eisenstein series (which is {\em not\/} a modular function). 
Using the abbreviation
$$
  D^i = {\rm cod}_{(2i-2)}\ldots{\rm cod}_{(2)}{\rm cod}_{(0)}\,,
$$
and some reasonable assumptions on the asymptotics of characters,
the modular differential equation for an $n$-dimensional representation
of the modular group takes the simple form
$$
  \sum_{k=0}^{n} \Phi_{n(n+1)-d-2k} D^k \chi_i = 0\,,\ \ 1\leq i\leq n\,,
$$
where $d=12(\sum_{i=1}^{n} h(i) - n c/24)$. Thus, the equation depends
on the conformal data, i.e.\ the central charge $c$ and all conformal
weights $h(i)$ of the $n$ representations. 
The coefficients $\Phi_n$ must be entire modular functions, i.e.\
must be given in terms of Eisenstein series,
$$
  \Phi_n = \sum_{{k,l\in\BZ_+\atop 4k+6l=n}} a_{n,l}(G_4)^k (G_6)^l\,. 
$$
Since there is no modular function of weight two, there can be no $\Phi_2$,
and for completeness one defines $\Phi_0\equiv a_{0,0}$ to be a constant.
Usually, the equation can used to infer the power series expansion of unknown
characters, if the central charge, all conformal weights, and at least
one character asymptotics are known. However, if it happens that
two characters have the same weight, $h(i)=h(j)$ for some pair $i\neq j$,
the equation degenerates (as every differential equation does), and
pure power series ans\"atze are not sufficient to get all solutions.

A more precise exposition of this technique and all the assumptions
one has to make on the characters is, unfortunately, beyond
the scope of these notes.
\eq
  
We would like to recall that one can formally read off the possible
representations from the conformal grid of minimal models in the
following way: The possible $h$-values of a minimal model with
$c = c_{p,q}$ are given by $h_{r,s} = \frac{(pr-qs)^2-(p-q)^2}{4pq}$
with $1\leq r<q$ and $1\leq s<p$. One obtains the $h$-values for a
$c_{p,1}$-model including all inequivalent representations to the same
highest weight from the conformal grid of $c_{3p,3}$.
  
For simplicity, we concentrate now on the case $c=-2$, i.e.\ $p=2$.
We first assume the usual form of the characters,
\begin{equation}
   \chi_i = q^{h_i-c/24}\sum_{l=0}^{\infty} b_{i,l}q^l\,,
\end{equation}
where $h_i$ is given by $h_{1,i} = \frac{i^2-2ip+2p-1}{4p}$.
Solving the modular differential equation yields up to multiplicative
prefactors the characters
\begin{equation}\begin{array}{rcl}
  \chi_1 & = & A\Lambda_{1,2} + B(\partial\Lambda)_{1,2}\,,\\
  \chi_2 & = & \Lambda_{0,2}\,,\\
  \chi_3 & = & A'\Lambda_{1,2} + B'(\partial\Lambda)_{1,2}\,,\\
  \chi_4 & = & \Lambda_{2,2}\,,\\
  \chi_5 & = & \frac{1}{2}\Lambda_{1,2} - \frac{1}{2}(\partial\Lambda)_{1,2}\,.
\end{array}\end{equation}
Therefore, $\chi_1$, $\chi_3$ and $\chi_5$ are linear dependent.
If $\chi_1$ is supposed to belong to the vacuum representation, its
coefficient to $q$ must vanish, i.e.\ $b_{1,1} = 0$. This forces
$A = B = 1/2$, if one also requires $b_{1,0} = 1$ (which essentially
means that the vacuum state is not degenerate).
  
We now need one further, linear independent solution. We make the ansatz
\begin{equation}
  \tilde{\chi}_3 = \log(q)q^{1/12}\sum_{l=0}^{\infty}\tilde{b}_{3,l}q^l\,.
\end{equation}
Inserting this into the modular differential equation, we get
\begin{equation}
  \tilde{\chi}_3 = (\nabla\Lambda)_{1,2}\,,
\end{equation}
where we define the characters as functions in $q$, i.e.\
$(\nabla\Lambda)_{\lambda,k} \equiv -\frac{\log(q)}{2\pi{\rm i}}
(\partial\Lambda)_{\lambda,k}$. Please note that we always mean by $\log(q)$
the branch of the logarithm given by $2\pi{\rm i}\tau$.
Indeed, our result is exactly the same as what we got from the explicit
calculation of the vacuum character and its $S$-transformation.

We collect our intermediate results:
  The LCFTs with
  $c = c_{p,1} = 13 - 6(p + p^{-1})$ and chiral symmetry
  algebra $\w(2,2p-1,2p-1,2p-1)$ admit
  precisely $3p-1$ HWRs with highest weights
  $h_{1,s}, 1\leq s\leq 3p-1$. Of them $2\cdot (p-1)$ HWRs have
  pairwise identical highest weights, further $p-1$ highest weights
  differ from these pairs by positive integers which are the levels of
  the corresponding singular vectors. A basis for the characters is given by
  ($\eta^{-1}$ times) the functions $\{\Theta_{\lambda,p},
  (\partial\Theta)_{\mu,p},(\nabla\Theta)_{\mu,p} |
  0\leq\lambda,\mu\leq(2p-1), \mu\neq 0,p \}$.
To distinguish the representations with identical conformal weights $h$,
we denote one of them a $[h]$, the other as $[\tilde h]$.
The $S$-matrix has determinant one and satisfies $S^2=\Bid$, which
one may expect, since $t\mapsto -1/\tau$ is an involution.
We already noted that the functions
$(\nabla\Theta)_{\mu,p}$ lead to a non diagonal
$T$-matrix. It decomposes into blocks similar to Jordan cells, but
which also mix characters whose corresponding highest weights differ by
integers. 
Nonetheless, this matrix satisfies together with the $S$-matrix
the relation $(ST)^3 = \Bid$. This condition is very important in
order to have modular invariance of the CFT, and resembles the
associativity condition of the OPE. 

But what are the ``physical'' characters? Note that due to the fact that
many conformal weights differ only by integers, the characters are only
determined upto linear combinations among such characters whose
formal expansions have the same fractional overall power modulo one.
The question will be answered
by enforcing modular invariance of the partition function. From
our discussion of the modular properties of the characters we know
that the following expression is modular invariant:
\begin{equation}\label{eq:zlog}
  Z_{{\rm log}}[p] = \sum_{\lambda=0}^{2p-1}|\Theta_{\lambda,p}|^2
                   + \alpha\sum_{{\mu=1\atop\mu\neq p}}^{2p-1}\frac{1}{2}\left(
                     (\partial\Theta)_{\mu,p}(\nabla\Theta)^*_{\mu,p} +
                     (\nabla\Theta)_{\mu,p}(\partial\Theta)^*_{\mu,p}
                     \right)\,,
\end{equation}
where $\alpha$ is a free constant. The normalization of the part of
$Z_{{\rm log}}[p]$ independent of $\alpha$ results from the requirement
that its expansion must have integer coefficients only to be
physical relevant. Furthermore, the coefficient yielding the multiplicity
of the ground state should be as small as possible, preferable one.
Note that we cannot impose such a condition on $\alpha$ since
this part of the partition function is not a power series in $q,\bar q$ which
could be interpreted as yielding multiplicities of states in Verma modules.

\bq
The task of finding the linear combinations which yield the physical correct
characters is not trivial. Even in the simplest $c_{p,1}$ model, the $c=-2$
theory, we only know two characters for sure, namely the characters for
the twist field sectors $[-\frac18]$ and $[\frac38]$. The reason that we
know these is simply that the functions $(\del\Theta)_{\lambda,k}$ and
consequently also $(\nabla\Theta)_{\lambda,k}$ vanish for $\lambda=0,k$.
That leaves us only with $\theta_{0,k}$ and $\theta_{k,k}$. The same holds
for all $c_{p,1}$ models, meaning that only the two characters to the
sectors $[h_{1,p}]$ and $[h_{1,2p}]$ can be
fixed a priori. All the other
sectors of the $c=-2$ theory have conformal weights differing by integers,
allowing arbitrary linear combinations among the functions
$\theta_{\lambda,k},(\del\Theta)_{\lambda,k},(\nabla\Theta)_{\lambda,k}$ for
fixed $\lambda$.

One needs some knowledge about the different representations in the
$c_{p,1}$ models. We know that all sectors besides the two twist sectors
come in triplets. The following picture within the Kac table emerges
$$
  \epsfig{file=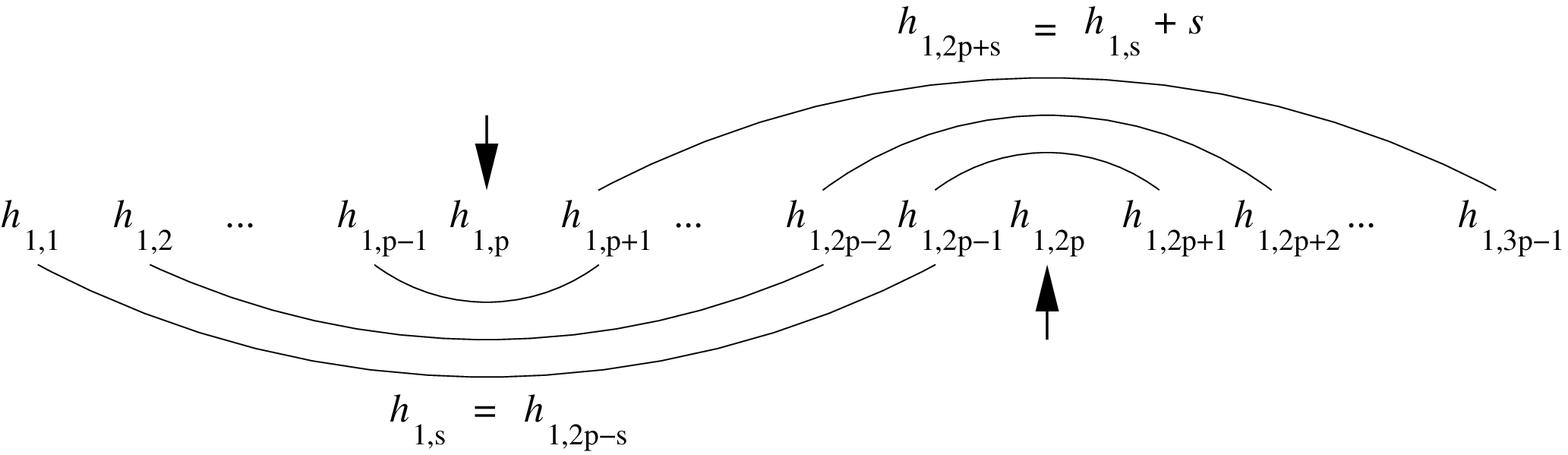,width=10cm}\,.
$$
We will denote the characters for the representations $[h_{1,s}]$, 
$1\leq s < p$, as $\chi^+_{s,p}$. These shall be the characters to the
irreducible sub-representations contained in the Jordan blocks.
The characters to the representations $[h_{1,2p+s}]$, $1\leq s<p$, are called
$\chi^-_{s,p}\equiv \chi_{-s,p}$. Note that $h_{1,2p+s}=h_{1,s}+s$.
Finally, we introduce for the remaining set of representations $[h_{1,2p-s}]$,
$1\leq s <p$, the characters $\tilde\chi_{s,p}$. We know for these LCFTs
that the representations of the triplets $(h_{1,s}=h_{1,2p-s},h_{1,2p+s})$ are 
linked with each other \cite{Gurarie:1993,Rohsiepe:1996}. They come from
one Jordan block built from a primary state $|h_{1,s}\ket$ and its
logarithmic partner $|h_{1,2p-s}\ket$, and another indecomposable 
block built on a highest-weight state $|h_{1,2p+s}\ket$, which also
contains a logarithmic part. However, we don't find something for it
in the Kac table. On the other hand, the logarithmic state of this
second module is not independent from the logarithmic state of the former
Jordan block.

A deeper analysis reveals that the characters $\tilde\chi_{s,p}$ should
be split into two parts, $\sqrt{2}\tilde\chi_{s,p}=\tilde\chi^+_{s,p}
+\tilde\chi^-_{s,p}$. The so-called quantum dimension of the original
$\tilde\chi_{s,p}$ character is zero, being the sum of the quantum 
dimensions of the split characters $\tilde\chi^{\pm}_{s,p}$.
This procedure is well known in the theory of quantum groups. Whenever
the quantum deformation parameters become roots of unity, additional
so-called exceptional representations appear in pairs, whose quantum
dimensions add up to zero \cite{GoSi91}. In fact, every rational CFT has an
underlying quantum group structure, and as it happens, one of the
corresponding quantum deformation parameters becomes precisely
$1=\exp(2\pi{\rm i}(p/q))$ for the $c_{p,q}$ minimal models with $q=1$.
It can be shown that this exceptional quantum group structure manifests
itself in the CFT itself, and suggests the above mentioned split of
characters. 

Armed with these rather involved results from more advanced algebraic
insights, we make the following general ansatz for all characters:
\bea\label{eq:lincomb}
  \chi^{}_{\lambda,p} &=& \frac{1}{\eta}\left[
  \alpha_{\lambda,p}\Theta_{\lambda,p} +
  \beta_{\lambda,p}(\partial\Theta)_{\lambda,p} +
  \gamma_{\lambda,p}(\nabla\Theta)_{\lambda,p}\right]\,,\ \ \ \
  -p<\lambda\leq p
  \,,\\ \label{eq:lincombtilde}
  \tilde{\chi}^{\pm}_{\mu,p} &=& \frac{1}{\eta}\left[
  \alpha^{\pm}_{\mu,p}\Theta_{\mu,p} +
  \beta^{\pm}_{\mu,p}(\partial\Theta)_{\mu,p} +
  \gamma^{\pm}_{\mu,p}(\nabla\Theta)_{\mu,p}\right]\,,\ \ \ \
  1\leq\lambda<p\,.
\eea
The first set also includes the characters for the two twist
representations, since the affine $\Theta$-functions vanish for $\lambda=0,p$. 
One may now write down an arbitrary sesqui-linear
combination of these characters and check whether it takes the form
of the partition function (\ref{eq:zlog}) for some $\alpha$.
This determines solutions for the coefficients $\alpha_{\lambda,p},
\alpha^{\pm}_{\lambda,p},\ldots$.

We also mention that a completely different approach is possible, and will
be presented in the lectures by Matthias Gaberdiel. In this approach,
one does not consider characters for the irreducible sub-representation of 
the indecomposable representations separately. This avoids many of the
difficulties with $\log(q)$ terms, singular $S$-matrices and fusion rules
with negative coefficients. However, it looses all information on
the inner structure of the indecomposable representations. This is the
reason why we stick to our approach, despite its many difficulties.
\eq
  
Since the complete deduction of the correct physical base of
characters is quite lengthy and involved, we can only quote the result here.
The industrious reader might try to recover it by making a suitable 
ansatz (as described in the above small print) and then ensuring all
conditions from physical requirement. These conditions are, for example, that
the parts of characters, which are pure power series, must have
non-negative integer coefficients, that the sesqui-linear combination must
reduce to the form (\ref{eq:zlog}) for some $\alpha$, that the sesqui-linear
combination may only combine characters with each other, whose fractional
overall power are congruent modulo one\footnote{If two characters are
given as $\chi=q^{\rho}[\sum_{n\in\mathbb{Z}} a_nq^n+\log(q)(\ldots)]$, 
$\chi'=q^{\sigma}[\sum_{n\in\mathbb{Z}}b_nq^n+\log(q)(\ldots)]$ with
$\rho,\sigma\in\mathbb{Q}$, they are congruent modulo one, if
$\rho=\sigma+\ell$ for an integer $\ell$.}, that the sesqui-linear
form must have non-negative integer coefficients only, and so on. We finally
note that we did compute the character for the $SL(2,\mathbb{C})$ invariant
vacuum representation explicitly in (\ref{eq:chi-vac}), which may be used
as additional input fixing one further character a priori. One can,
in principle, also compute the characters of other representations from
first principles, and it is often helpful to do so for the first few
levels. This has been accomplished in \cite{Gaberdiel:1996a,Kausch:1995,
Rohsiepe:1996}, and the results suggest the following general form of
the characters:
\bea\label{chi:1}
  \chi^{}_{0,p}      & = & \frac{1}{\eta}\Theta_{0,p}\,,\\
  \chi^{}_{p,p}      & = & \frac{1}{\eta}\Theta_{p,p}\,,\\
  \chi^+_{\lambda,p} & = & \frac{1}{p\eta}\left[(p-\lambda)\Theta_{\lambda,p}
                       + (\partial\Theta)_{\lambda,p}\right]\,,\\
  \chi^-_{\lambda,p} & = & \frac{1}{p\eta}\left[\lambda\Theta_{\lambda,p}
                       - (\partial\Theta)_{\lambda,p}\right]\,,\\
  \tilde{\chi}^+_{\lambda,p} & = & \frac{1}{\eta}\left[\Theta_{\lambda,p}
                  + {\rm i}\alpha\lambda(\nabla\Theta)_{\lambda,p}\right]\,,\\
  \label{chi:6}
  \tilde{\chi}^-_{\lambda,p} & = & \frac{1}{\eta}\left[\Theta_{\lambda,p}
                  - {\rm i}\alpha(p-\lambda)(\nabla\Theta)_{\lambda,p}\right]
                       \,,
\eea
where $0<\lambda<p$. The conformal weights are (in the same order)
$h(p,1)_{1,p}$, $h(p,1)_{1,2p}$, $h(p,1)_{p-\lambda}$, $h(p,1)_{3p-\lambda}$,
and $h(p,1)_{p+\lambda}$. The last set refers to both $\tilde\chi^{\pm}_{
\lambda,p}$ characters together, which incorporate the effect of the
logarithmic operators. 
All the ``non-tilde'' characters are free of any
$\log(q)$ terms and have an immediate physical interpretation.
They are the characters of irreducible representations $[h_{\lambda,p}]$
The $\tilde\chi$ characters cannot be considered as characters of
representations in the usual sense, but we will denote the
corresponding modules which could be associated to the power series part of the
$\tilde\chi$ characters by $[\tilde h_{\lambda,p}]$.
One easily sees that the partition function
\bea
  \lefteqn{Z_{{\rm log}}[p,\alpha]
    = |\chi^{}_{0,p}|^2 + |\chi^{}_{p,p}|^2
    + \sum_{\lambda=1}^p\left[
    \chi^+_{\lambda,p}{\tilde{\chi}^{+^{\scriptstyle *}}_{\lambda,p}} +
    {\chi^{+^{\scriptstyle *}}_{\lambda,p}}\tilde{\chi}^+_{\lambda,p} +
    \chi^-_{\lambda,p}{\tilde{\chi}^{-^{\scriptstyle *}}_{\lambda,p}} +
    {\chi^{-^{\scriptstyle *}}_{\lambda,p}}\tilde{\chi}^-_{\lambda,p}\right]
  }\\
  & = &\frac{1}{\eta\eta^*}\left\{|\Theta_{0,p}|^2 + |\Theta_{p,p}|^2 +
    \sum_{\lambda=1}^{p} \left[ 2|\Theta_{\lambda,p}|^2
    +{\rm i}\alpha\left((\partial\Theta)_{\lambda,p}(\nabla\Theta)_{\lambda,p}^*
    -(\partial\Theta)_{\lambda,p}^*(\nabla\Theta)_{\lambda,p}\right)
    \right]\right\}\nonumber
\eea
is modular invariant for all $\alpha\in\BR$ and coincides with
(\ref{eq:zlog}). Notice the important fact that
the partition function remains modular invariant even for $\alpha=0$ and
then equals the standard $c=1$ Gaussian model partition function
$Z(\sqrt{p/2})$. This in particular means that we have a modular invariant
partition function even in the case where the characters do not form a
closed finite dimensional representation of the modular group by themselves.

However, if we wish to compute an $S$-matrix from this set of characters,
we run into the problem that the full set is not linearly independent.
In order to find the $S$-matrix we have to forget about the split of
the $\tilde\chi$ characters. Thus, we choose as linear independent set
$\{\chi^{}_{0,p},\chi^{}_{p,p},\chi^{\pm}_{\lambda,p},
[(p+x-\lambda)\tilde{\chi}^+_{\lambda,p}+
(\lambda-x)\tilde{\chi}^-_{\lambda,p}]\}$. The resulting $S$-matrix, and
therefore also the fusion rules \cite{Ver88}, 
depend on the value of $\alpha$. Clearly,
the $S$ matrix becomes singular for $\alpha\rightarrow 0$, but it turns out
that the fusion
rules remain well defined. The limes $\alpha\rightarrow 0$ just puts several
of the fusion coefficients to zero. One can show that in general only the
fusion rules in the limit $\alpha\rightarrow 0$ are consistent and integer
valued.

As an example let us again consider the case $c=c_{2,1}=-2$. The $S$ matrix
reads
\be
  S_{(2,\alpha)} = \left(\begin{array}{ccccc}
    \frac{1}{2\alpha} & \frac{1}{4} & \frac{1}{2\alpha} & -\frac{1}{4} &
      -\frac{1}{4\alpha} \\
    1 & \frac{1}{2} & 1 & \frac{1}{2} & 0 \\
    -\frac{1}{2\alpha} & \frac{1}{4} & -\frac{1}{2\alpha} & -\frac{1}{4} &
      \frac{1}{4\alpha} \\
    -1 & \frac{1}{2} & -1 & \frac{1}{2} & 0 \\
    -2\alpha & 1 & 2\alpha & -1 & 0
  \end{array}\right)\,.
\ee
The $S$-matrix $S_{(p,\alpha)}$ in general is neither symmetric nor unitary,
which is a remarkable difference to the
case of generic non-logarithmic rational CFT.
But at least it fulfills $S_{(p,\alpha)}^2 = \Bid$. 
The general expression for the
$S$-matrix is cumbersome, but can easily be obtained from the explicit form
of the characters in terms of $\Theta$, $(\del\Theta)$, and $(\nabla\Theta)$
functions (cf.\ eqs.\ \ref{chi:1}--\ref{chi:6})
and the known modular transformation behavior of the latter (cf.\ eqs.\
\ref{eq:theta}, \ref{eq:dtheta}, and \ref{eq:nlambda}).

In order to compute the fusion rules, we will use a modification of the
Verlinde formula (\ref{eq:verlinde-formula}). Since $S_{(p,\alpha)}$ depends
on $\alpha$ in a continuous way, we cannot expect that the fusion coefficients
are independent of $\alpha$. On the other hand, the characters don't have a
physical evident meaning, as long as $\log(q)$ terms are present which,
however, are necessary to get a closed finite-dimensional representation
under the modular group. Thus, we define fusion coefficients
$N_{ij}^{\ k}(\alpha)$ according to the Verlinde formula with respect to
the $S$-matrix $S_{(p,\alpha)}$, and then take the ``physical'' limit
\be
  N_{ij}^{\ k} = \lim_{\alpha\rightarrow 0}N_{ij}^{\ k}(\alpha)\,.
\ee
So far, so good. This procedure yields integer valued fusion rules, but
unfortunately sometimes negative signs. Moreover, since the set of
characters becomes linearly dependent in the limit $\alpha\rightarrow 0$,
the right hand side of the fusion rules is not necessarily uniquely
determined. Care must be taken in interpreting the right hand sides,
since we may have character identities which translate to relations among the
representations. For example, we have for $\alpha=0$ at least that
$2[h(p,1)_{1,p-\lambda}]+2[h(p,1)_{1,3p-\lambda}] = [h(p,1)_{1,p+\lambda}]$,
$0<\lambda<p$, e.g.\ the relation $[\tilde 0] = 2[0] + 2[1]$ in
the $c=-2$ theory.
The fusion rules now read
\be
  \begin{array}{rclcl}
    {}           [0]&   *  &[\Phi]        &=&[\Phi]                      \,,\\
    {}[-\frac{1}{8}]&   *  &[-\frac{1}{8}]&=&2[0]+2[1] = [\tilde{0}]     \,,\\
    {}[-\frac{1}{8}]&   *  &[\frac{3}{8}] &=&2[0]+2[1] = [\tilde{0}]     \,,\\
    {}[-\frac{1}{8}]&   *  &[1]           &=&[\frac{3}{8}]               \,,\\
    {}[-\frac{1}{8}]&   *  &[\tilde{0}]   &=&2[-\frac{1}{8}]+2[\frac{3}{8}]
                                             \,,\\
    {} [\frac{3}{8}]&   *  &[\frac{3}{8}] &=&2[0]+2[1] = [\tilde{0}]     \,,
  \end{array}\ \ \ \
  \begin{array}{rclcl}
    {} [\frac{3}{8}]&   *  &[1]           &=&[-\frac{1}{8}]              \,,\\
    {} [\frac{3}{8}]&   *  &[\tilde{0}]   &=&2[-\frac{1}{8}]+2[\frac{3}{8}]
                                             \,,\\
    {}           [1]&   *  &[1]           &=&[0]                         \,,\\
    {}           [1]&   *  &[\tilde{0}]   &=&4[0]+4[1]-[\tilde{0}]
                                             = [\tilde{0}]               \,,\\
    {}   [\tilde{0}]&   *  &[\tilde{0}]   &=&8[0]+8[1] = 4[\tilde{0}]    \,.
  \end{array}
\ee
We see that one negative sign occurs. Not accidentally, it happens where
the only ``representation'' $[\tilde 0]$ whose character 
has a $\log(q)$ term appears on both sides. Recall that for $\alpha=0$, 
the characters $\tilde\chi^{\pm}_{1,2}$ coincide. 
We can reintroduce the split representations $[\tilde{h}^{\pm}]$ back
into the fusion rules by hand. This yields the following modifications
\be\label{eq:gaka}
  \begin{array}{rclcl}
    {}[-\frac{1}{8}]&   *  &[-\frac{1}{8}]      &=&[\tilde{0}^+]     \,,\\
    {}[-\frac{1}{8}]&   *  &[\frac{3}{8}]       &=&[\tilde{0}^-]     \,,\\
    {} [\frac{3}{8}]&   *  &[\frac{3}{8}]       &=&[\tilde{0}^+]     \,,
  \end{array}\ \ \ \
  \begin{array}{rclcl}
    {}           [1]&   *  &[\tilde{0}^{\pm}]   &=&[\tilde{0}^{\mp}] \,,\\
    {}[\tilde{0}^{\pm}]&   *  &[\tilde{0}^{\pm}]&=&2[\tilde{0}^+]
                                                   +2[\tilde{0}^-]   \,,\\
    {}[\tilde{0}^{\pm}]&   *  &[\tilde{0}^{\mp}]&=&2[\tilde{0}^+]
                                                   +2[\tilde{0}^-]   \,.
  \end{array}
\ee
To see this, one has to make an ansatz where each occurence of $[\tilde 0]$
is replaced by either $[\tilde 0^+]$ or $[\tilde 0^-]$ (if $[\tilde 0]$
appears with multiplicity $n>1$, then this is to be replaced by
$n_+[\tilde 0^+]+n_-[\tilde 0^-]$ with $n_-+n_-=n$), and check that
associativity of the fusion rules is satisfied. It makes sense to consider
$[\tilde 0^+]$ to stand for the complete indecomposable representation
for $h=0$, while $[\tilde 0^-]$ stands for the indecomposable module
with the $h=1$ primary. As explained in \cite{Gaberdiel:1996a}, these
two modules are equivalent, which nicely agress with the fact that
our characters $\tilde\chi^{\pm}_{1,2}$ coincide for $\alpha=0$.
Although it is cumbersome to insert the split representations by hand,
our approach has the advantage to be able to distinguish between the
two equivalent indecomposabel representations. This holds also true in the
general $c_{p,1}$ case, where this has to be applied to all the
triplets $([h(p,1)_{1,r}],
[\tilde h(p,1)_{1,2p-r}],[h(p,1)_{1,2p+r}])$.

\subsection{Moduli space of $c_{p,1}$ LCFTs}

We have seen that the modular invariant partition function 
$Z_{{\rm log}}[p,\alpha]$ of a $c_{p,1}$
model (\ref{eq:zlog}) has a part which is independent of $\alpha$.
The approach of \cite{Gaberdiel:1996a} yields precisely this part, i.e.\
concides with our approach for $\alpha=0$.
This part is well known, it is nothing else than the partition function
of the CFT of a single free boson compactified on a circle with radius
$R=\sqrt{p/2}$. It is customary to denote the free boson partition function
by $Z(R)$, but we will chose the slightly different notation
$Z[2R^2]$. The benefit of this will become clear below. For completeness,
we note that
$$
  Z(R)\equiv Z[2R^2] = \frac{1}{\eta\bar\eta}\sum_{m,n\in\mathbb Z}
  q^{\frac12(mR+\frac12n/R)^2}\bar q^{\frac12(mR-\frac12n/R)^2}\,,
$$
which for $2R^2=p/q\in\mathbb Q$ may be expressed as $(\eta\bar\eta)^{-1}$ 
times a sesqui-linear form in $\theta_{\lambda,pq}$ functions. The
partition function $Z(R)$ is also called the standard Gaussian $U(1)$
partition function.

The case $p=1$ is trivial, there are no logarithmic representations. It is
just $\widehat{\euf{su}(2)}$, the simplest non-abelian
infinite dimensional Lie algebra $A_1^{(1)}$, with $c=1$. In particular,
$Z_{{\log}}[1] = Z[1]$.
This means that our logarithmic
CFT reduces to the Gauss
model at the multi-critical point of radius $1/\sqrt{2}$.
But R.~Dijkgraaf and E.~\&~H.~Verlinde \cite{DVV88} have proven that
there are {\em no\/} marginal deformations, which can lead out of the
known moduli space of $c=1$ CFTs. There is one field of marginal
dimension, $\phi_{2,p-1}$ with $h_{2,p-1} = 1$, which belongs to the
(extended) conformal grid of section 1. Since the first label is even,
it has vanishing self coupling, which is necessary for a marginal
operator to be integrable. But this field does not exist for $p=1$,
since all fields $\phi_{r,s}$
with $r=0$ or $s=0$ decouple completely from the physical Hilbert space
due to annihilation by the BRST operator. Thus, we indeed cannot go from
the moduli space of regular $c=1$ CFTs to the logarithmic CFTs with
$c_{{\rm eff}}=1$ via marginal deformations.\footnote{The so-called
{\em effective\/} central charge is defined as 
$c_{{\rm eff}}=c-24h_{{\rm min}}$ for a rational CFT. Here 
$h_{{\rm min}} = {\rm min}\{h | h\in{\cal R}\}$ is the minimal 
eigenvalue of $L_0$. For unitary theories, $c_{{\rm eff}}= c$. For non-unitary
theories, where $c\leq 0$, is $c_{{\rm eff}}$ always $\geq 0$. The $c_{p,1}$
models all have $c_{{\rm eff}}=1$.} 
If we finally note that the
partition function (\ref{eq:zlog}) also allows non diagonal
decompositions, we have the following statement:

The moduli space of logarithmic CFTs with
  $c_{{\rm eff}} = 1$ is generic one dimensional and not connected to
  the moduli space of regular $c=1$ CFTs.
  The partition function of a logarithmic CFT is for $(p,q)=1$ given by
  \bea
    Z_{{\rm log}}[p/q] &=& 
    \frac{1}{\eta\bar{\eta}}\left[|\chi_{0,pq}|^2
                       + |\chi_{pq,pq}|^2\phantom{
                         \sum_{1\leq s\leq pq-1}}\right.\\
                      &+&\left.\sum_{1\leq s\leq pq-1}\left(
                         {\chi_{s,pq}^+}{\tilde\chi_{s',pq}^{+^*}}
                       + {\chi_{s,pq}^{+^*}}{\tilde\chi_{s',pq}^+}
                       + {\chi_{s,pq}^-}{\tilde\chi_{s',pq}^{-^*}}
                       + {\chi_{s,pq}^{-^*}}{\tilde\chi_{s',pq}^-}
                         \right)\right]\,,\nonumber
  \eea
  where $s=pn-qm$ mod $2pq$ implies $s'=pn+qm$ mod $2pq$.
  
The connected part of the moduli space of $c=1$ theories has an exact
copy of logarithmic theories in the following manner: First, one writes
\begin{equation}
  Z_{{\rm log}}[x] = \left(1 + \frac{2x^2}{\pi{\rm i}}
  \frac{\partial}{\partial x}\right)Z[x]\,,
\end{equation}
which by the way defines $Z_{{\rm log}}[x]$ for arbitrary, not
necessarily rational $x$. In the same way we obtain the partition
function of the $\BZ_2$-orbifolds of the logarithmic theories by
applying $(1+\frac{2x^2}{\pi i}\partial_x)$ to $Z_{{\rm orb}}[x]$,
\begin{equation}
  Z_{{\rm log,orb}}[x] = \left[
  \left(1+\frac{2x^2}{\pi{\rm i}}\frac{\partial}{\partial x}\right)Z[x]
  %+ 2Z[4] - Z[1])/2\,.
  + \left.\left(1+\frac{2y^2}{\pi{\rm i}}\frac{\partial}{\partial y}
  \right)Z[y]\right|_{y=4} - Z[1]\right]/2\,.
\end{equation}
The corresponding $\w$-algebras, which exist at points of enhanced
symmetry analogous to the regular case, are the following:
To $Z_{{\rm log}}[p], p\in\BN$ belongs a $\w(2,(2p-1)^{\otimes 3})$, whose
$\BZ_2$-orbifold contains a $\w(2,6p-2)$, the $\BZ_2$-orbifold of
$\w(2,2p-1)$ where the singlet field is given by $W = W_0 + W_+ + W_-$ and
the orbifold is obtained by identifying $W$ with $-W$. Since the structure
constant $C_{WW}^W$ does not vanish for the triplet, the $\BZ_2$-orbifold
of the triplet should be given by the identifications
$W_0\leftrightarrow -W_0$, $W_+\leftrightarrow -W_-$, and
$W_-\leftrightarrow -W_+$ such that one field, e.g.\ $\tilde W =
W_+ - W_-$ survives. The orbifold would then be a $\w(2,2p-1,6p-2)$.
If $p$ is a complete square,
$p = n^2$, these algebras can be extended by a field of dimension
$h_{2n+1,1} = p(n^2+n)-n = n^4 + n^3 - n$.
%This must remain true
%for $p\neq n^2$, since $\w(2,6p-2)$ is not maximally extended in the way
%that it yields a (logarithmic) rational CFT \cite{EHH93b}.
In the same manner
one can write down the logarithmic analogs of the three exceptional
$c=1$ partition functions. Setting $D_x=\frac{2x^2}{\pi{\rm i}}\partial_x$,
the exceptional logarithmic partition functions simply read
\begin{eqnarray}
  Z_{{\rm log},E_6} &=& \frac{1}{2}\left(\sum_{x\in\{4,9,9\}}(1+D_x)Z[x]
    -Z[1]\right)\,,\\
  Z_{{\rm log},E_7} &=& \frac{1}{2}\left(\sum_{x\in\{4,9,16\}}(1+D_x)Z[x]
    -Z[1]\right)\,,\\
  Z_{{\rm log},E_8} &=& \frac{1}{2}\left(\sum_{x\in\{4,9,25\}}(1+D_x)Z[x]
    -Z[1]\right)\,.
\end{eqnarray}
In this way, the full $c=1$ moduli space is
recovered in the ``logarithmic'' regime. There are no other linear
combinations possible, since the non-logarithmic part of the partition
function has to satisfy the usual requirements to be physical relevant,
which only yield the known $c=1$ solutions.
  \par
Of course, there could be higher powers of logarithmic terms.
All expressions of the form $(\sum_{n\in\BZ_+}a_n D^n_x)Z[x]$ are
modular invariant. Fortunately, as mentioned above, this presumably cannot
happen for theories with $c_{{\rm eff}} \leq 1$ (see also last ref.\ in
\cite{Flo93}).
  \par
We conclude with a remark on $N=1$ supersymmetric theories. The explicit
known examples as well as the general results on
the modular properties of characters make it clear that $N=1$ CFTs will
have the same structure. One finds again logarithmic theories
(with $c_{{\rm eff}} = 3/2$), which have a completely analogous
representation theory. This analogy extends the similarity of the
representation
theory of the already known $N=0,1$ rational CFTs \cite{Flo93}.
Some works dealing with $N=1$ supersymmetric LCFTs are \cite{
Kheirandish:2001a,
Khorrami:1998a,
Mavromatos:1999a}
But as already observed in other cases, such results do not extend to
$N=2$, since there no rational like structure can be found for non-unitary
theories.
It remains the conjecture that for
$N=2$ rationality of a CFT implies its unitarity.

\section{Conclusion}

These notes by no means provide a comprehensive introduction to the
vast theme of logarithmic conformal field theory. 
Many topics of great importance
have been skipped completely, or mentioned only in a half-sentence.
In particular, a thorough and mathematical rigorous discussion of the
algebraic aspects of LCFT is given in the lectures by Matthias Gaberdiel.
We did not mention anything about boundary states in LCFTs, since these
are discussed in the lectures by Y.~Ishimoto and S.~Kawai. Many other issues
such as the logarithmic partners of the stress energy tensor as well as
LCFT of current algebras, presented by Alex Nichols, or applications
such as disorder, the topic of Reza Rahimi Tabar's lectures, are left out
here. Notes of the other lectures are to appear on the web as well, and
we encourage the (still) interested reader, to consult these for further
information on the young and exciting field of LCFT.

These notes pretty much consist of the material presented in the actual
lectures, which were mainly designed to address an audience, which not only
was new to the subject, but which also did not have experience with
ordinary common conformal field theory.
Therefore, the selection of covered material was made along the lines of
this course. 
The nature of the course, to provide a preliminary survey on logarithmic
conformal field theory as well as a basic introduction to some parts of 
standard conformal field theory, is reflected in the incompleteness
of these notes. Moreover, since LCFT is still a field in its infancy,
there are still many open topics. Of course, these notes often reflect
foremost the authors point of view, in particular concerning such not yet
fully understood issues. Here, and also with regard to the bibliography,
the author apologizes for any omissions made, and there certainly are many.
The bibliography might help the reader to find some
more comprehensive and detailed works on the topics touched upon or covered
by these notes. Again, also the
bibliography does not attempt to be thorough in any sense, but 
is intended to list easily accessible papers on logarithmic conformal
field theories as well as some of its applications. Fortunately, since
this is a young topic, most of the papers can be found on the arXive
servers.
A few papers on particularly important aspects of and results in general
conformal field theory, especially those needed in some of the arguing in
our text, have been listed for completeness.

\bigskip
{\sc Acknowledgments:} There are many colleagues with whom I had 
stimulating discussions on
logarithmic conformal field theory and related topics throughout
the last years. In particular I would like to thank Matthias Gaberdiel,
Ian Kogan and Phillipe Ruelle for numerous discussions and comments.
The author heartily thanks the local organizers, staff and participants
of the first
{\em School and Workshop on Logarithmic Conformal Field Theory and
its Applications\/} at the IPM in Tehran, Iran, for their warm 
hospitality, for the wonderful atmosphere at the school and workshop,
and for the deeply fascinating experience of their home country.
My deepest gratitude goes in particular to Shahin Rouhani, without
whom this event would never have become reality, as well as to M.~Reza Rahimi
Tabar and Shirin Davapanah.

\end{document}